\documentclass[reqno,12pt]{amsart}

\usepackage{fancyhdr}
\usepackage{datetime}
\settimeformat{ampmtime}
\usepackage[svgnames]{xcolor}
\usepackage{setspace}
\usepackage{pgf}
\usepackage{tikz}
\usepackage{mathpazo} 
\usepackage[scaled]{helvet} 
\usepackage{courier} 
\normalfont
\usepackage[T1]{fontenc}

\usepackage{pdflscape}
\usetikzlibrary{patterns}
\usepackage{soul}
\usepackage{mathrsfs}
\usepackage{stmaryrd}
\usepackage[hidelinks=true]{hyperref}
\usepackage{natbib}
\usepackage{xfrac}
\usepackage[color=Orange]{todonotes}
\usepackage{layout}
\usepackage{microtype}
\usepackage{fancyhdr}
\usepackage{mathtools}

\newtheorem{lemma}{Lemma}

\newtheorem{theorem}{Theorem}

\usepackage{cleveref}

\crefname{figure}{figure}{figures}
\creflabelformat{figure}{#2#1#3}
\crefname{equation}{equation}{equations}
\creflabelformat{equation}{#2(#1)#3}
\crefname{lemma}{lemma}{lemmas}
\creflabelformat{lemma}{#2#1#3}
\crefname{theorem}{theorem}{theorems}
\creflabelformat{lemma}{#2#1#3}
\crefname{condition}{condition}{conditions}
\creflabelformat{condition}{#2#1#3}
\crefname{assumption}{assumption}{assumptions}
\creflabelformat{assumption}{#2#1#3}
\crefname{appendix}{appendix}{appendices}
\creflabelformat{appendix}{#2#1#3}
\crefname{enumi}{}{}
\creflabelformat{enumi}{(#2#1#3)}

\usepackage{amssymb}

\renewcommand{\Pr}{\mathbb{P}}

\newcommand\abstraction{This paper estimates individual treatment effects in  a triangular model with binary--valued endogenous treatments. Following the identification strategy established in \cite{vuong2014counterfactual},  we propose a two--stage estimation approach. First, we estimate the counterfactual outcome and hence the individual treatment effect (ITE) for every observational unit in the sample.  Second, we estimate the density of individual treatment effects in the population.  Our estimation method does not suffer from the ill--posed inverse problem associated with inverting a non--linear functional.   Asymptotic properties of the proposed method are established. We study its finite sample properties in Monte Carlo experiments. We also illustrate our approach with an empirical application assessing the effects of 401(k) retirement programs on personal savings.  Our results show that there exists a small but statistically significant proportion of individuals who experience negative effects, although the majority of ITEs is positive. \\ 

\noindent
\textbf{Keywords:} Nonseparable triangular models, binary endogenous variable, counterfactual mapping, individual treatment effects, 401(k) retirement programs
\\

}

\usepackage{subfig}

\begin{document}

\thispagestyle{empty}

\title{Estimation of heterogeneous individual treatment effects with endogenous treatments}

\thanks{$^*$We thank Jason Abrevaya, Isaiah Andrews,  Robert Leili, Leigh Linden, Matt Masten, Andres Santos as well as seminar participants at NYU, University of Texas at Austin,  Texas Econometrics Camp 2015, and 2016 CEME conference at Duke University for their helpful comments.}

\author{
\href{mailto:qianfeng.qf@utexas.edu}{Qian Feng$^*$}}
\thanks{$^*$Department of Economics, University of Texas at Austin, Austin, TX, 78712, 
\href{mailto:qianfeng.qf@utexas.edu}{qianfeng.qf@utexas.edu}}
\author{
\href{mailto:qvuong@nyu.edu}{Quang Vuong$^\dag$}}
\thanks{$^\dag$(corresponding author) Department of Economics, New York University, 19 W. 4th Street, 6FL, New York, NY, 10012,
\href{mailto:qvuong@nyu.edu}{qvuong@nyu.edu}}
\author{
\href{mailto:h.xu@austin.utexas.edu}{Haiqing Xu$^{\ddag}$}}
\thanks{$^{\ddag}$Department of Economics, University of Texas at Austin, \href{mailto:h.xu@austin.utexas.edu}{h.xu@austin.utexas.edu}}

\date{\today}

\maketitle

\begin{abstract}
 \abstraction
\end{abstract}

\vspace{5ex}

\clearpage
\section{Introduction}
Nonseparable triangular models have been studied extensively in the recent econometric literature, thereby allowing researchers to understand the nature of instrumental variables in the presence of endogeneity. See e.g. \cite{chesher2003identification,chesher2005nonparametric} and \cite{imbens2009identification}. One  appealing feature of nonseparable models is that the non-additive error in the causal relationship implies that the \textit{ceteris paribus} effects of covariates on the outcome variable  ``vary across individuals that, measured by covariates, are identical,'' \cite{chesher2003identification}.  Such heterogeneous causal effects are referred as \textit{``individual treatment effects''}(ITE) in the literature. See e.g. \cite{rubin1974estimating}, \cite{heckman1997making} and \cite{heckman2005structural}.

Estimating ITE and its distribution is crucial for evaluating a social program, especially in  view of the political issues associated with it \citep[see][]{heckman1997making}. From an individual's perspective, however, her ITE is more helpful for evaluating her treatment participation decision than an average effect. While the ``average person'' may benefit from a particular treatment, some individuals may experience little benefit or even some loss from participating,  in which case alternative treatment options may be preferred. Indeed, while the individual treatment effects of 401(k) retirement programs on personal savings are mostly positive in our sample, our empirical analysis indicates that there are individuals who experience negative benefits from participating to 401(k) retirement programs.



In this paper, we consider  a triangular model with a binary endogenous regressor.  Because of the self--selection issue, individuals who are treated are different from those who choose not to be treated. We address this issue with a binary valued instrumental variable \citep[see e.g.][]{imbens1994identification}. Limited variations of instrumental variables have been emphasized in the recent treatment effect literature. Moreover, natural experiments \citep[e.g.][]{angrist1998children,post2008deal} and eligibility for treatment participation \citep[e.g.][]{angrist1990lifetime,abadie2003401kplan} provide  commonly used binary--valued instrumental variables.

The distribution of heterogeneous treatment effects  has also been studied using quantiles. For instance, \cite{abadie2002instrumental} and \cite{froelich2013unconditional} estimate the quantile treatment effects (QTE) for  the complier group, a subpopulation defined by  \cite{imbens1994identification} under binary--valued instruments. For the population  QTE,  \cite{cherno2004401kplan}   propose a GMM--type approach in a linear quantile specification. Subsequently, \cite{cherno2006instrumental,chernozhukov2008instrumental} generalize \cite{cherno2004401kplan}'s estimation procedure by using quantile regression methods.  In a fully nonparametric setting, \cite{horowitz2007nonparametric} and \cite{gagliardini2012nonparametric} modify  \cite{cherno2004401kplan}'s moment conditions using the Tikhonov regularization to deal with the ill--posed inverse problem for deriving asymptotic properties of their estimators.

Our approach is novel and simple to implement. Instead of solving the moment conditions in \cite{chernozhukov2005iv},  we use the quantile invariance condition to match  the realized outcome with its counterfactual outcome for every observational unit in the sample through a so-called counterfactual mapping.  Specifically, our approach recovers the  ITE for every individual in the sample and does not suffer from the ill--posed inverse problem associated with inverting a non--linear functional.  In particular, we show that the ITEs are estimated uniformly at the parametric rate.  Given the recovered ITEs,  we estimate the density by kernel methods and establish its asymptotic properties.  Though it might be possible to obtain a density estimate from QTE estimates, this would involve a more complicated two--stage procedure and a delicate trimming scheme \cite[see e.g.][]{marmer2012quantile}.

We apply our approach to study the effects of 401(k) retirement programs on personal savings.  Introduced in the early 1980s, the 401(k) retirement programs aim to increase savings for retirement.  Endogeneity arises as individuals with a higher preference for savings are more likely to participate and also have higher savings than those with lower preferences \citep[see, e.g.,][]{poterba1996401kplan}.   Following e.g. \cite{abadie2003401kplan} and \cite{cherno2004401kplan}, we use 401(k) eligibility as an instrumental variable for 401(k) participation. We estimate the ITEs  for every individual in the sample as well as its density.   Our results show that there exists a small but statistically significant proportion (about 8.77\%) of individuals who experience negative effects,  although the majority of ITEs is positive.  It has been argued in the literature \citep[see e.g.][]{engen1996illusory} that some individuals could suffer from the  program due to the {\it Crowding Out Effect}.  We offer a complementary explanation as  individuals with negative ITEs are more likely to be younger, single, from smaller and lower income  families but with higher family net financial assets than the rest of the sample. 




The structure of the paper is organized as follows.  In Section 2, we introduce the triangular model and discuss its  identification and estimation. Section 3 provides Monte Carlo experiments to illustrate the performance of our proposed estimator. Section 4 derives its asymptotic properties.  Section 5 applies our estimation method to assess the effects of 401(k) retirement programs on personal savings.  Proofs of our results are collected in the Appendix.


%
%
%
%
%
%
%
%

%
%
%
%
%
%

\section{Model, Identification and Estimation}
\label{section: model}

\subsection{The triangular model}
Following \cite{chesher2005nonparametric}, we consider a nonseparable triangular model with an outcome equation and a selection equation: 
\begin{align}
\label{eq_1}
& Y= h(D,X,\epsilon),\\
\label{eq_2}
& D= \mathbb 1 \{\nu\leq m(X,Z)\}.
\end{align}Here $Y\in\mathbb R$ is the outcome variable, $D\in\{0,1\}$ is an endogenous dummy that indicates the treatment status, $X\in\mathbb S_X\subseteq \mathbb R^{k}$ is a vector of observed covariates (not necessary exogenous) and $Z\in \{0,1\}$ is a binary instrumental variable for $D$, i.e., $Z\bot (\epsilon,\nu)|X$. The two latent random variables $\epsilon$ and $\nu$ are scalar valued disturbances. Moreover, the function $h$ and $m$ are unknown structural relationships. In particular, $h$ is continuous and strictly increasing in $\epsilon$.


The key feature in the above triangular model is the nonseparability of $h$ in the error term $\epsilon$. With a nonseparable $h$, the \textit{ceteris paribus} effects on the outcome variable from covariates ``vary across individuals that, measured by covariates, are identical,'' \cite{chesher2003identification}. In the treatment effect literature, such heterogeneous causal effects are referred as \textit{``individual treatment effects''}(ITE), i.e., 
\[
\Delta\equiv h(1,X,\epsilon)-h(0,X,\epsilon).
\] 
See e.g. \cite{rubin1974estimating} and \cite{heckman1997making}. After controlling for $X$,  the ITE $\Delta$ is still a random object since it depends on the latent variable $\epsilon$. Our interest is to recover the ITE for each individual from her observables $(Y,D,X)$, and  to estimate the probability density function of ITE in the population. In particular, a decision-maker can use the former to evaluate an individual's participation choice, while the latter characterizes the distribution of treatment effects, which has been central in the program evaluation literature \citep[see e.g.][]{heckman1997making}.

 We now provide two examples to illustrate the nonseparability of the structural relationship $h$. 

\noindent
Example 2.1 (Additive error with generalized heteroscedasticity): Let
\[
Y=h^*(D,X)+\sigma^*(D,X)\cdot \epsilon,
\] where $h^*$ is a real-valued function, $\sigma^*$ is a positive function that captures the heteroscedasticity in the disturbance, and $\epsilon\in\mathbb R$ has zero mean and unit variance, unconditionally. This model is a generalization of a nonparametric regression model with heteroskedastic errors studied by e.g. \cite{andrews1991asymptotic}. The difference is that the heteroscedasticity term $\sigma^*$ depends on the endogenous binary variable $D$. In particular, when $\sigma^*$ is a constant, the above specification becomes an additive nonparametric regression with some endogenous regressor as studied by e.g. \cite{newey2003instrumental} and \cite{darolles2011nonparametric}. 

\noindent
Example 2.2 (Semiparametric transformation model): Consider
\[
\Gamma(Y)=X'\beta+\gamma D +\epsilon,
\] where $(\beta',\gamma)'\in \mathbb R^{k+1}$  and  $\Gamma:\mathbb R\rightarrow \mathbb R$ is an unknown monotone function. See \cite{horowitz1996semiparametric} when $(X,D)$ is exogenous.  A parametric example of the monotone function $\Gamma$ is the Box--Cox transformation when $Y$ is positive:
\[
\Gamma(y)= \left\{\begin{array}{cc}\frac{y^\lambda-1}{\lambda}, & \text{if } \lambda\neq 0; \\ \log y, & \text{if } \lambda=0,\end{array}\right.
\]where $\lambda\in\mathbb R$ is a model parameter. Such a transformation is useful when the dependent variable has a limited support. Indeed, the transformed dependent variable can have an unlimited support thereby ensuring a linear model specification with its usual assumptions. Various extensions of the Box--Cox transformation have been developed in the literature \citep[see e.g.][]{sakia1992box}, where monotonicity is a common feature in all these transformations.  Recently, \cite{chiappori2015nonparametric} have studied the case  where some variables such as $D$ is endogenous.



\subsection{Identification}
 \cite{vuong2014counterfactual} establish identification of the triangular  model (1)-(2) in a constructive way and show that it only requires binary variations of the instrumental variable $Z$. Given the monotonicity of $h$, the ITE can be written as a function of the observables $(Y,D,X)$:
\begin{equation}
\label{ITE}
\Delta = D\times (Y-\phi_{0X}(Y)) + (1-D) \times (\phi_{1X}(Y)-Y),
\end{equation}where $\phi_{dX}(\cdot)$ for $d=0,1$ are defined as the counterfactual mappings that depend on covariates $X$ and the value of $d$, namely,\footnote{The function $h^{-1}(d,x,\cdot)$ denotes the inverse of $h(d,x,\cdot)$.  Hereafter, for a generic random variable $W$ with distribution $F_W$, we denote its support by $\mathscr S_W$, defined as the closure of the open set $\mathscr S^o_W \equiv \{ w : F_W(w) \text{ is strictly increasing in a neighborhood of }  w \}$.} 
\begin{align*}
&\phi_{0X}(y)=h(0,X, h^{-1}(1, X,y)),\ \ \ \forall \ y\in\mathscr S_{h(1,X,\epsilon)|X},\\
&\phi_{1X}(y)=h(1,X, h^{-1}(0, X,y)),\ \ \ \forall \ y\in\mathscr S_{h(0,X,\epsilon)|X}.
\end{align*}  
By definition, $\phi_{dX}$ are monotone functions mapping $\mathscr S_{h(d',X,\epsilon)|X}$ onto $\mathscr S_{h(d,X,\epsilon)|X}$, where $d'=1-d$, and we have $\phi_{0X}=\phi^{-1}_{1X}$.

To obtain the ITE for an individual with $(Y,D,X)=(y,d,x)\in\mathscr S_{YDX}$, it suffices to identify the counterfactual mapping $\phi_{d'x}(y)$, where $d'=1-d$. Let $p(x,z)=\Pr(D=1|X=x,Z=z)$ be the propensity score function. For expositional simplicity, suppose $\mathscr S_{XZ}=\mathscr S_X\times \{0,1\}$ and $p(x,0)\neq p(x,1)$ for all $x\in\mathscr S_X$. W.l.o.g., throughout we assume $p(x,0)<p(x,1)$. Moreover,  for any  $y\in\mathbb R$ and $d=0,1$, let
\begin{equation}
\label{eq4}
C_{dx}(y)\equiv \frac{\Pr (Y\leq y; D=d|X=x, Z=0)-\Pr (Y\leq y; D=d|X=x, Z=1)}{\Pr(D=d|X=x,Z=0)-\Pr(D=d|X=x,Z=1)}.
\end{equation} \cite{imbens1997estimating} show that $C_{dx}(\cdot)$ is the conditional distribution function of $h(d,X,\epsilon)$ given the complier group, namely, $\{X=x, m(x,0)<\nu\leq m(x,1)\}$.  Let  $\mathscr C_{dx}$ be the support of $C_{dx}(\cdot)$. It is straightforward to see that $\mathscr C_{dx}\subseteq \mathscr S_{h(d,X,\epsilon)|X=x}$. Next, we present the identification of $\phi_{dx}$ established in \cite{vuong2014counterfactual}.

\begin{theorem}\citep{vuong2014counterfactual}
\label{th1}
In the triangular model (1)-(2), suppose (i) $h$ is continuous and strictly increasing in $\epsilon$; (ii) $Z$ is conditionally independent of $(\epsilon,\nu)$ given $X$, i.e., $Z\bot (\epsilon, \nu)|X$ with $p(x,0)\neq p(x,1)$ for all $x\in\mathscr S_X$; (iii) conditional on $X$, the joint c.d.f.  $F_{\epsilon\nu|X}$ is  continuous; (iv)  $\mathscr C_{dx}=\mathscr S_{h(d,X,\epsilon)|X=x}$ for $d=0,1$ and $x\in\mathscr S_X$. Then,  $\mathscr S_{h(d,X,\epsilon)|X=x}=\mathscr S_{Y|D=d,X=x}$, and  the counterfactual mapping  $\phi_{dx}$ is identified by
\[
\phi_{dx} (y) =C_{dx}^{-1}\big(C_{d'x}(y)\big), \ \ \forall \ y\in\mathscr S_{Y|D=d',X=x}
\]
where $C_{dx}(\cdot)$ is continuous on  $\mathbb R$ and strictly increasing on $\mathscr C^\circ_{dx} \equiv \mathscr S^\circ_{Y|D=d,X=x}$ for $d=0,1$, and $d'=1-d$.
\end{theorem}

In \Cref{th1}, condition (i) -- (iii) are standard in the triangular model literature. The support condition (iv) requires that, conditional on $X=x$, the subpopulation $m(x,0)<\nu\leq m(x,1)$, i.e., the complier group introduced in  \cite{imbens1994identification},  contains the same information on individual treatment effects as the whole population.  It is weak as it is satisfied as soon as  $(\epsilon,\nu)$ has a rectangular support given $X$.  See \cite{vuong2014counterfactual}.  It is testable since $C_{dx}$ is identified by \eqref{eq4}. When (iv) fails to hold,  the counterfactual mappings are partially identified on intervals. It is worth pointing our that (iv) is needed for identification of ITE even if one assumes the error term $\epsilon$ was observed in the data.

With $\phi_{dx}$ identified, we can use \eqref{ITE}  to construct the counterfactual outcome for any individual in the population from her observables $(Y,D,X)$. Moreover, the probability distribution of ITE is also identified under the conditions in \Cref{th1}.

\subsection{Estimation}\label{section: ne}
We now develop nonparametric estimators of the counterfactual mappings $\phi_{dx}$ for $d=0,1$ and the probability density function $f_{\Delta}$ of ITE.  On one hand, $\phi_{dx}$ can be used to construct the ITE for any individual in the population from her observables $(Y,D,X)$. On the other hand, the probability density function is a convenient way to characterize the  distribution of the ITE when the ITE is continuously distributed.\footnote{Under Condition (i)--(iii), the ITE can have a mass point when $\phi_{dx}$ has slope one in some intervals contained in its support, i.e., $\phi_{1x}(y)=g(x)+y$ on some $[a,b]\subseteq \mathscr S_{h(0,x,\epsilon)|X=x}$. Then, conditional on $X=x$, ITEs take the same value $g(x)$ for all $\epsilon\in \{e: h(0,x,e)\in [a,b]\}$. Hence, ITE has a mass point at $g(x)$. Such a case, however, can be detected given the identification of $\phi_x$.}  
Our estimation approach is fully nonparametric. To present the basic ideas, we assume that the covariates $X$ are discrete random variables with a finite support. Our analysis can be extended using e.g. the kernel method to the case where $X$ are continuous at the cost of exposition. 


Let $\{(Y_i,D_i,X'_i, Z_i)': i=1,\cdots, n\}$ be an i.i.d. sample generated from the underlying structure of the triangular model.  Our proposed estimation procedure takes two steps: First, for a given value of $(y,d,x)\in\mathscr S_{YDX}$, we estimate the counterfactual mapping $\phi_{d'x}(y)$ by a simple estimator that minimizes a convex population objective function. In the second step, we construct a pseudo sample of the counterfactual outcomes for all individuals in the sample and then nonparametrically estimate the density function $f_\Delta$  using the kernel method.  We introduce some notation.   Fix $x\in\mathscr S_{X}$. For simplicity, we suppress the dependence on $X=x$ in the following discussion.  For each $(y_0,y_1)\in\mathbb R^2$ and $z\in\{0,1\}$, let  
\begin{align*}
& \rho_0(y_0,y_1; z)=\mathbb E \big[|Y-y_0| (1-D)\big|X=x,Z=z\big]-\mathbb E\big[ \text{sign}(Y-y_1)\cdot D\big|X=x, Z=z\big]\cdot y_{0},\\
 &\rho_1(y_0,y_1; z)=\mathbb E \left[|Y-y_1| D\big|X=x, Z=z\right]-\mathbb E\big[ \text{sign}(Y-y_0) \cdot (1-D)\big|X=x, Z=z\big]\cdot y_{1}.
\end{align*} 
where $\text{sign}(u)\equiv 2\times \mathbb 1 (u > 0)-1$. 
 
For $d=0,1$, let 
 \[
 Q_d(y_0,y_1)= (-1)^d\times\big[\rho_d(y_0,y_1; 0)-\rho_d(y_0,y_1;1)\big]
 \] 
be the population objective function.  Such an objective function is motivated by the quantile regression method in  \cite{koenker1978regression}.   To see this, note that the quantile invariant condition in \cite{chernozhukov2005iv} implies that for  $(y_0,y_1)\in\mathbb R^2$ satisfying $y_1=\phi_{1x}(y_0)$ (equivalently, $y_0=\phi_{0x}(y_1)$),  we have
\begin{multline}
\label{chh}
\Pr(Y\leq y_1;D=1|X=x, Z=0)+\Pr(Y\leq y_0;D=0|X=x, Z=0)\\
=\Pr(Y\leq y_1;D=1|X=x, Z=1)+\Pr(Y\leq y_0;D=0|X=x, Z=1).
\end{multline} 
In the next lemma, we show that \eqref{chh} is indeed the first--order condition of the population objective function $Q_0(\cdot,y_1)$, which is continuously differentiable and weakly convex on $\mathbb R$.  We also show that $Q_0(\cdot,y_1)$ is strictly convex on $\mathscr S^\circ_{Y|D=0,X=x}$ and minimized uniquely on $\mathbb R$ at $y_0=\phi_{0x}(y_1)$  whenever $y_1\in\mathscr S^\circ_{Y|D=1,X=x}$. A similar argument also holds for the population objective function $Q_1(y_0,\cdot)$. 
\begin{lemma}
\label{lemma1}
Suppose the conditions in \Cref{th1} hold. Then, for $d=0,1$ and $y_{d}\in\mathbb R$, the function $Q_{d'}(y_0, y_1)$  is continuously differentiable and weakly convex in $y_{d'}\in\mathbb R$  where $d'=1-d$.  Moreover,  if  $y_d\in\mathscr S^\circ_{Y|D=d,X=x}$, then $Q_{d'}(y_0, y_1)$ is strictly convex in $y_{d'}\in\mathscr S^\circ_{Y|D=d',X=x}$, and uniquely minimized  on $\mathbb R$ at $\phi_{d'x}(y_d)$.
\end{lemma}

\noindent 
\Cref{lemma1} provides a basis for our nonparametric estimation of the counterfactual mappings $\phi_{0x}(\cdot)$ and $\phi_{1x}(\cdot)$.  It is worth pointing out that each minimization is a one--dimensional optimization problem. 

We are now ready to define our estimator. For expositional simplicity, let $\mathscr S_{Y|D=d,X=x}$ be a compact interval $[\underline{y}_{dx}, \overline{y}_{dx}]$.  For $d=0,1$, $(y_0,y_1)\in\mathbb R^2$ and $z\in\{0,1\}$, let $d'=1-d$ and 
\begin{eqnarray*}
\hat \rho_d(y_0,y_1;z)&=&\frac{\sum_{j=1}^n|Y_j-y_d|\times \mathbb 1 (D_j=d;X_j=x; Z_j=z)}{\sum_{j=1}^n \mathbb 1 (X_j=x; Z_j=z)}\\
&-&\frac{\sum_{j=1}^n \text{sign} (Y_j-y_{d'})\times\mathbb 1( D_j=d';X_j=x; Z_j=z)}{\sum_{j=1}^n \mathbb 1 (X_j=x; Z_j=z)}\times y_d.
\end{eqnarray*}  
Moreover, let
\[
\hat \phi_{d'x}(y_{d})=\underset{y_{d'}\in[\underline y_{d'x},\overline y_{d'x}]}{\arg\min} \ \hat Q_{d'}(y_0,y_1), \ \ \forall \ y_{d}\in\mathscr S_{Y|D=d,X=x}.
\]
where $\hat Q_{d'}(y_0,y_1)= (-1)^{d'}\times\big[\hat \rho_{d'}(y_0,y_1; 1)-\hat \rho_{d'}(y_0,y_1;0)\big]$. For simplicity,  we assume the support $[\underline{y}_{d'x}, \overline{y}_{d'x}]$ is known. See e.g. \cite{guerre2000optimal} for nonparametric estimation of the support $[\underline{y}_{d'x}, \overline{y}_{d'x}]$ if it is unknown.

Given the sample $\{(Y_i,D_i,X'_i,Z_i)': i=1,\cdots,n\}$, we can construct the counterfactual outcome for every individual in the sample from her observables $(Y_i,D_i,X_i)$. Namely,
\[
\left\{\begin{array}{cc}\hat h(0,X_i,\epsilon_i)=\hat \phi_{0X_i}(Y_i), & \text{ if } D_i=1; \\\hat h(1,X_i,\epsilon_i)=\hat \phi_{1X_i}(Y_i),  & \text{ if } D_i=0.\end{array}\right.
\]
Thus, we can  estimate the ITE by \eqref{ITE}, i.e., for $i=1,\cdots,n$, 
\begin{equation}
\label{ITE_est}
\hat \Delta_i=\left\{\begin{array}{cc}Y_i-\hat h(0,X_i,\epsilon_i), & \text{ if } D_i=1; \\\hat h(1,X_i,\epsilon_i)-Y_i,  & \text{ if } D_i=0.\end{array}\right.
\end{equation}
 In particular, we can construct a pseudo sample  $\{\hat \Delta_i: i=1,\cdots,n\}$  from the observed sample $\{(Y_i,D_i,X'_i, Z_i)': i=1,\cdots, n\}$.

It is worth pointing out that the first--stage estimation is computationally simple and does not suffer from an ill--posed inverse problem \citep[see e.g.][]{horowitz2007nonparametric}. In particular,  to solve the one--dimensional  optimization problem for each individual's counterfactual outcome,   the practitioner can use a grid search algorithm that is simple but highly robust. As is shown below, the first--stage estimation bias $\hat \phi_{dx} (\cdot)- \phi_{dx}(\cdot)$ uniformly converges to zero at the parametric rate of $\sqrt n$, given that all the covariates $X$ are discrete variables.\footnote{If $X_i$ contains continuous random variables, then we need to smooth over $X_i$ as otherwise there may not be enough observations for which $X_j=X_i$.}

Next, we follow \cite{guerre2000optimal} to estimate the density function $f_{\Delta}$  by the kernel method. To clarify ideas, let $[\underline \delta, \overline \delta]$ be a subinterval of the ITE's support. Then, we define the density estimator: 
\[
\hat f_{\Delta}(\delta)= \frac{1}{nh}\sum_{i=1}^n K\left(\frac{\hat \Delta_i-\delta}{h}\right), \ \ \forall \delta\in [\underline{\delta}+h,\overline{\delta}-h],
\]where $h$ is a bandwidth and $K$ is a  kernel with a compact support. Because the kernel estimator $\hat f_{\Delta}$ suffers from boundary issues, then we restrict the estimation of $ f_{\Delta}$ to the inner subset $[\underline{\delta}+h,\overline{\delta}-h]$.

\section{Monte Carlo Experiments}
To illustrate the finite sample performance of the proposed estimator, we conduct a Monte Carlo study.  For simplicity, we do not include other covariates $X$ in the specification. Following the conditions in \Cref{th1},  the data generating process is given by
\begin{align*}
Y= h(D,\epsilon),\ \ \ \ \ \ \ \ \  
D=\mathbb 1(\gamma_0+\gamma_1 \cdot Z+\nu\geq0),
\end{align*}where $h(d,\epsilon)=(\epsilon+1)^{2+d}$ for $d=0,1$,\footnote{We also consider other functional forms for $h(d,\cdot)$, e.g.,  $h(0,\epsilon)=\ln (\epsilon+1)$ and $h(1,\epsilon)=(\epsilon+1)^2$. The results are qualitatively similar.} and $(\epsilon, \nu)$ conforms to a joint distribution with uniform marginal distributions on $[0,1]$ and Gaussian copula with correlation coefficient $0.3$.\footnote{A copula is a multivariate probability distribution of random variables, each of which is marginally uniformly distributed
on $[0,1]$. The Gaussian copula is constructed from a multivariate normal distribution. See e.g. \cite{nelsen2007introduction}.} Because $h(d,\cdot)$ is continuous and strictly increasing in $\epsilon$, Condition (i) in \Cref{th1} is satisfied. We set $\gamma_0=-0.7$ and $\gamma_1=0.1, 0.2$ and $0.3$, respectively.   The value of $\gamma_1$ determines the size of the compliers group, i.e., $-\gamma_0-\gamma_1\leq \nu<-\gamma_0$. Hence, the larger $\gamma_1$, the more ``effective''  the instrumental variable $Z$. In our setting, $\Delta=\epsilon(\epsilon+1)^2$ is distributed on $[0,4]$ with mean $1.417$ and median 1.125 in the population. Moreover, we set $Z = \mathbb 1 \{\xi\geq 0\}$ where $\xi\sim N(0,1)$ is  independent of $(\epsilon,\nu)$. Conditions (ii)--(iv) in \Cref{th1} are satisfied.   In particular, condition (iv) holds since $F_{\epsilon\nu|X}$ has a rectangular support as noted in \cite{vuong2014counterfactual}.

\Cref{est: tab1} reports the  finite sample performance of our ITE estimates in terms of  the Root Mean Squared Error (RMSE). Specifically, for each size $n=1000,2000,4000$ we draw $\{(\epsilon_i,\nu_i,\xi_i): i=1,\cdots,n\}$ to obtain  a sample $(Y_i,D_i,Z_i)$ of size $n$. We then compute the true ITE $\Delta_i$ by  $h(1,\epsilon_i)-h(0,\epsilon_i)$ and its estimate $\hat \Delta_i$ by \eqref{ITE_est} for each individual $(Y_i,D_i,Z_i)$.  To obtain the RMSE for each such individual's ITE, we draw another 200 samples $\{(Y_i^{(r)},D_i^{(r)},Z_i^{(r)}): i=1,\cdots,n\}$ from $\{(\epsilon_i^{(r)},\nu_i^{(r)},\xi_i^{(r)}): i=1,\cdots,n\}$ for $r=1,\cdots, 2000$. These are used to repeatedly estimate the ITEs for the individuals in the original sample by $\hat {\Delta}_{i}^{(r)} = [Y_i- \hat \phi_{0}^{(r)}(Y_i) ]D_i + [\hat \phi_{1}^{(r)}(Y_i) - Y_i ](1-D_i)$ where $\hat \phi_{d}^{(r)}$ is the estimate of $\phi_{d}$ using the $r$--th new drawn sample. Thus, we obtain the RMSE of $\hat {\Delta}_i$ by $\sqrt{\frac{1}{200}\sum_{r=1}^{200}\big[\hat {\Delta}_{i}^{(r)}-\Delta_i\big]^2}$. For comparison, we also provide the RMSE of the LATE over the 200 replications/samples within curly brackets as proposed by \cite{imbens1994identification}.\footnote{For our Monte Carlo setting, the LATE reduces to $\left[\mathbb E(Y|Z=1)-\mathbb E(Y|Z=0) \right]/\left[p(1)-p(0) \right]=1.5351, 1.4912, 1.4449$ for $\gamma_1=0.1,0.2,0.3$, respectively. Moreover, the LATE is estimated by $\left[\overline{Y}(1)-\overline{Y}(0)\right]/\left[\hat{p}(1)-\hat{p}(0)\right]$
for a given sample, where $\overline{Y}(z)$ and $\hat{p}(z)$ are the sample means of $Y$ and $D$ given $Z=z$, respectively, for $z=0,1$.  In particular, unlike ITE and its estimate, LATE and its estimate do not vary across individuals by definition.}
By comparing their RMSEs from \Cref{est: tab1}, a surprising result is that estimating treatment effects at individual level (i.e. ITE)  is not more difficult than to estimate treatment effects at aggregated level (e.g. LATE) for every sample size.  As sample size increases, both the bias and standard error decrease at the expected $\sqrt n$--rate.   The estimation error (i.e. its size and standard deviation) depends on the sample size $n$ and the compliers group's proportion $\gamma_1$. Specifically in the different designs, the finite sample performance of the ITE estimator depends on the value of  $n\cdot \gamma_1^2$.  For example, the performance of our estimator under $(n,\gamma_1)=(1000, 0.2)$ is similar to that under $(n,\gamma_1)=(4000,0.1)$. This observation is consistent with our asymptotic properties established in the next section.


\begin{table}[h] \small
\caption{Finite sample performance of ITE}
\begin{tabular}{cl|cccccccc}\hline\hline
 Sample size &                      & $\gamma_1=0.1$& $0.2$&$0.3$\\ \hline
                     &Ave. RMSE    &  1.2918                &   0.6076               &0.4071\\
   1,000         &Std. RMSE     &(0.5279)               &   (0.2912)             &(0.2231)\\
                     &LATE  RMSE  &\{1.0448\}             &     \{0.5159\}         &\{0.3619\}   \\\hline
                     &Ave. RMSE     &0.9343                 &0.4381                   &0.2670\\
   2,000         &Std. RMSE      &(0.4289)              &(0.2122)                 & (0.1511)\\
                     &LATE  RMSE  &\{0.6639\}              &    \{0.3759\}          &\{0.2532\}     \\\hline
                     &Ave. RMSE     &0.6059                 &0.3245                   &0.18313\\
   4,000         &Std. RMSE      &(0.2839)              &(0.1455)                 &(0.0985) \\
                     &LATE  RMSE  &\{0.5057\}              &    \{0.2220\}           &\{0.1790\}\\  \hline              
\end{tabular}
\label{est: tab1}
\end{table}



\begin{figure}[h] 
   \centering
   \includegraphics[width=2in]{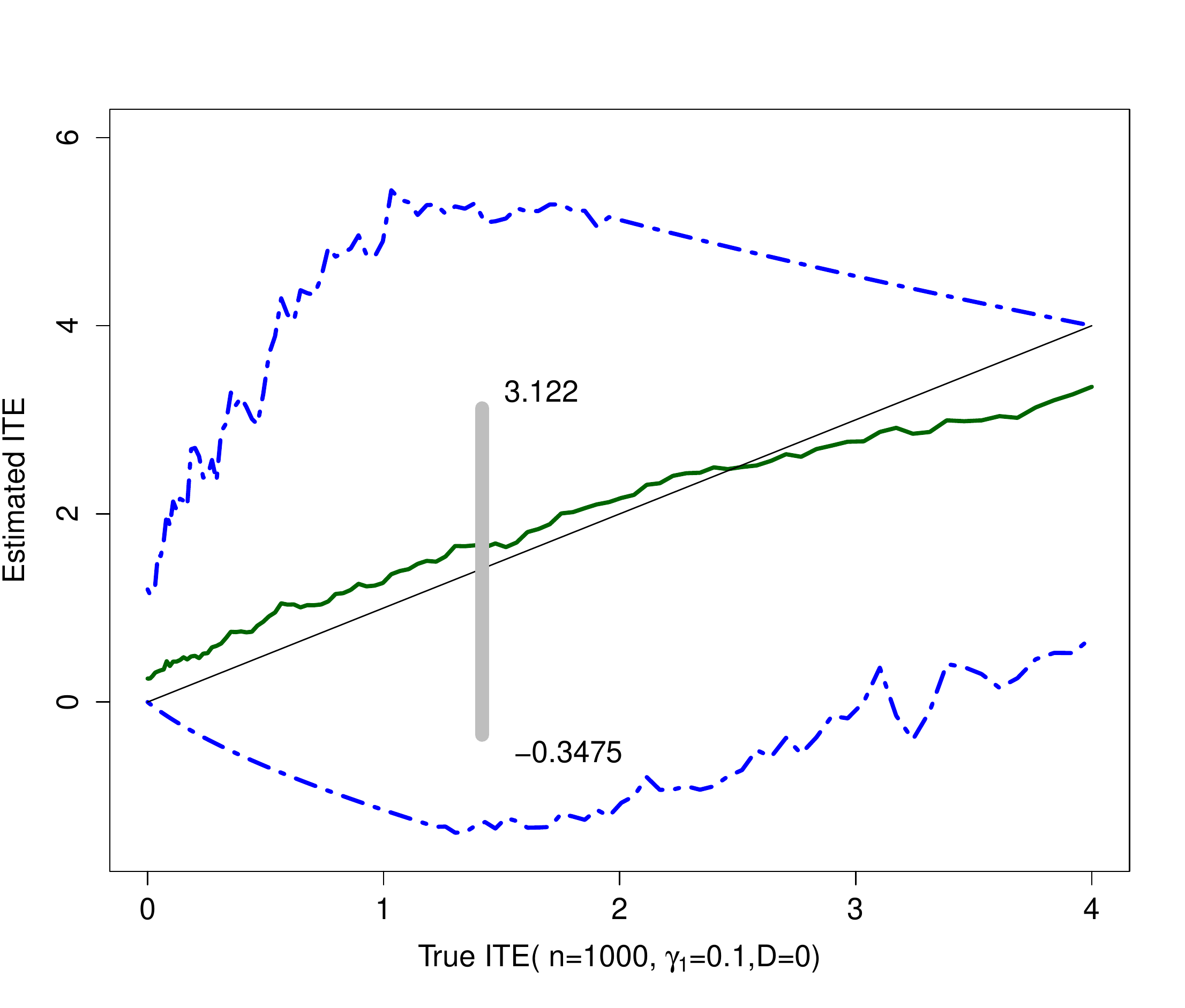} 
   \includegraphics[width=2in]{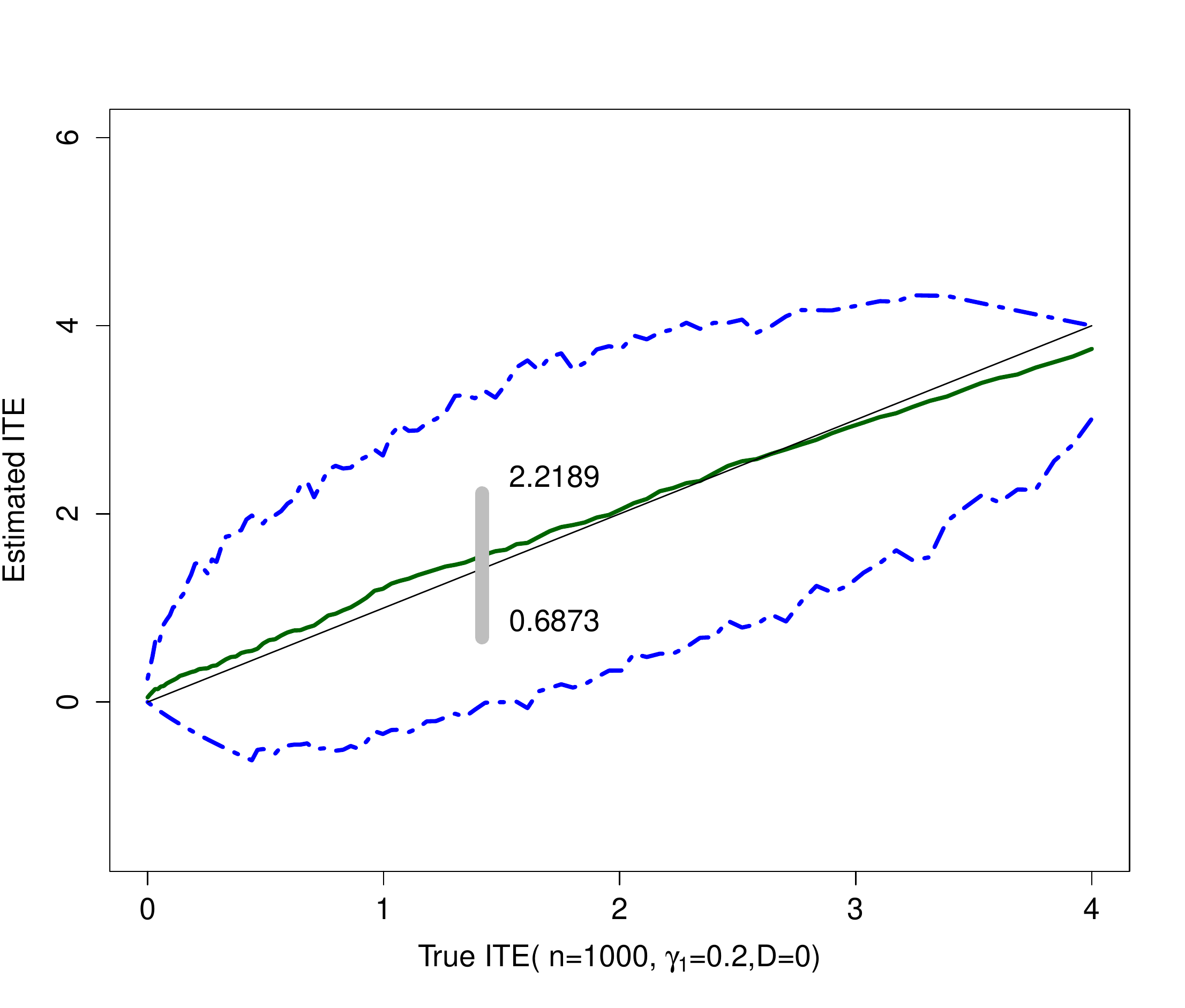} 
   \includegraphics[width=2in]{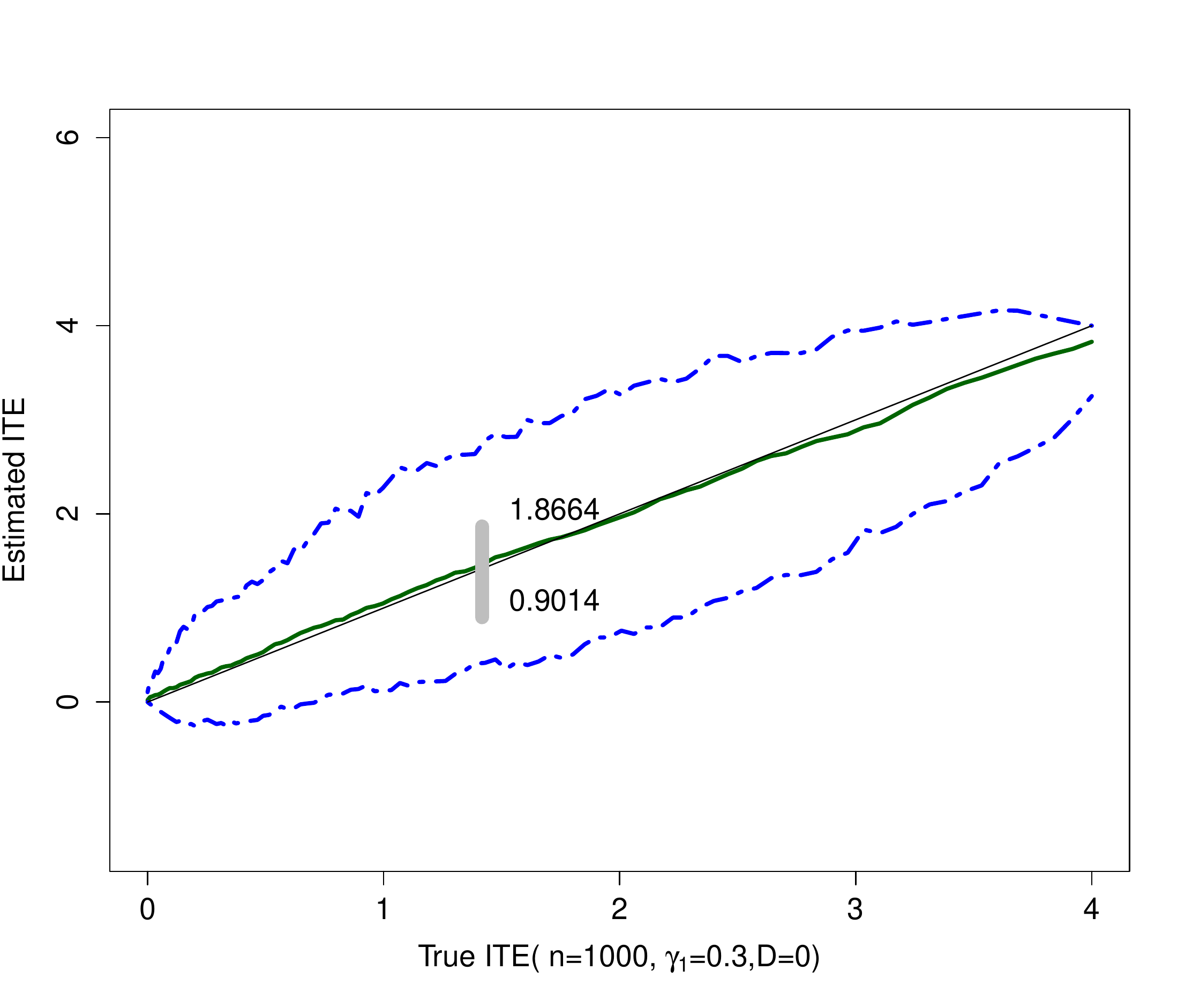} \\
      \includegraphics[width=2in]{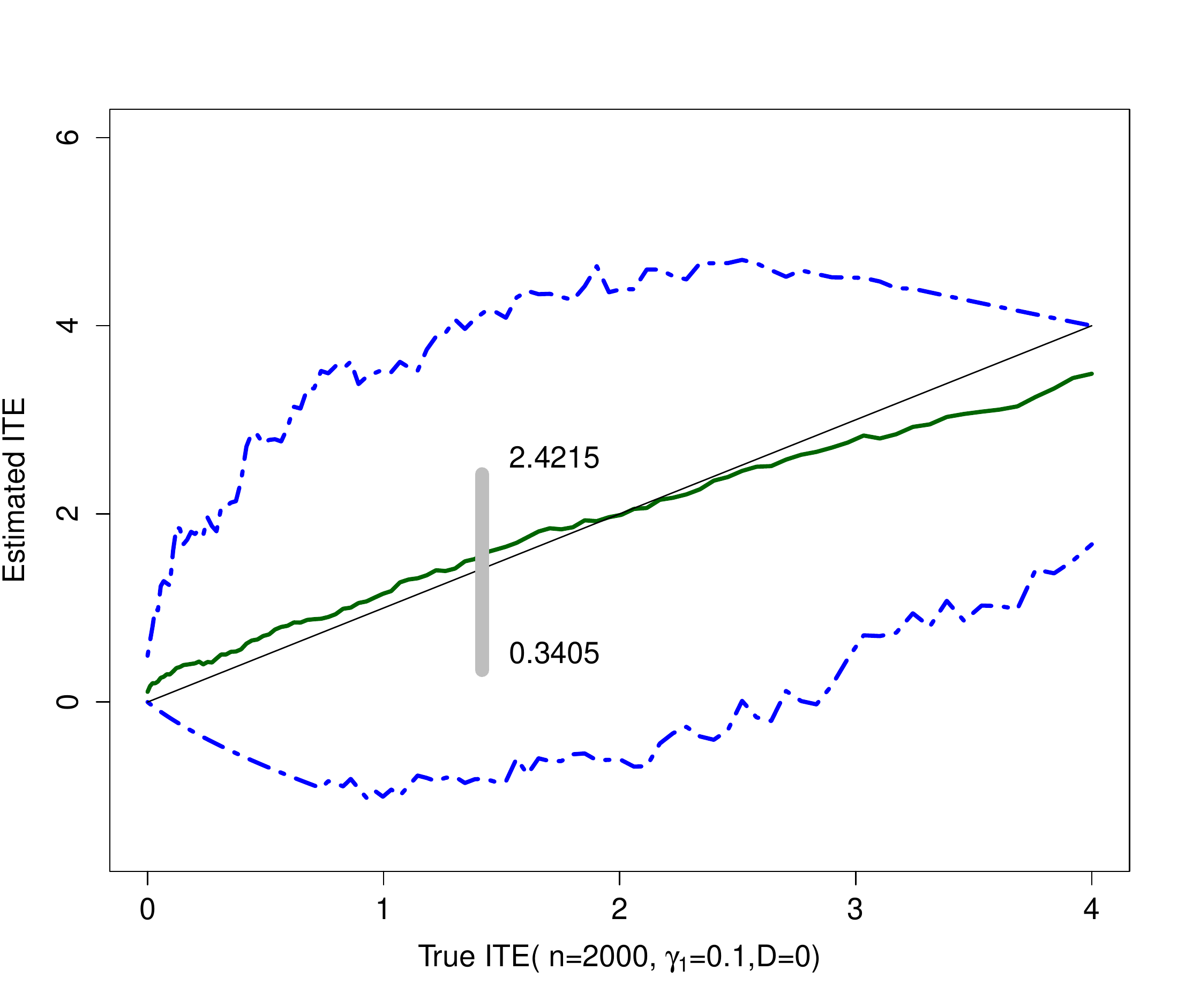} 
   \includegraphics[width=2in]{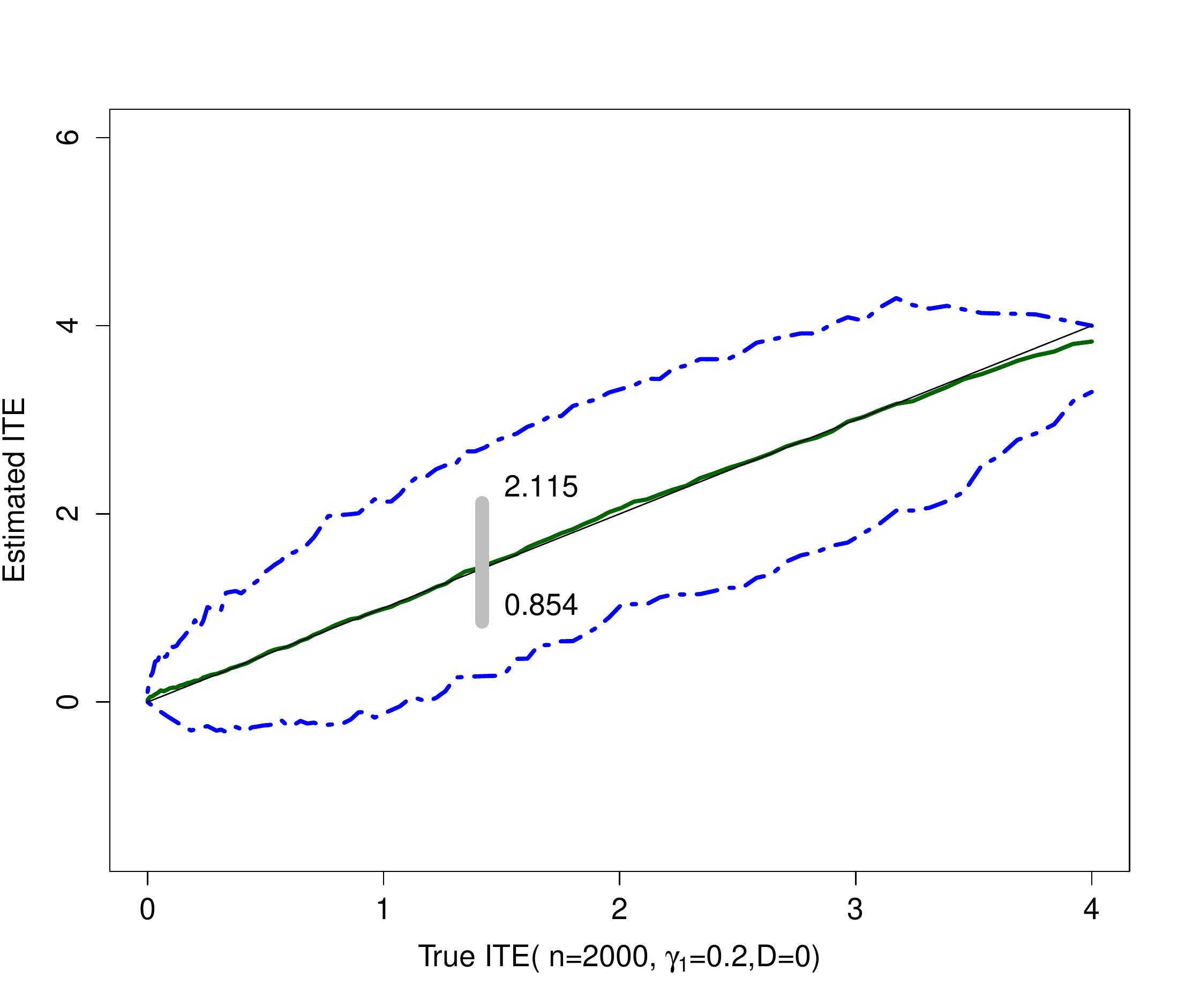} 
   \includegraphics[width=2in]{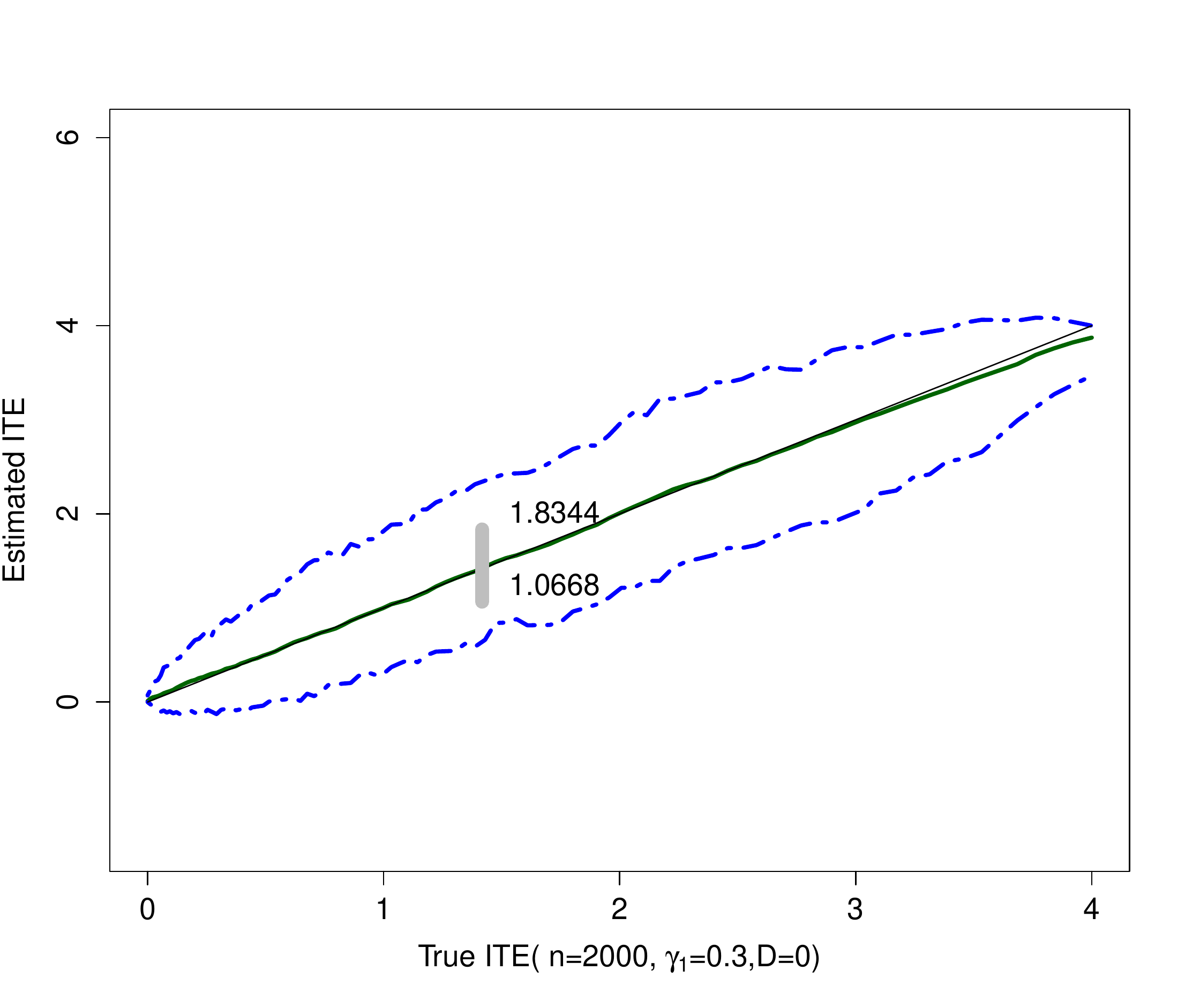} \\
      \includegraphics[width=2in]{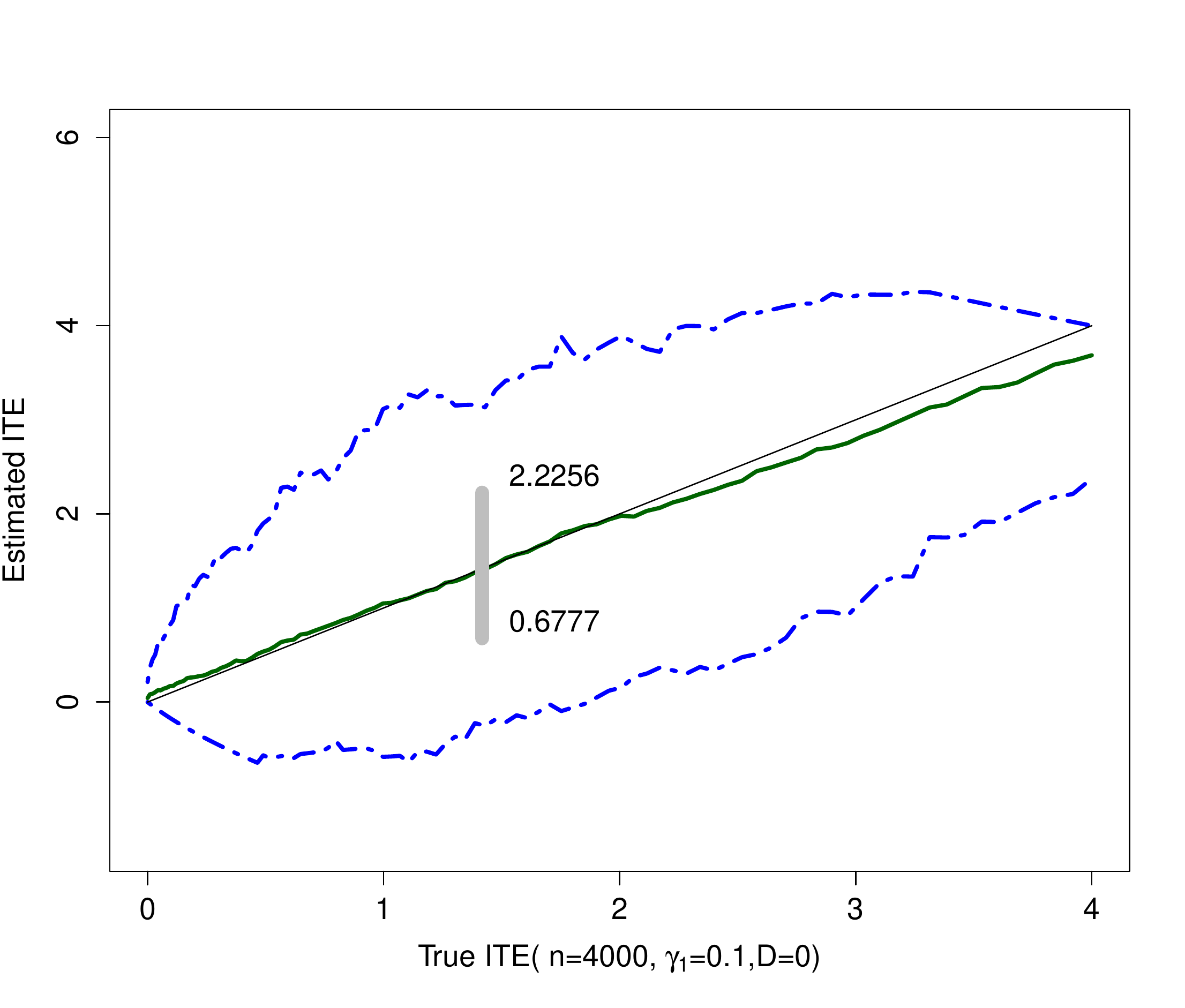} 
   \includegraphics[width=2in]{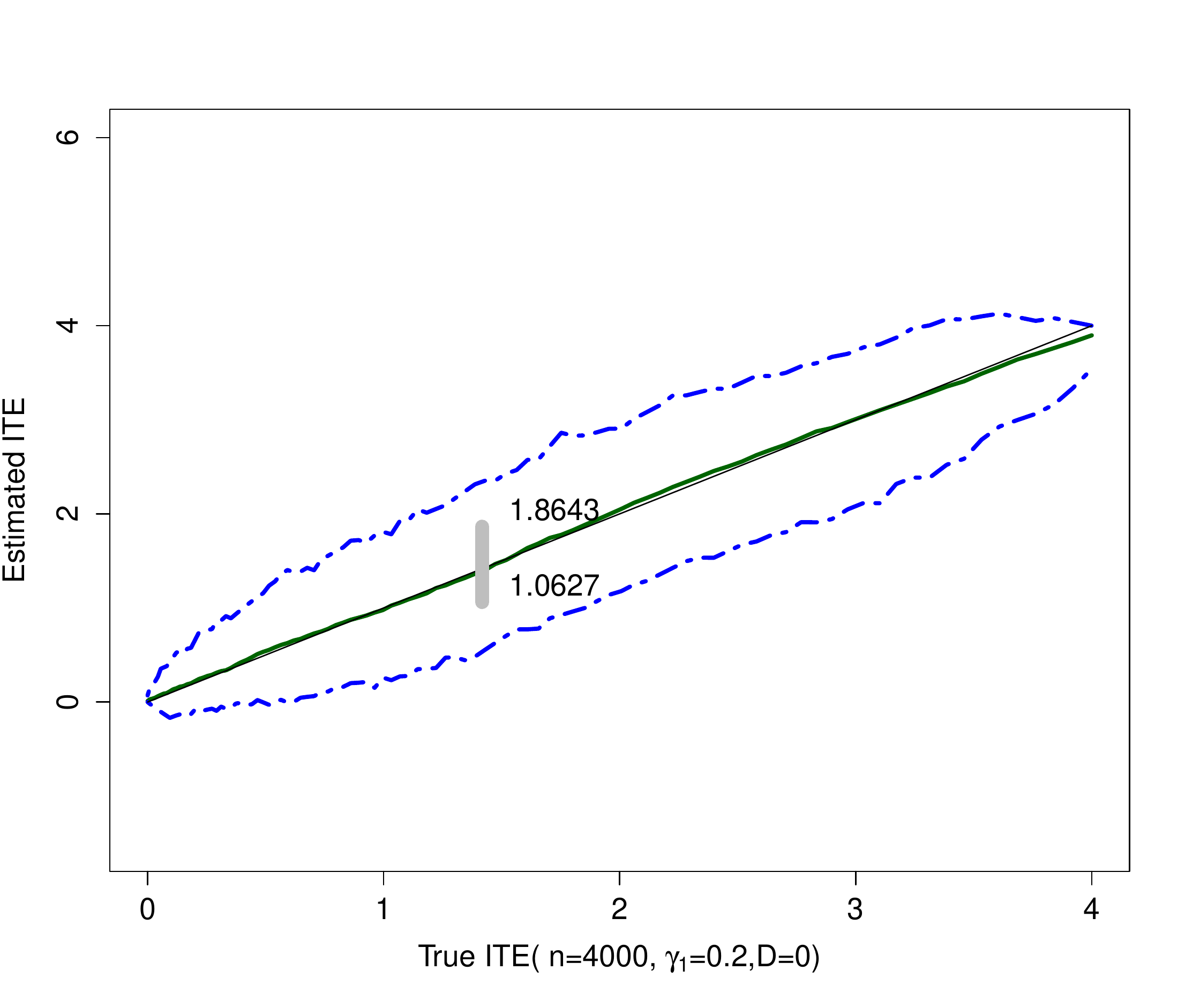} 
   \includegraphics[width=2in]{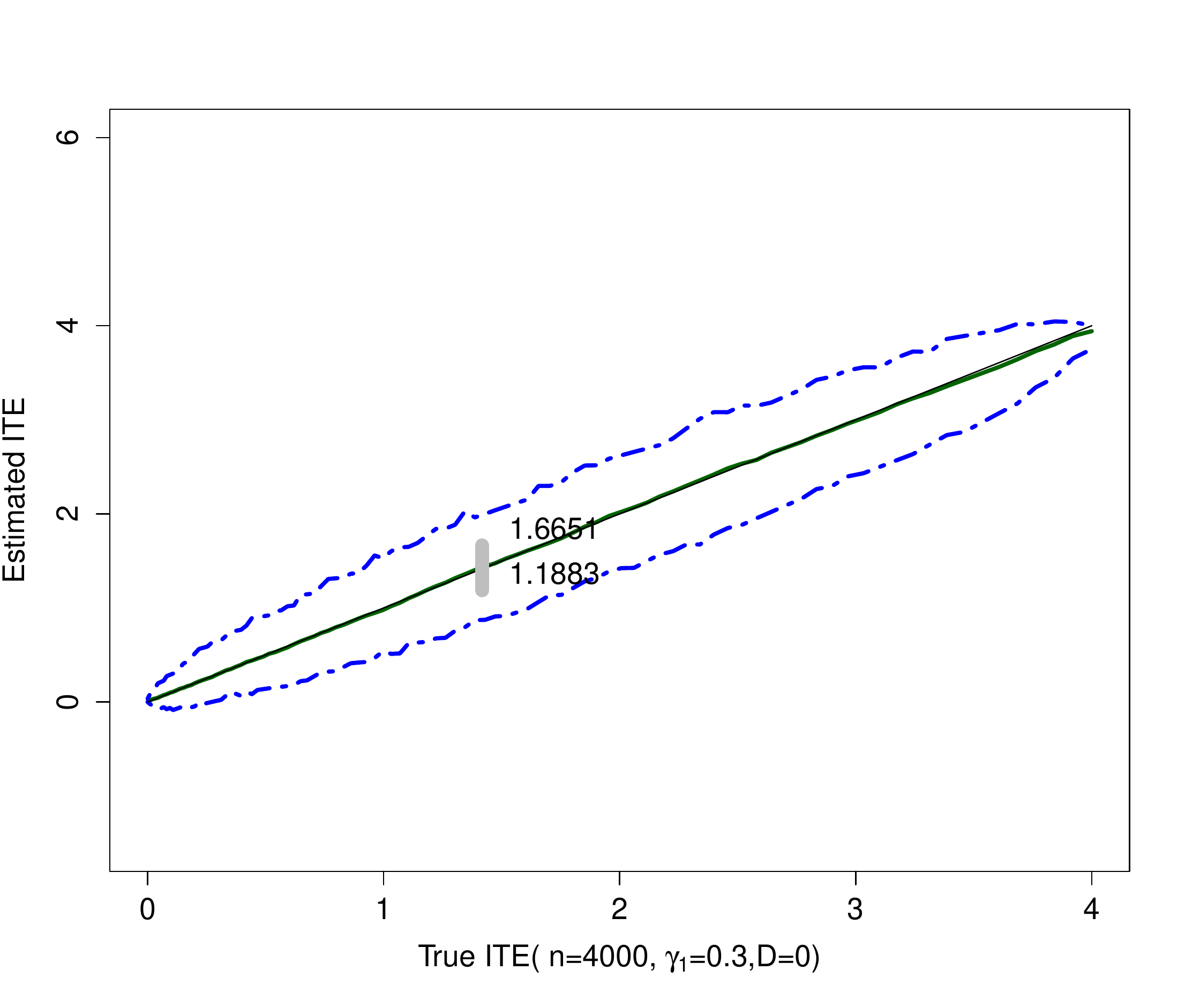} \\
   \caption{True and estimated ITE for $ D=0$}
   \label{fig2_d0}
\end{figure}

\begin{figure}[h] 
   \centering
   \includegraphics[width=2in]{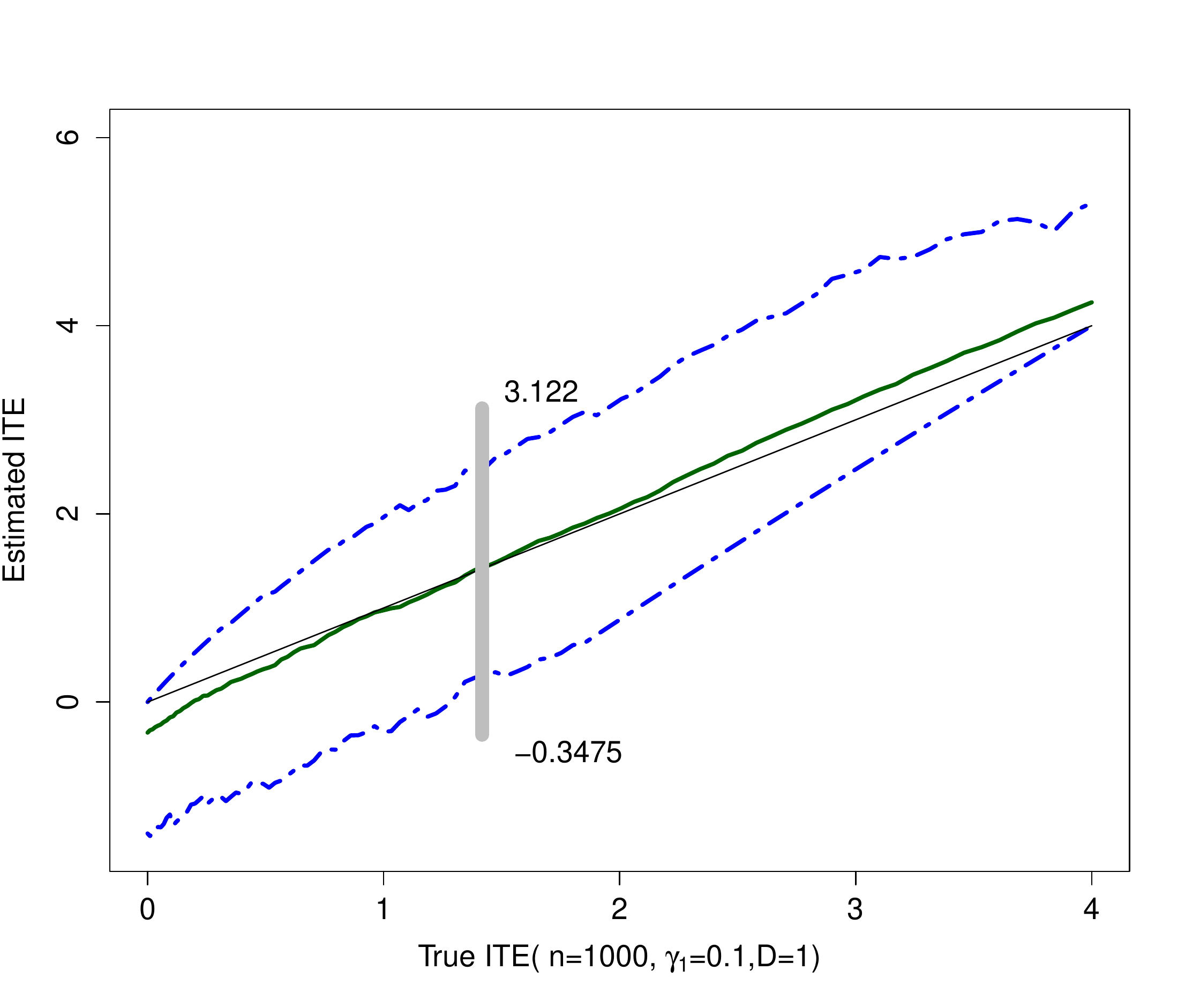} 
   \includegraphics[width=2in]{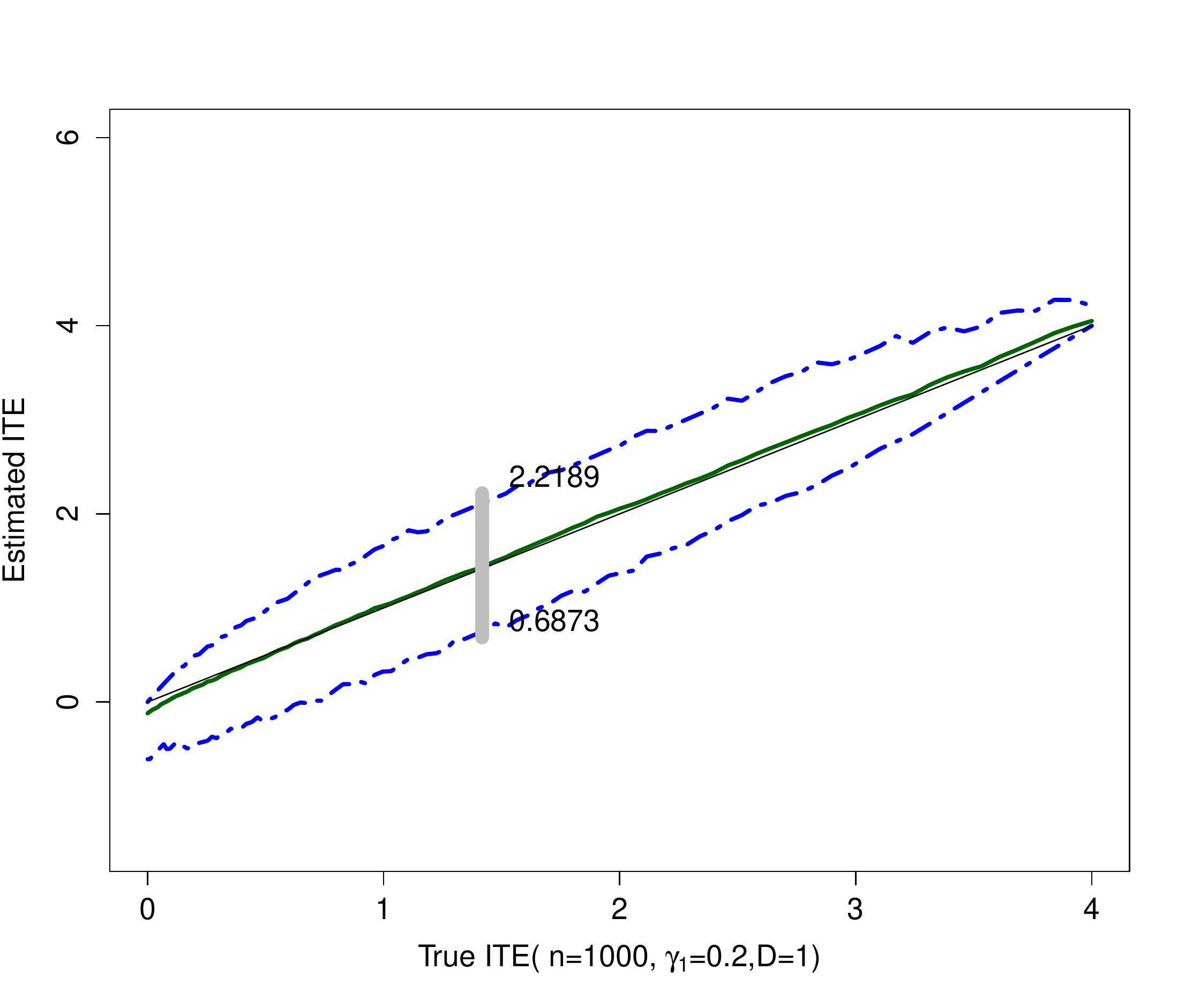} 
      \includegraphics[width=2in]{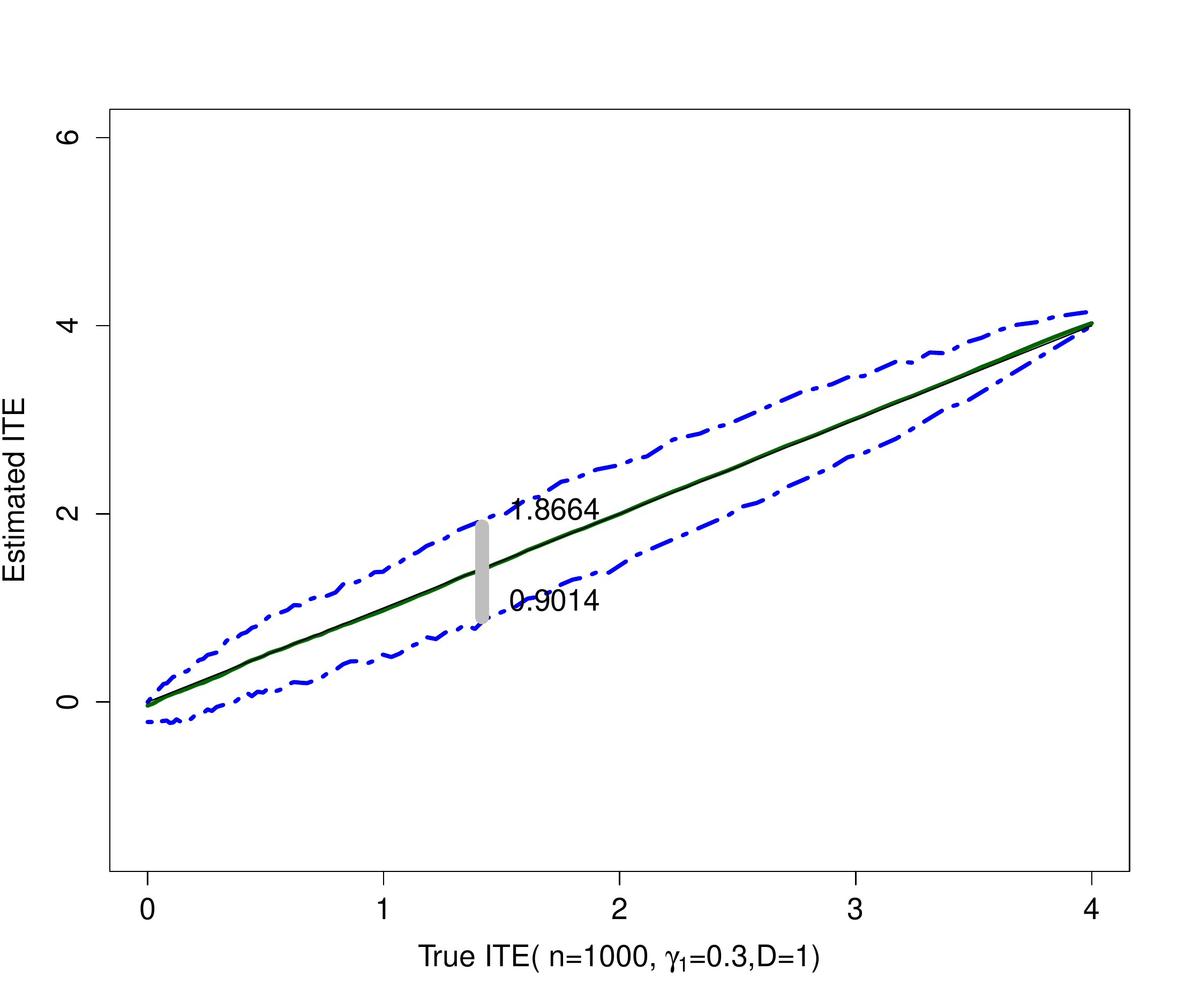} \\
         \includegraphics[width=2in]{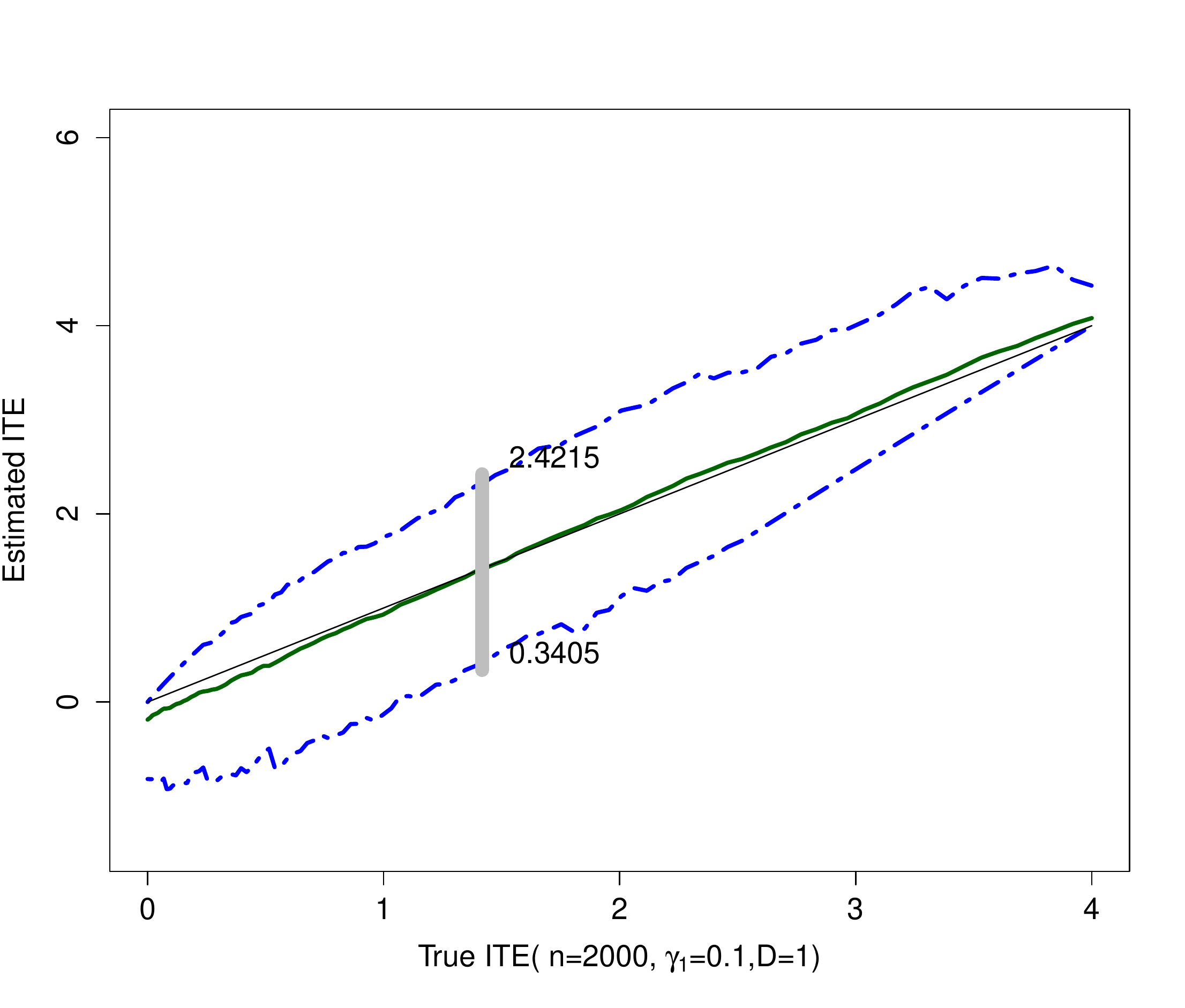} 
   \includegraphics[width=2in]{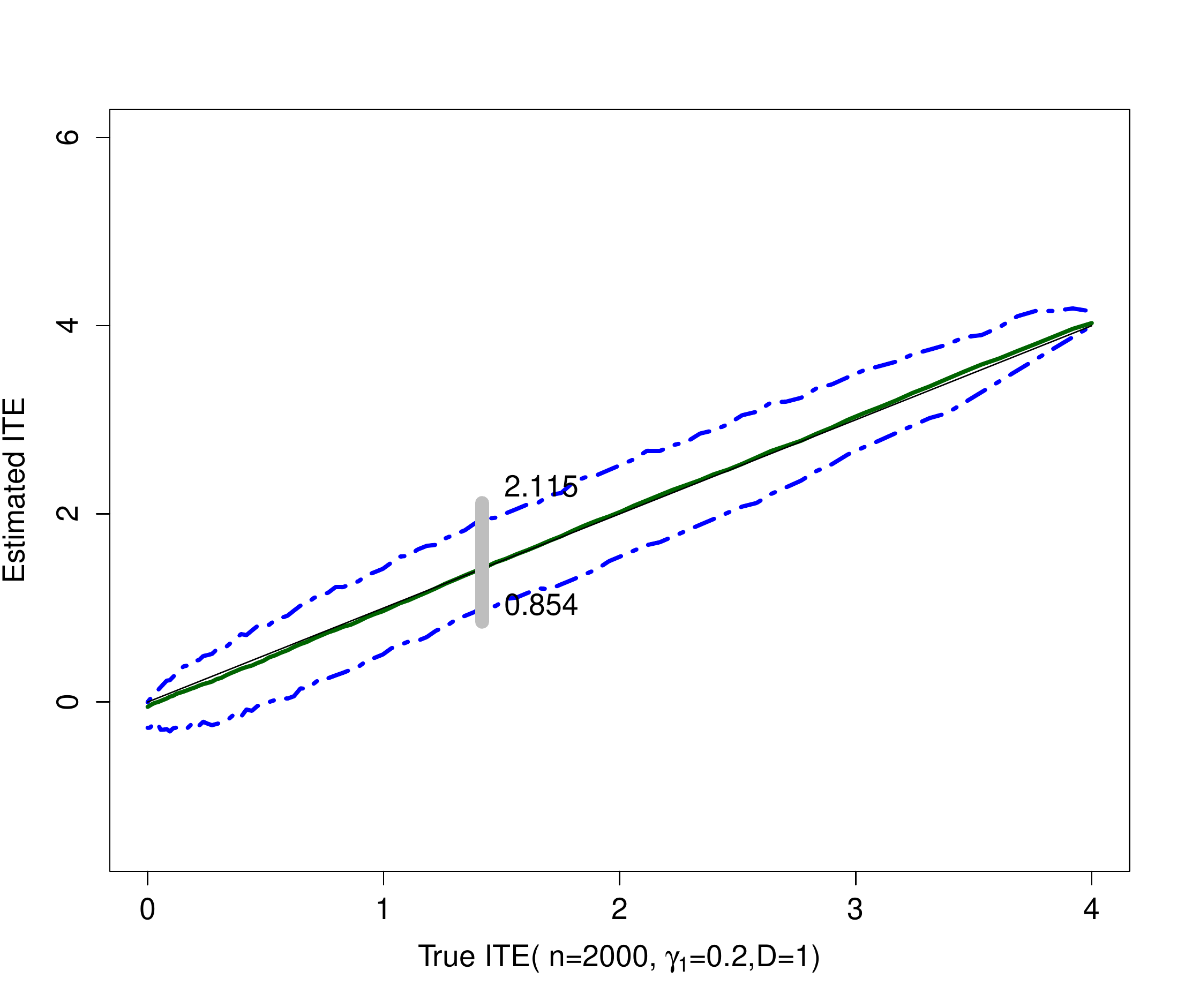} 
      \includegraphics[width=2in]{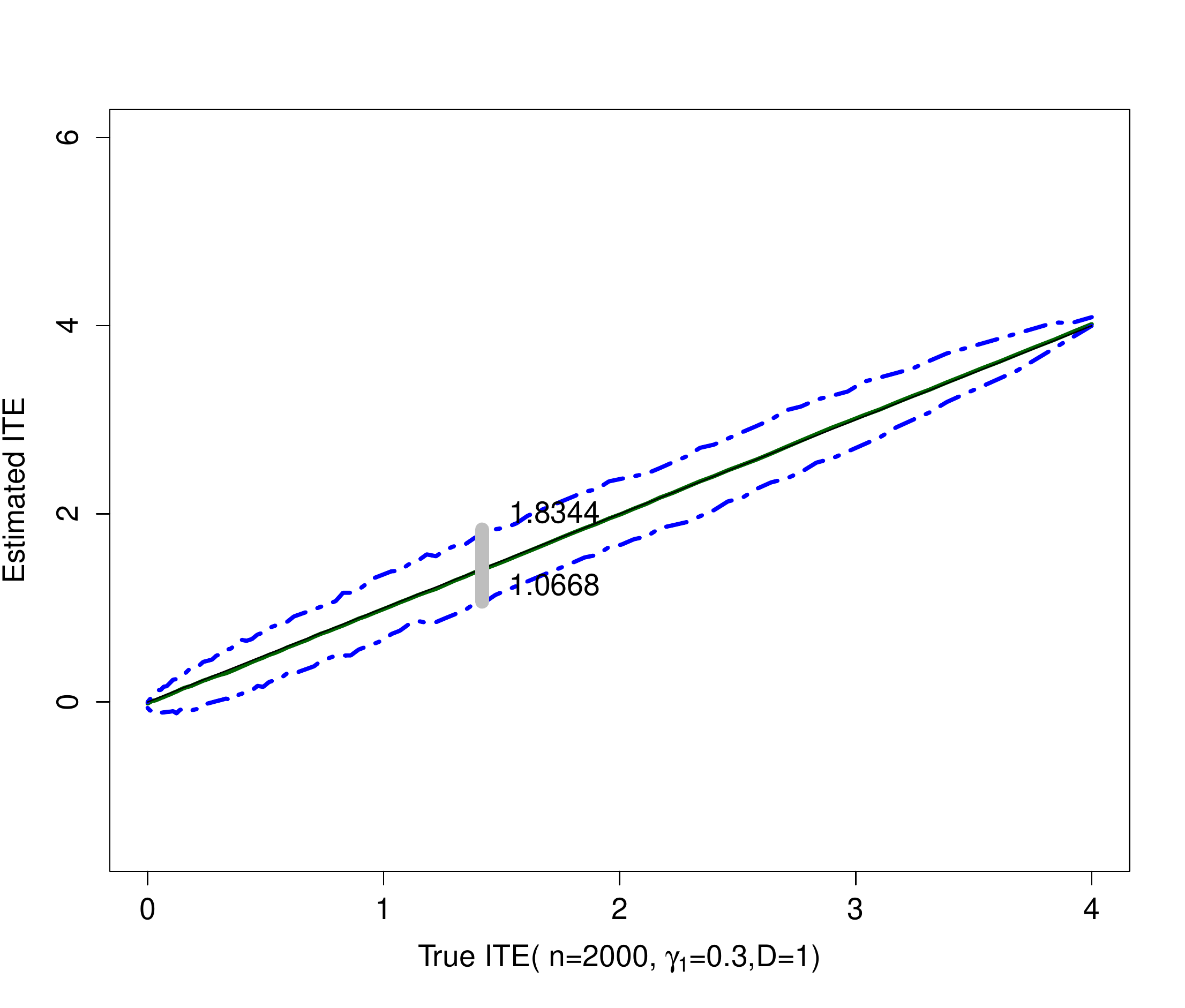} \\
         \includegraphics[width=2in]{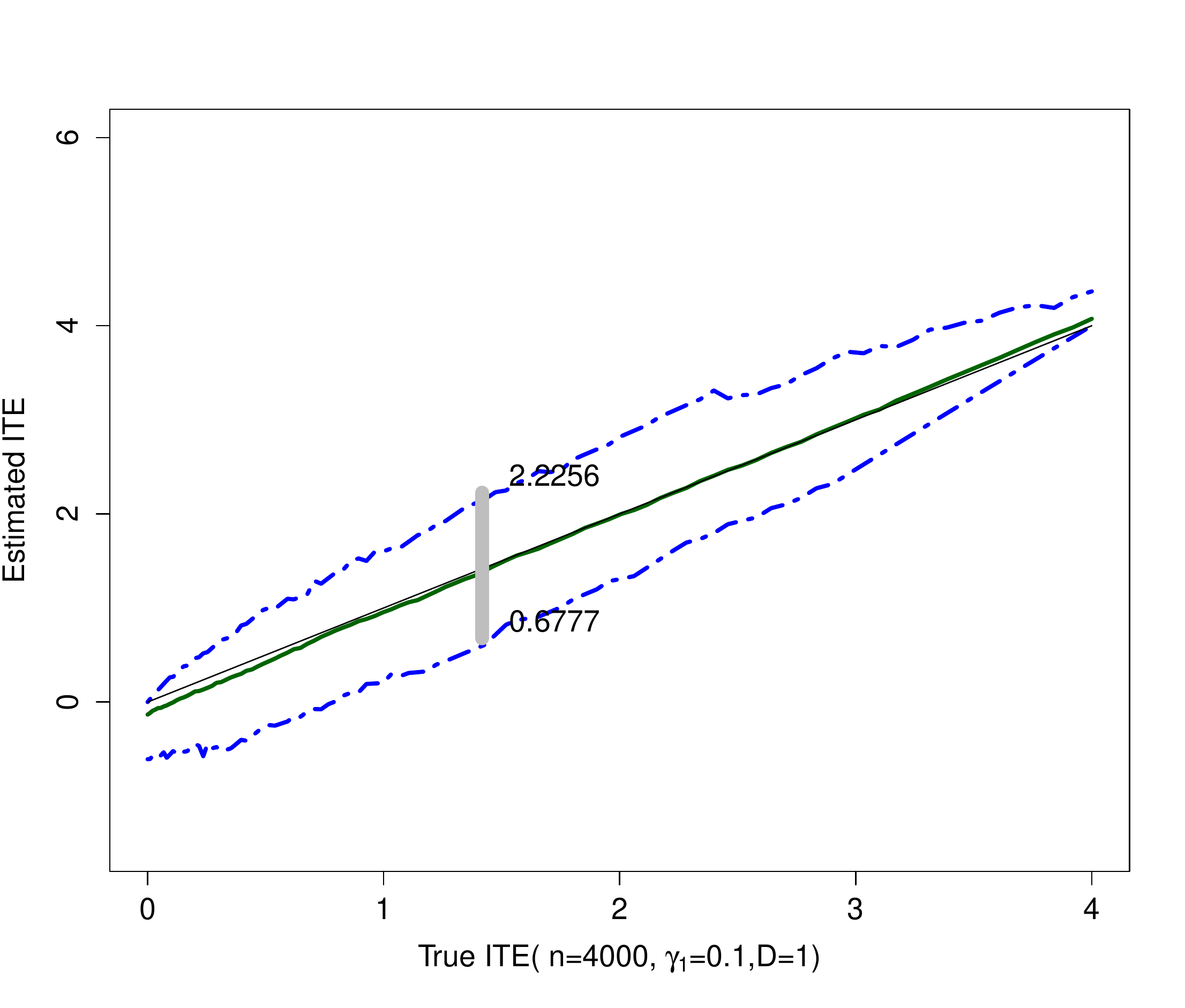} 
   \includegraphics[width=2in]{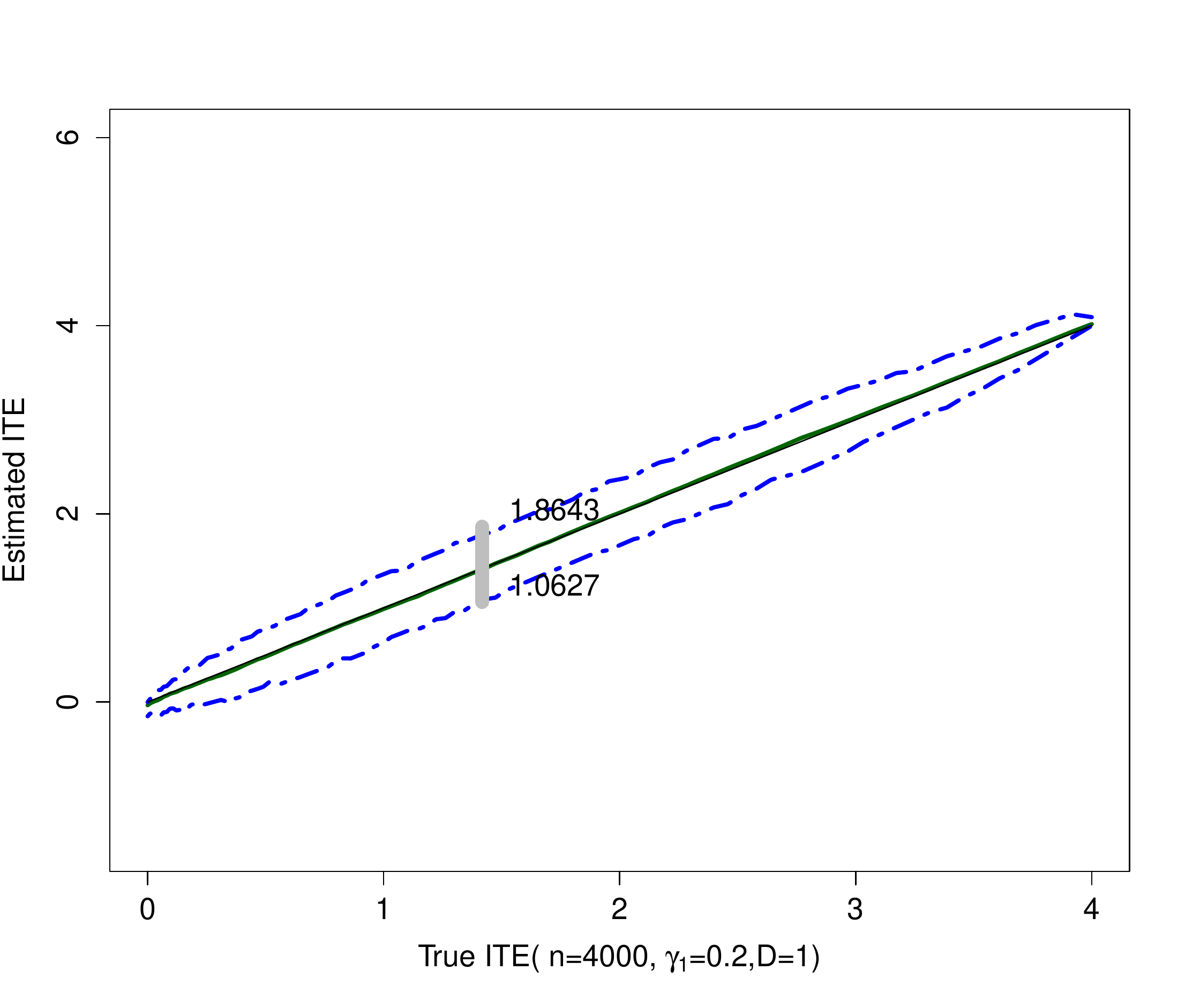} 
      \includegraphics[width=2in]{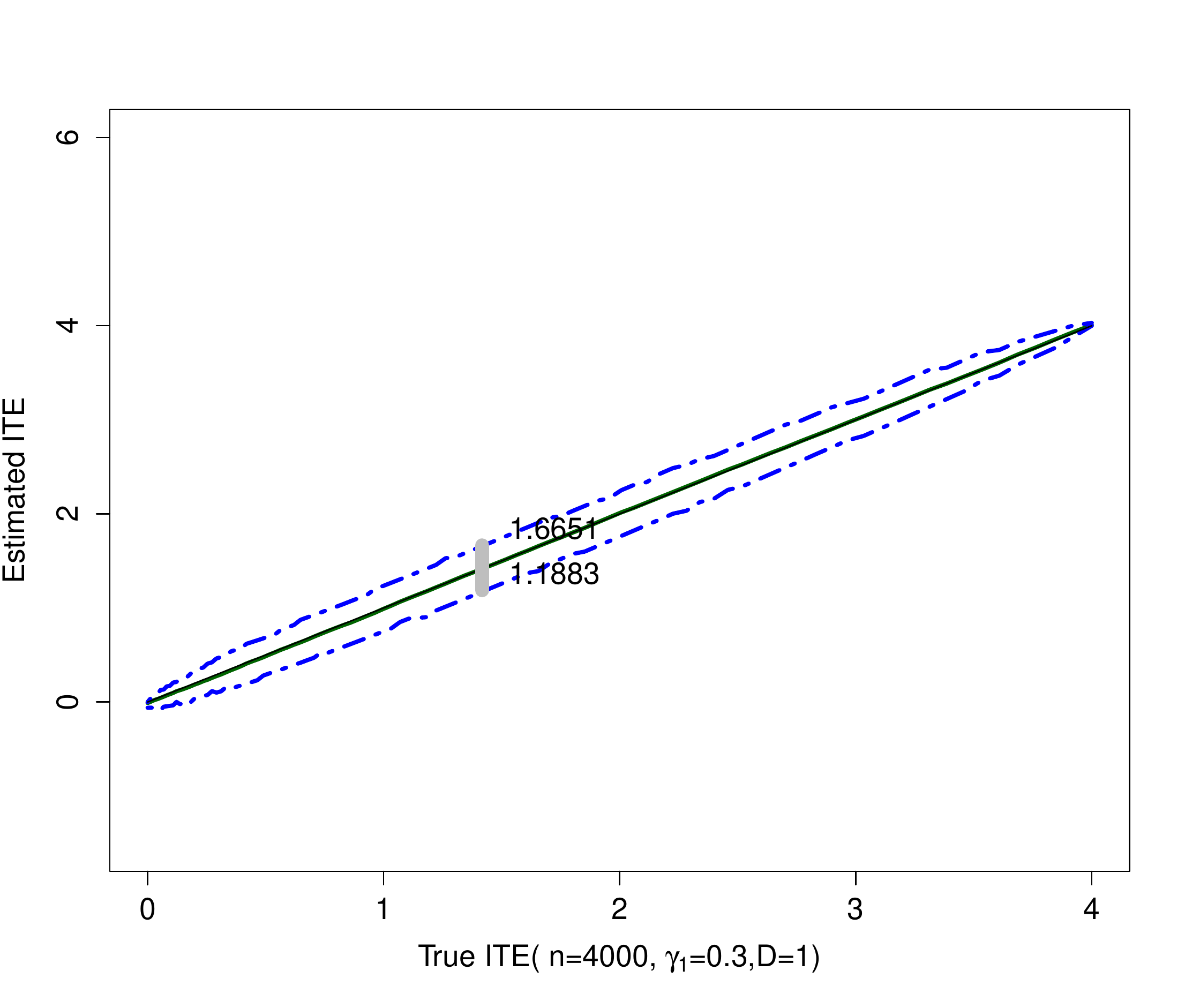} \\
   \caption{True and estimated ITE for $D=1$}
   \label{fig2_d1}
\end{figure}

\Cref{fig2_d0,fig2_d1} illustrate the performance of the ITE estimates for the $n$ individuals with $D=0$ and $D=1$, respectively. In particular, we   plot  the ITE estimates versus the true ITE.  The green solid line is the mean and the dotted lines give the $90\%$ confidence interval computed from the 200 repetitions. The grey solid line is the 45--degree diagonal.  The ITE estimates for the group $D=1$ behave better than the estimates for $D=0$. This observation is also consistent with our asymptotic results in the next section:  The performance of $\hat \Delta$ of an individual with $D=d$ depends on the density function of $h(d',x,\epsilon)$, evaluated at her quantile in the distribution, conditional on the compliers group (and $X=x$ as well). In our setting,  the conditional density of $h(0,\epsilon)$ given the compliers group is larger uniformly at all quantiles than that of $h(1,\epsilon)$, which leads to a more accurate estimator $\hat \Delta$ for the group $D=1$.
For comparison, we also plot the true value of LATE with the 90\% confidence interval of its estimate in grey color columns. Overall, estimates of ITE  and LATE behave similarly. Note that for any individual in the group $D=1$, our estimator of the ITE  behaves better than LATE. 

\begin{figure}[h] 
   \centering
   \includegraphics[width=2in]{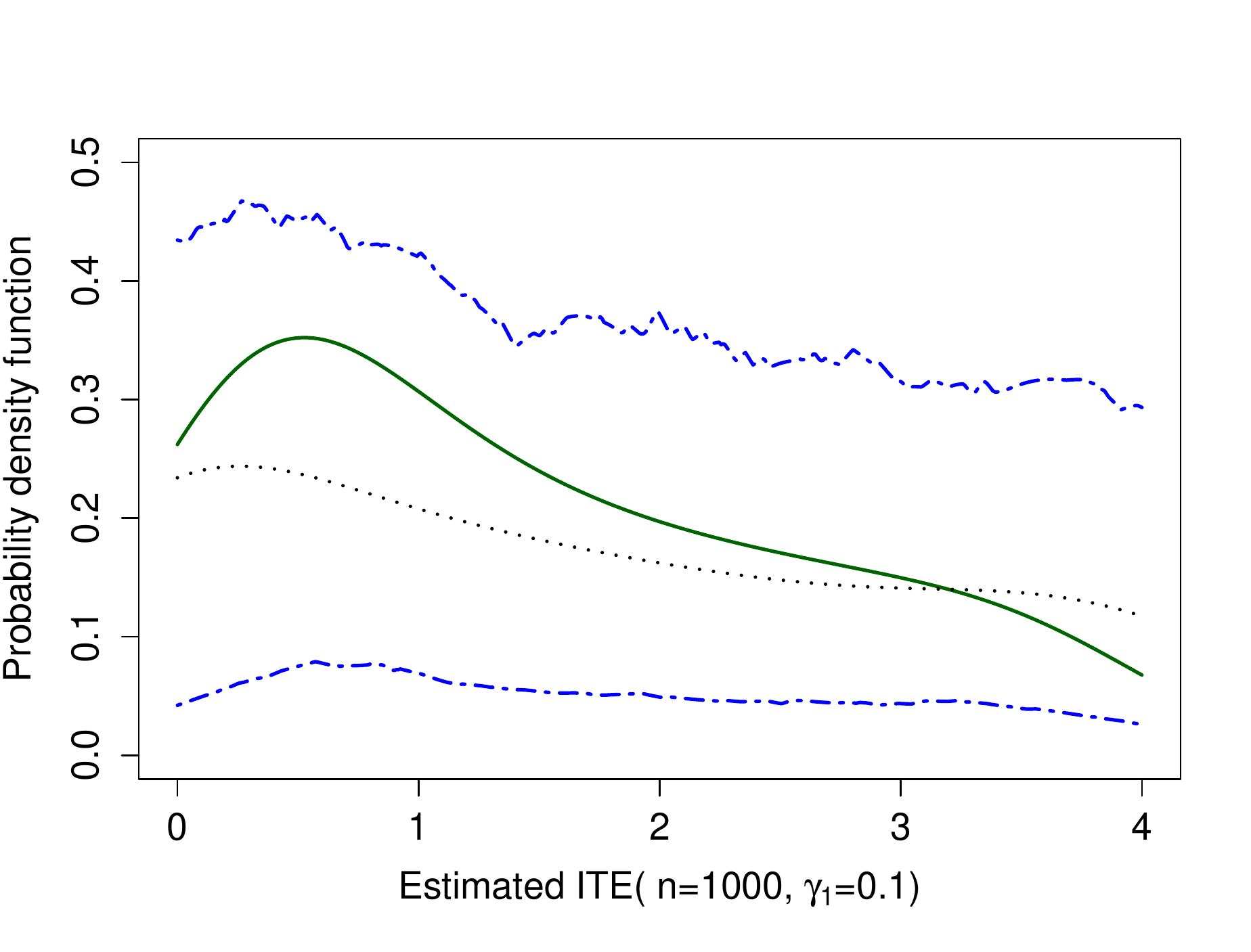} 
     \includegraphics[width=2in]{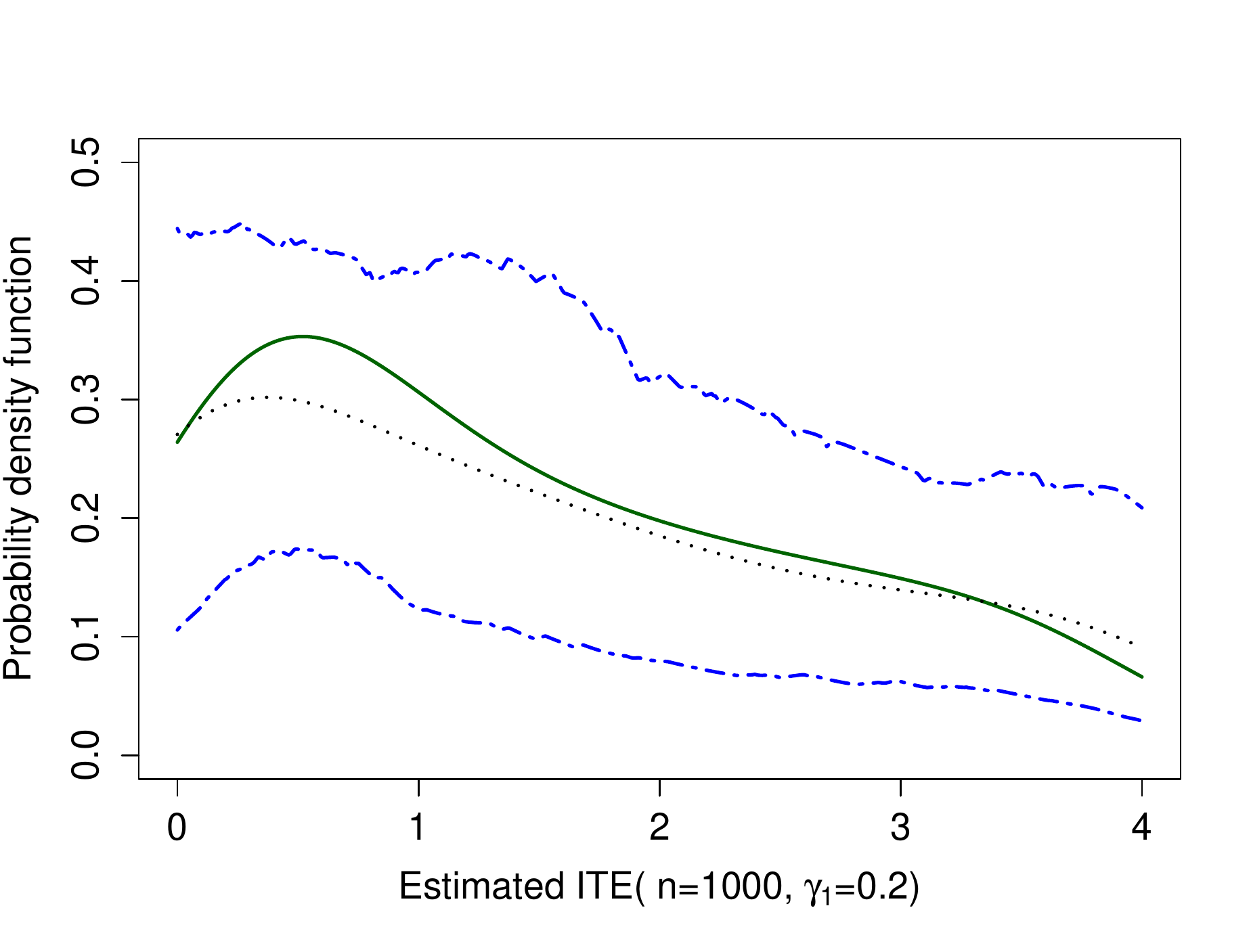} 
       \includegraphics[width=2in]{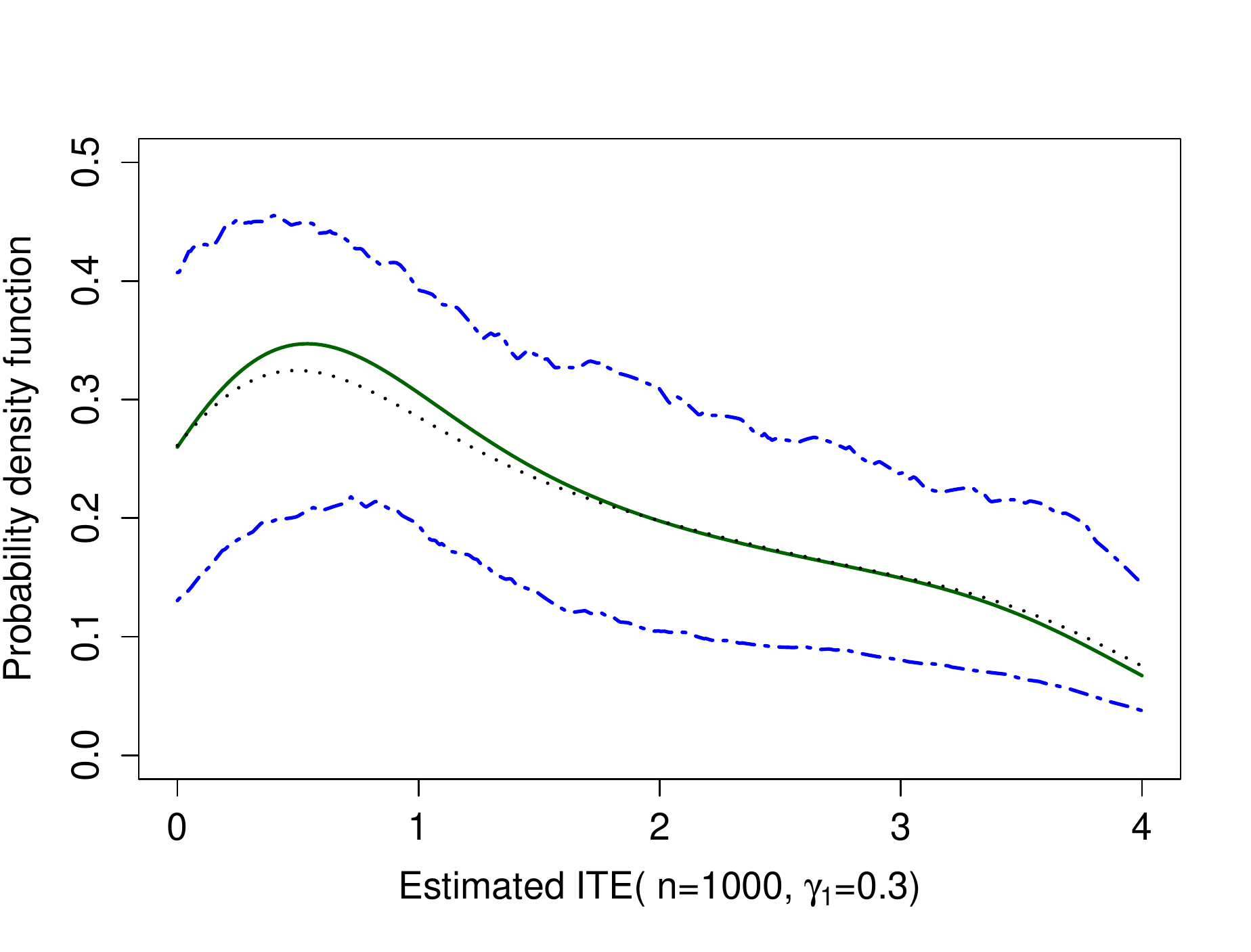} \\
         \includegraphics[width=2in]{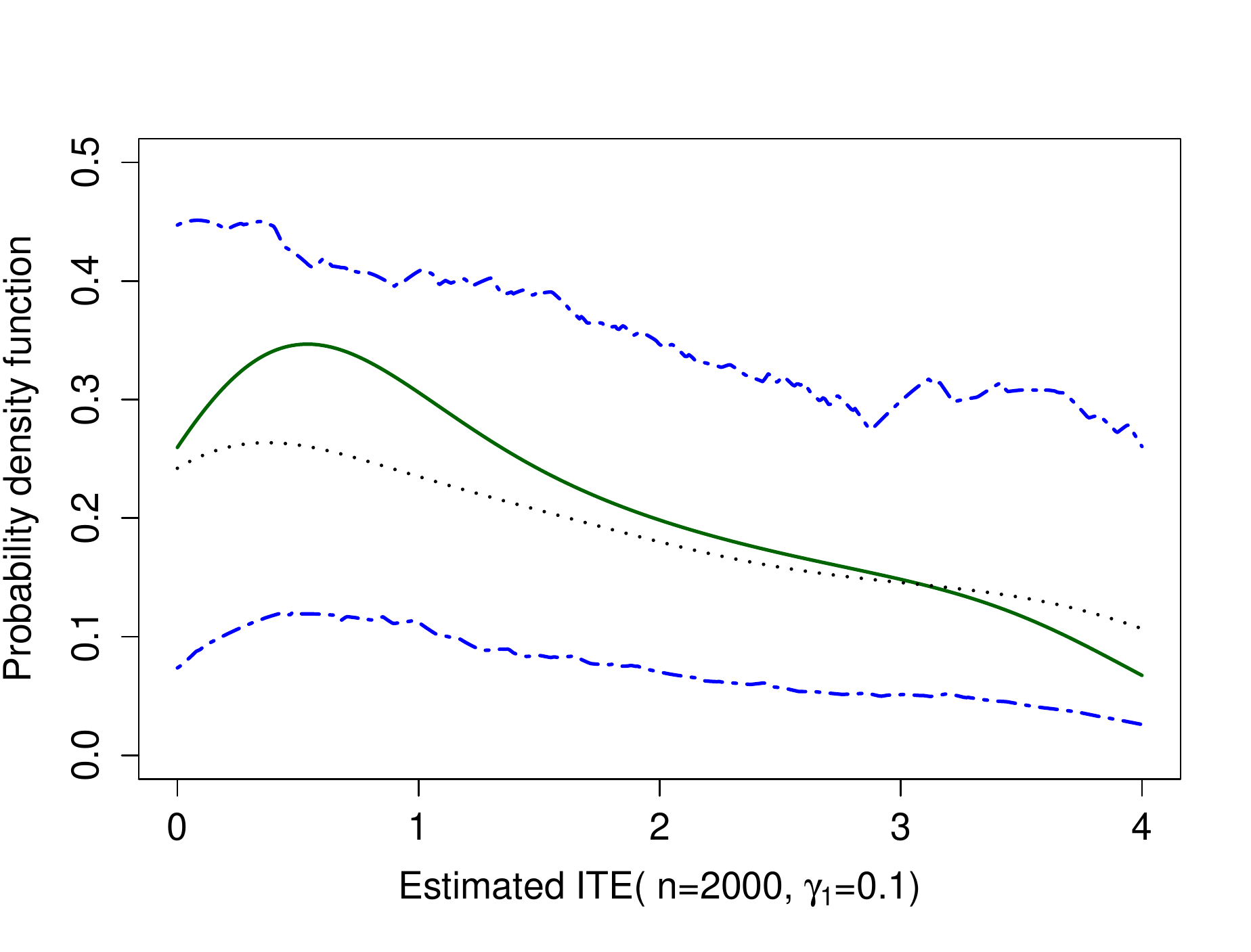} 
           \includegraphics[width=2in]{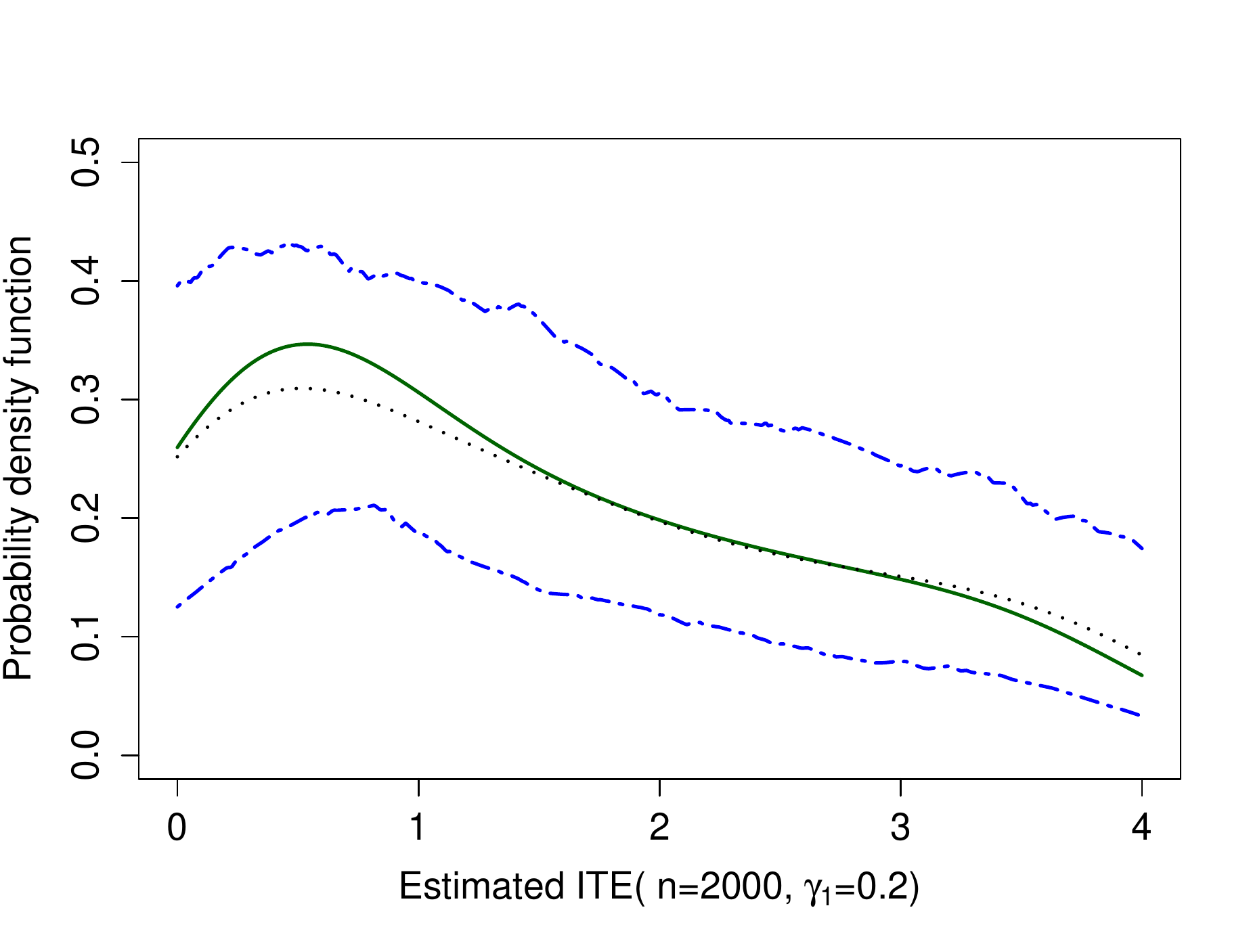} 
             \includegraphics[width=2in]{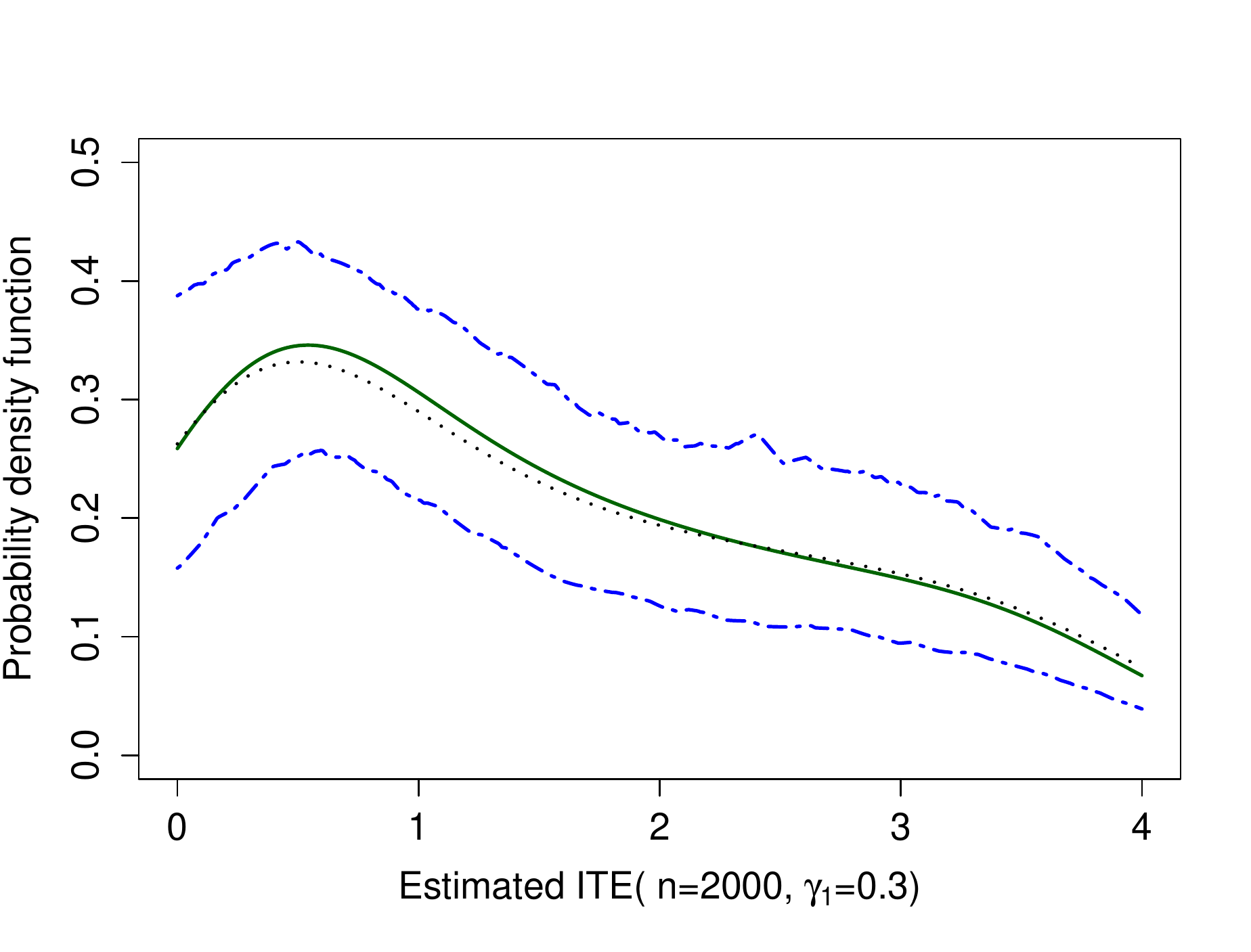} \\
               \includegraphics[width=2in]{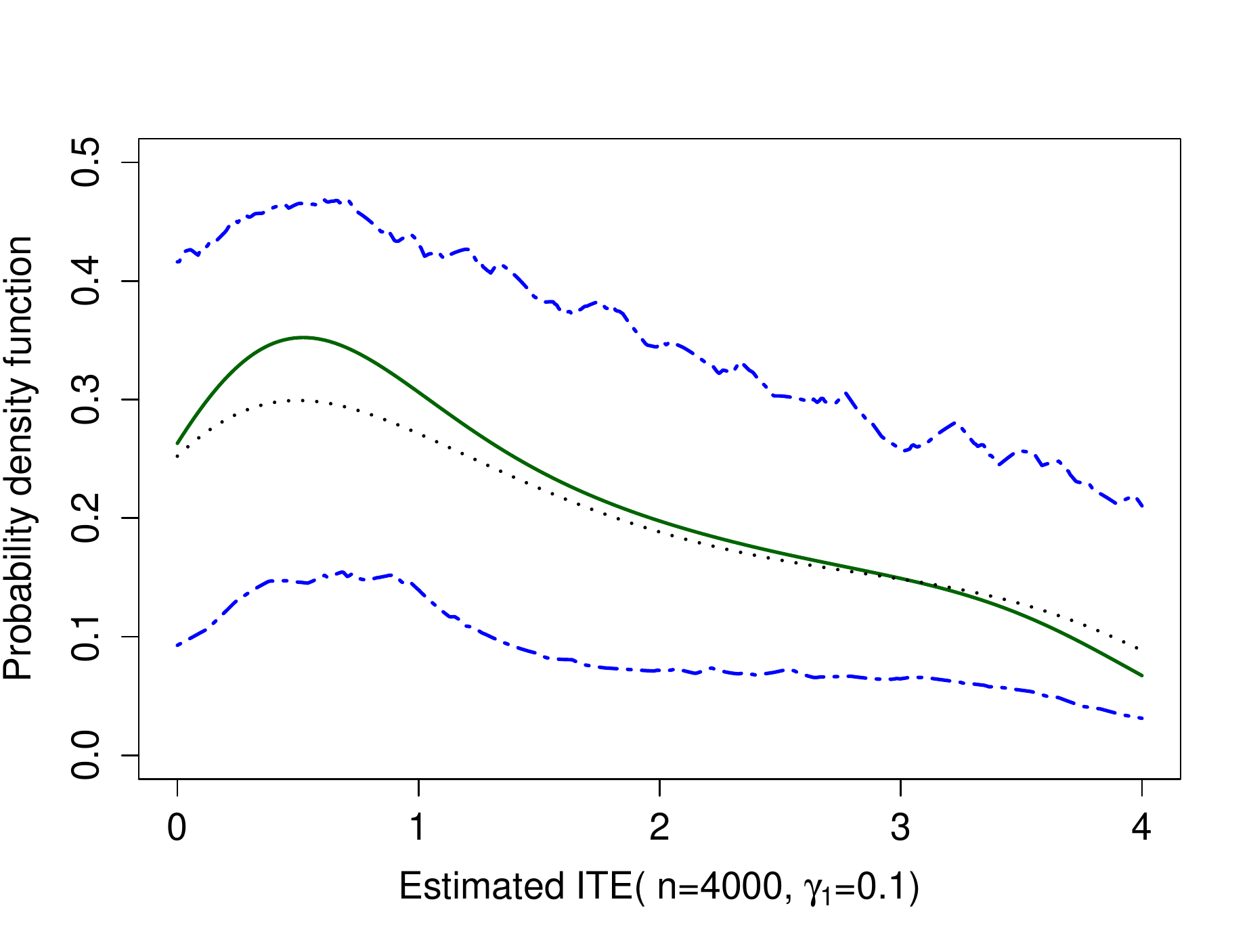} 
                 \includegraphics[width=2in]{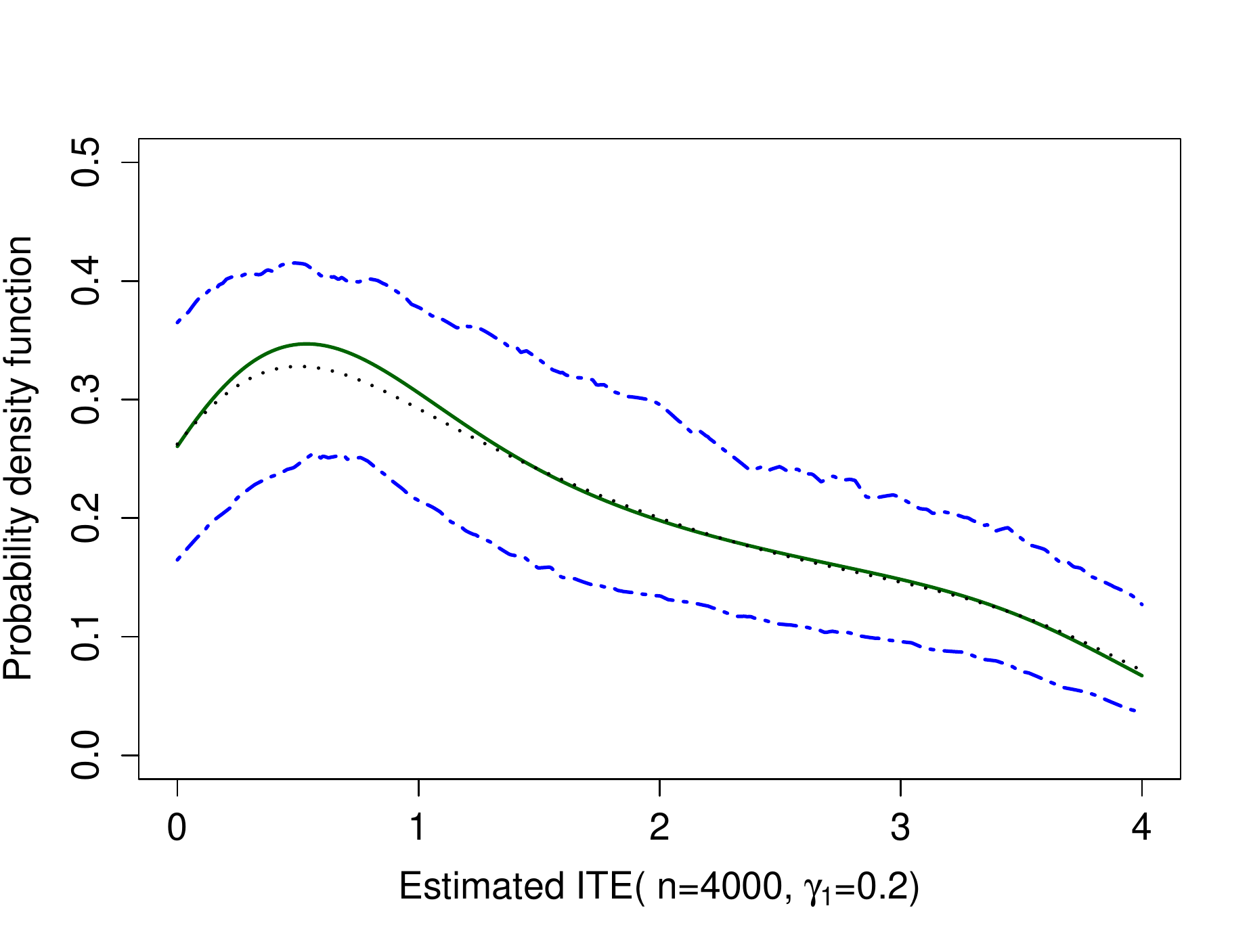} 
                   \includegraphics[width=2in]{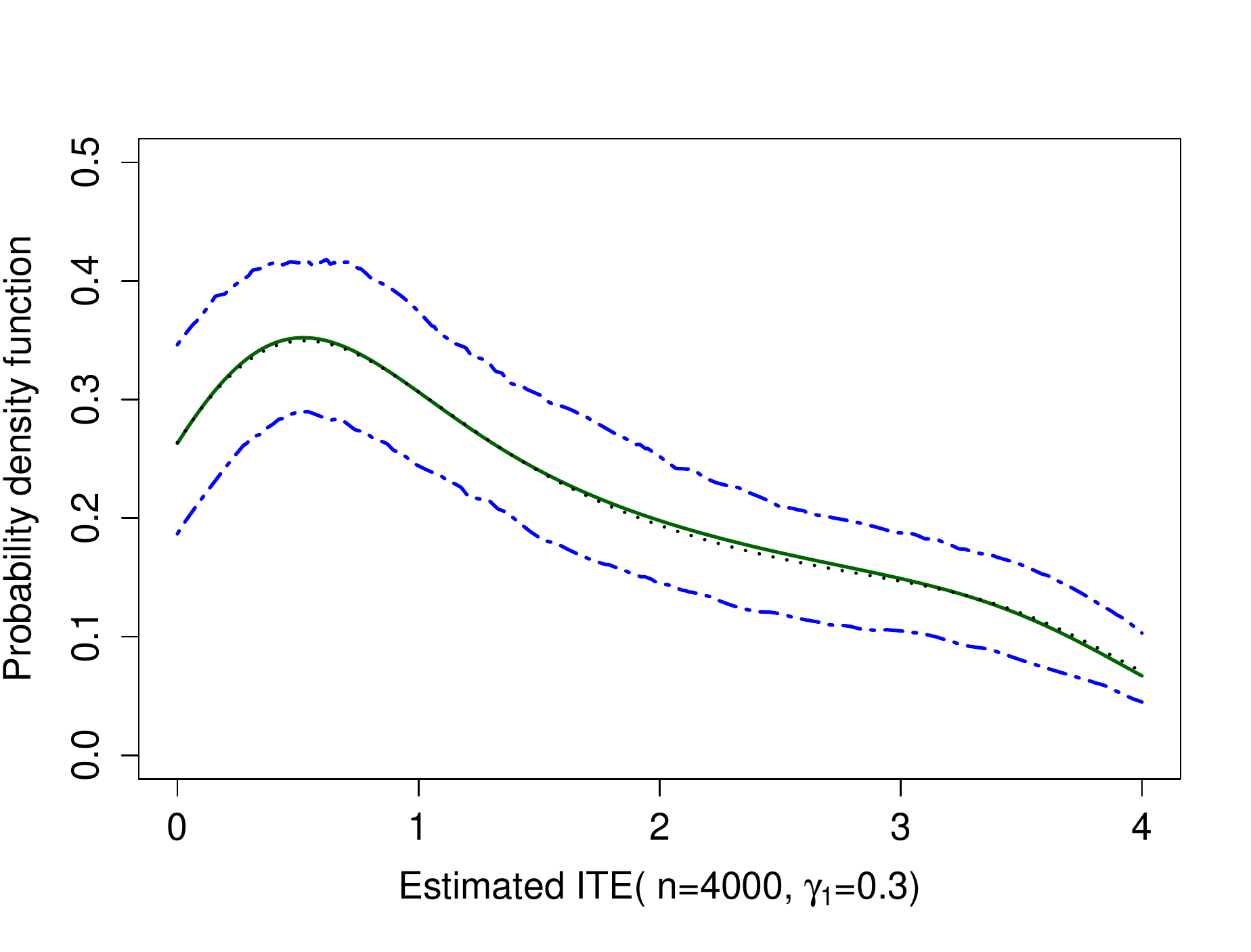} 
   \caption{Estimated density of ITE}
   \label{fig3}
\end{figure}

For the density estimator,  we choose the bandwidth $h= (\ln n/n)^{1/7}$ and the pdf of the standard normal as the kernel function. \Cref{fig3} shows the performance of our density estimator $\hat f_\Delta$.  The black dotted line  is the true density of the ITE and the green one is the average of our density estimates $\hat f_{\Delta}$ over the 200 repetitions. We also provide the $5\%$ and $95\%$ percentiles of estimated densities using blue dotted lines, which  gives the (pointwise) 90\% confidence band. \Cref{fig3} shows again the importance of the size of the complier group through $n\gamma^2_1$. 

\section{Asymptotic Properties}
We now establish the asymptotic properties of our proposed nonparametric estimators. We first show the uniform $\sqrt n$--consistency  of  the counterfactual mapping estimator $\hat \phi_{dx}$, and we give its limiting distribution.  We then establish the asymptotic properties of our density estimator $\hat f_\Delta$ taking into account the first-step estimation of $\Delta$.

For estimation, we strengthen Conditions (i) and  (iii) in \Cref{th1}, respectively,  to

\medskip
\noindent
\textbf{Condition (i)'}:  $h$ is continuously differentiable  and strictly increasing in $\epsilon$. 

\medskip
\noindent
\textbf{Condition (iii)'}: The conditional distribution of  $(\epsilon, \nu)$ given $X$ is absolutely continuous with respect to Lebesgue measure.  Moreover,  the conditional density function $f_{\epsilon|X}(\cdot|x)$ is continuous for all $x\in\mathscr S_X$.

\medskip\noindent
 Under Conditions (i)', (ii) and  (iii)', the conditional distribution $F_{Y|DXZ}$ is absolutely continuous with respect to Lebesgue measure and its density $f_{Y|DXZ}$ is also continuous. Therefore,  the complier distribution  $C_{dx}(\cdot)$ defined by \eqref{eq4} is also  absolutely continuous with respect to Lebesgue measure for $d=0,1$.  Let  $c_{dx}(\cdot)$ be its density. 
 
To simplify the exposition, we introduce the following assumption.

\medskip\noindent
\textbf{Assumption 1}: For every $(d,x)\in\mathscr S_{D,X}$, (i) $\mathscr S_{Y|D=d,X=x}=[\underline{y}_{dx},\overline{y}_{dx}]$, where $ \underline{y}_{dx}$ and $\overline{y}_{dx}$ are finite, and (ii) $\inf_{y\in\mathscr C_{dx}} c_{dx}(y)>0$. Moreover, (iii) $X$ is a vector of discrete random variables with a finite support.

\medskip\noindent
When $\mathscr S_{Y|D=d; X=x}$ has an unbounded support, we can always apply a known strictly increasing bounded continuous transformation to $Y$ to satisfy Assumption 1-(i).  Assumption 1-(ii) requires that the density $c_{dx}$ be bounded away from zero on its support.  It can be relaxed at the cost of technical complications due to e.g. some trimming.  As indicated earlier, Assumption 1-(iii) can be relaxed to allowed for continuous variables in $X$ by introducing some smoothing methods such as kernel ones. 

The next theorem establishes the uniform consistency of the counterfactual mapping estimator $\hat \phi_{dx}(\cdot)$ on its full support.  It also gives its $\sqrt{n}$--asymptotic distribution.
For $d=0,1$ and $y\in\mathscr C_{dx}$, let $c^*_{dx}(y)= c_{dx}(y)\cdot[p(x,1)-p(x,0)]$ be the scale--adjusted complier density and $R_{dx}(y)=\Pr (h(d,X,\epsilon)\leq y|X=x)$ be the probability rank of $y$ in the distribution of $h(d,x,\epsilon)$ given $X=x$. Under the monotonicity of $h$ and the definition of $\phi_{d'x}$, we have 
\[
R_{dx}(y)=\Pr (Y\leq y;D=d|X=x)+\Pr (Y\leq \phi_{d'x}(y); D=d'|X=x)
\]
where $d'=1-d$. 

\begin{theorem}
\label{theorem2}
Suppose the conditions in  \Cref{th1}, Conditions (i)', (iii)' and Assumption 1 hold. Then,  for $d=0,1$ and $d'=1-d$, we have 
\[
\sup_{y\in\mathscr S_{Y|D=d; X=x}} |\hat \phi_{d'x}(y)-\phi_{d'x}(y)|=o_p(1).
\] 
Moreover, the empirical process $c^*_{d'x}(\phi_{d'x}(\cdot))\times \sqrt n \big(\hat \phi_{d'x}(\cdot)-\phi_{d'x}(\cdot)\big)$ converges in distribution to a zero--mean Gaussian process with covariance kernel  
\[
\Sigma_{d'x}(y,y')=\frac{R_{dx}(\min\{y,y'\})-R_{dx}(y)\times  R_{dx}(y')}{\Pr(Z=0|X=x) \Pr(Z=1|X=x)}.
\]
\end{theorem}
\medskip
\noindent
The uniform convergence of $\hat \phi_{d'x}$ includes the boundaries, which is due to  Assumption 1-(ii).   Moreover, letting $y=y'$ in $\Sigma_{d'x}$ gives the asymptotic variance of $c^*_{d'x}(\phi_{d'x}(y))\times\sqrt n \ \hat\phi_{d'x}(y)$ as follows:
\[
\sigma^2_{d'x}(y)\equiv \frac{R_{dx}(y)-R^2_{dx}(y)}{\Pr(Z=0|X=x) \Pr(Z=1|X=x)}.
\] 
As $y$ approaches its boundaries, the asymptotic variance decreases to zero. Therefore, we obtain a more accurate estimate of the counterfactual outcome when it is closer to the boundary points. We also note that the asymptotic variance of $\hat \phi_{d'x}(y)$ is inversely proportional to $c^{*2}_{d'x}(\phi_{d'x}(y)) = c_{d'x}^2(\phi_{d'x}(y))\times [p(x,1)-p(x,0)]^2$, but is independent of the magnitude of ITE. 

\Cref{theorem2} is important for several reasons. First, given an arbitrary triplet $(y,d,x)$, we can provide a $ \sqrt n$--consistent estimate $\hat \phi_{d'x}(y)$ of  the counterfactual outcome $\phi_{d'x}(y)$ whenever $y\in \mathscr S_{Y|D=d,X=x}$ and $x\in \mathscr S_{X}$. Its standard error is given by 
\[
\frac{1}{\sqrt n\times   \hat c_{d'x}(\hat \phi_{d'x}(y)) [\hat p(x,1)-\hat p(x,0)]}\sqrt{\frac{\hat R_{dx}(y)-\hat R^2_{dx}(y)}{\hat \Pr(Z=0|X=x)\hat \Pr(Z=1|X=x)}}.
\]where $\hat R_{dx}(y)$ and $\hat \Pr (Z=z|X=x)$ are sample frequencies,   and 
\begin{multline}
\label{complier_density_est}
\hat c_{dx}(\cdot )\times [\hat p(x,1)-\hat p(x,0)]= (-1)^{d}\hat f_{Y|DXZ}(\cdot |d,x,z)\ \hat \Pr (D=d|X=x,Z=0)\\
-(-1)^{d} \hat f_{Y|DXZ}(\cdot|d,x,z) \hat \Pr (D=d|X=x,Z=1),
\end{multline}
in which $\hat f_{Y|DXZ}(y|d,x,z)$ is a kernel density estimator and $\hat \Pr (D=d|X=x,Z=z)$ are sample frequencies. Equation \eqref{complier_density_est} follows from differentiating \eqref{eq4}.
Second, given the uniform $\sqrt n$--consistency of $\hat \phi_{d'x}$, it follows that $\hat \Delta_i$ also uniformly converges to $\Delta_i$ at the $\sqrt n$--rate.

Next, we  turn to the asymptotic properties of our density estimator $\hat f_{\Delta}$. 

\medskip\noindent
\textbf{Assumption 2}: (i) On some interval $[\underline \delta, \overline \delta]$ of $\mathscr S_{\Delta}$, the density function $f_{\Delta}$  admits up to $P$--th  continuous bounded derivatives with $P\geq 1$. Moreover, $\inf_{\delta\in [\underline \delta, \overline \delta]} \ f_{\Delta}(\delta)>0 $. (ii) The kernel $K(\cdot)$ is a  symmetric $P$-th order kernel with support $[-1,+1]$ and twice continuously bounded derivatives.\footnote{A $P$-th order kernel is a function integrating to one and satisfying 
$\int u^{p} K(u)d u=0$ if $1\leq p\leq P-1$ and
$<\infty$  if $p=P$. }
(iii) The bandwidth $h\propto (\ln n/n)^{1/(2P+2)}$.

\medskip\noindent
The first part of Assumption 2-(i) is a high level condition requiring that the random variable $h(1,X,\epsilon)-h(0,X,\epsilon)$ has a smooth density function  conditional on $X=x$.  It is satisfied if $h(d,x,\cdot)$ for $d=0,1$ and the density of $\epsilon$ given $X$ are$P$--th continuously differentiable.   The second part of Assumption 2-(i) is standard for kernel estimation.  Assumptions (ii) and (iii) relate to the choice of the kernel function $K$ and bandwidth $h$, respectively. In particular, following \cite{guerre2000optimal}, the bandwidth in (iii) leads to oversmoothing relative to the optimal bandwidth, i.e., $h^*\propto (\ln n/ n)^{\frac{1}{2P+1}}$ \citep[see][]{stone1982optimal}.

Given Assumption 2 and the uniform convergence of $\hat \Delta$  to $\Delta$ at the $\sqrt n$--rate, we show in the Appendix that the first--step estimation error is asymptotically negligible in $\hat f_\Delta$. Thus, we obtain the following result.

\begin{theorem}
\label{theorem3}
Suppose the conditions in \Cref{theorem2} and Assumption 2 hold. Then, 
\[
\sup_{\delta\in [\underline \delta+h, \overline \delta-h]}|\hat f_{\Delta}(\delta)-f_{\Delta}(\delta)|=O_p\big((\ln n/n)^{\frac{P}{2P+2}}\big).
\]
\end{theorem}

\medskip\noindent
Note that the convergence rate in \Cref{theorem3} is uniform over the expanding interval $[\underline \delta+h,\overline \delta-h]$. It is slower than the optimal convergence rate if the ITEs were observed, which is
$(\ln n/n)^{\frac{P}{2P+1}}$ \citep[see][]{stone1982optimal}.

\section{Individual Effects of 401(k) Programs}
In this section we apply our estimation method to study the effects of 401(k) retirement programs on personal savings.  The 401(k) retirement programs were introduced in the early 1980s to increase savings for retirement. Since then, they became increasingly popular in the US. It has been argued in the literature  that participants might self--select into the programs non-randomly  \citep[see, e.g.,][]{poterba1996401kplan}.  People with a higher preference for savings are more likely to participate and have higher savings than those with lower preferences. 

Following e.g. \cite{abadie2003401kplan} and \cite{cherno2004401kplan}, we use 401(k) eligibility as an instrumental variable for 401(k) participation. This is because 401 (k) plans are provided by employers. Hence, only workers in firms that offer such programs are eligible so that the monotonicity in \eqref{eq_2} is satisfied.\footnote{\cite{imbens1994identification} define monotonicity as: $D_i(z_1)\leq D_i(z_2)$ for all $i$, where $D_i(z)$ is the potential treatment status at  $Z=z$. In our application, $Z$ is 401(k) eligibility and $D_i(0)=0$. Therefore, $D_i(0)\leq D_i(1)$ a.s., i.e., \cite{imbens1994identification}'s monotonicity condition holds. Moreover, \cite{vytlacil2002independence} show that such a condition is observationally equivalent to the functional monotonicity in \eqref{eq_2}.}

\subsection{Data}
The dataset consists of 9,275 observations from the Survey of Income and Program Participation (SIPP) of 1991 as in  \cite{abadie2003401kplan}. The observational units are household reference persons aged 25-64 and spouse if present. The included households are those with at least one member employed, with Family Income in the $\$10$k -- $\$200$k interval.   Eligibility for 401(k) outside the interval is rare as noted by \cite{poterba1996401kplan}.


\begin{table}[h]\small
   \centering
      \caption{Summary statistics}
   \begin{tabular}{lccccc}
\hline\hline
                  & Entire sample & \multicolumn{2}{l}{By 401(k) participation}& \multicolumn{2}{l}{By 401(k) eligibility} \\
                  &                        & Participants & Non-participants & Eligibles & Non-eligibles\\\hline
        Treatment &&&&& \\
        401(k) Participation & $0.2762$ &&& $0.7044$ & 0.0000 \\
                                         & (0.4472) &&& (0.4564) & (0.0000) \\
         Instrument &&&&&\\
         401(k) Eligibility & 0.3921 & 1.0000 & 0.1601 && \\
                  & (0.4883) & (0.0000) & (0.3668) &&\\
          Outcome variable &&&&&\\
          FNFA & 19.0717 & 38.4730 & 11.6672 & 30.5351 & 11.6768 \\
          (in thousand \$)&(63.9638)&(79.2711)&(55.2892)&(75.0190)&(54.4202)\\
          Covariates:&&&&&\\
          Family income & 39.2546 & 49.8151&35.2243&47.2978&34.0661\\ 
          (in thousand \$)&(24.0900)&(26.814.2)&(21.6492)&(25.6200)&(21.5106)\\
          Age & 41.0802&41.5133&40.9149&41.4845&40.8194\\
          &(10.2995)&(9.6517)&(10.5323)&(9.6052)&(10.7163)\\
          Married& 0.6286&0.6956&0.6030&0.6772&0.5972\\
          &(0.4832)&(0.4603)&(0.4893)&(0.4676)&(0.4905)\\
          Family size&2.8851&2.9204&2.8716&2.9079&2.8703\\
          &(1.5258)&(1.4681)&(1.5472)&(1.4770)&(1.5565)\\  \hline
   \end{tabular}
   \label{summary}
  \end{table}

\Cref{summary} presents the summary statistics of the full sample as well as by eligibility and participation status. The dependent variable is the Family Net Financial Assets (FNFA), the treatment variable is the participation in 401(k), and the instrumental variable is the eligibility for 401(k). About $28\%$ in the sample participate in the program and $39\%$ are eligible for it. Other covariates include family income, age, marital status and family size. Similar to \cite{cherno2004401kplan}, age and income are grouped into categorical variables 0, 1, 2 and 3 by using  the 1st, 2nd and 3rd quartiles.


\begin{table}[h]\small
   \centering
      \caption{ Average FNFA (in thousand \$) sorted according to covariates}
   \begin{tabular}{lccc|cccc}
 \hline\hline
&    & Family income &  Age & & & Married &  Family size\\\hline
By percentile &  <0.25   &2.29&4.29          & By value & 0  & 12.83  \\
                       &              &(18.83)& (21.08)&                &     &(50.55)  \\
                      &   0.25--0.5     & 7.68 &14.49 &                & 1&22.76     &13.59   \\
                      &                      &(29.16) &(62.78) &           &   & (70.45)  &(47.59)\\
                      &   0.5--0.75     & 16.63&21.43 &                & 2&              &29.11 \\
                      &                       &(53.15)&(67.33) &           &   &               &(82.70)\\
                      &   >0.75           & 49.76&36.86 &                & 3&              & 19.17  \\
                      &                        &(104.87)&(87.31) &         &   &              &(66.86) \\
                      &                        &             &             &         & 4&              &17.53 \\
                      &                        &             &             &         &   &              &(56.83) \\
                      &                        &             &             &         &>4&             & 12.51  \\
                      &                        &             &             &         &    &             &(52.46)\\
                                            \hline
   \end{tabular}

   \label{table3}
\end{table}

%
%
%
%
%
%

\Cref{table3} provides the mean and standard error (in parentheses) of the outcome variable FNFA by percentiles sorted according to covariates. Clearly, FNFA is monotone increasing in family income and age. According to family size, FNFA is maximized at family size 2 and decreases with family size when it is larger than 2. Moreover, married households have higher FNFA than unmarried ones on average.

In \Cref{ols_2sls}, we  provide OLS and 2SLS estimates as a benchmark for comparison  with our ITE estimates. Our results replicate the estimates in \cite{abadie2003401kplan}. The OLS estimates in column (1) show a significantly positive association between participation in 401(k) and net financial assets given covariates. Furthermore, the 2SLS estimates in column (3) confirms the positive, but attenuated treatment effects after controlling for endogeneity of participation.  It turns out that FNFA increases rapidly with family income and age, and is lower for married couples and larger families. 

\begin{table}[h]\small
   \centering
   \caption{OLS and 2SLS estimates  of 401(k) participation} 
   \begin{tabular}{lcccc} 
  &&&&\\ \hline\hline
 \multicolumn{2}{c}{}&OLS& \multicolumn{2}{c}{2SLS}\\
    \cline{4-5}
   &&&First stage&Second stage\\ \hline
    Participation in 401(k) && 13.5271 &&9.4188 \\
    & &(1.8103) &&(2.1521)\\
    Constant && 10.0421 &0.0567&9.0076\\
    &&(10.9142) &(0.0464)&(10.9559)\\
    Family income (in thousand \$) &&0.9769 &0.0013&0.9972\\
    &&(0.0833)&(0.0001)&(0.0838)\\
    Age && -2.3100 &-0.0048&-2.2386\\
    &&(0.6177) &(0.0023)&(0.6201)\\
    Age squared &&0.0387 &0.0001&0.0379\\
    &&(0.0077) &(0.0000)&(0.0077)\\
    Married && -8.3695 &-0.0005&-8.3559\\
    && (1.8299) &(0.0079)&(1.8290)\\
    Family size && -0.7856 &0.0006&-0.8190\\
    &&(0.4108) &(0.0024)&(0.4104)\\
    Eligibility for 401(k) &&&0.6883&\\
    &&&(0.0080)&\\ \hline
   \end{tabular}
   \caption*{\footnotesize Note: The dependent variable is family net financial assets (in thousand \$). Family income and age enter into the regression as continuous variables. The sample includes 9,275 observations from the SIPP of 1991. The observational units are household
reference persons aged $25$-$64$, and spouse if present, with Family Income in the \$$10$k-$\$200$k interval. Heteroscedasticity robust standard errors are given in parentheses.}
   \label{ols_2sls}
\end{table}

\subsection{ITE Estimates}
To begin with, we first check the support condition for identification, i.e. Condition (iv) in \Cref{th1}. Because those who are not eligible for 401(k)  (i.e. $Z=0$)  cannot participate  in the program,  then  $C_{1x}(\cdot)=\Pr(Y\leq \cdot|D=1,X=x,Z=1)$ by \eqref{eq4} and  $F_{Y|D=1,X=x}=F_{Y|D=1,X=x,Z=1}$. It follows that  $\mathscr C_{1x}=\mathscr S_{Y|D=1,X=x}$ for  all $x\in\mathscr S_X$. Hence, to check Condition (iv), it suffices to verify the support condition for $d=0$. To do so, we estimate the density function $c_{0x}$ by \eqref{complier_density_est} and the density function  $f_{Y|DX}(\cdot|0,x)$ directly from the data. 

\begin{figure}[h] 
   \centering
    \caption{Verifying the support condition}
   \includegraphics[height=3.2in,width=4in]{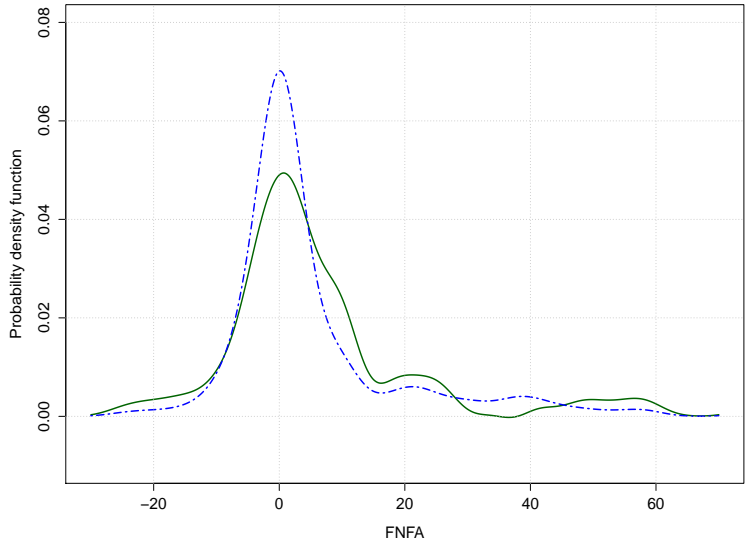} 
   \label{supp}
\end{figure}

Fix the subgroup of individuals whose income is between the 25\% and 50\% percentile, age  between 40 and 48 years old, and family size smaller than 3.\footnote{We repeat this for other values of covariates. The results are qualitatively similar.} \Cref{supp} plots the density estimate $\hat c_{0x}$ using the green solid line, and the density estimate $\hat f_{Y|DX}(\cdot |0,x)$  using the blue dotted line.  From \Cref{supp}, the two distributions roughly share the same support. 

\begin{figure}[h] 
   \centering
    \caption{The model restriction}
   \includegraphics[height=2.5in,width=3in]{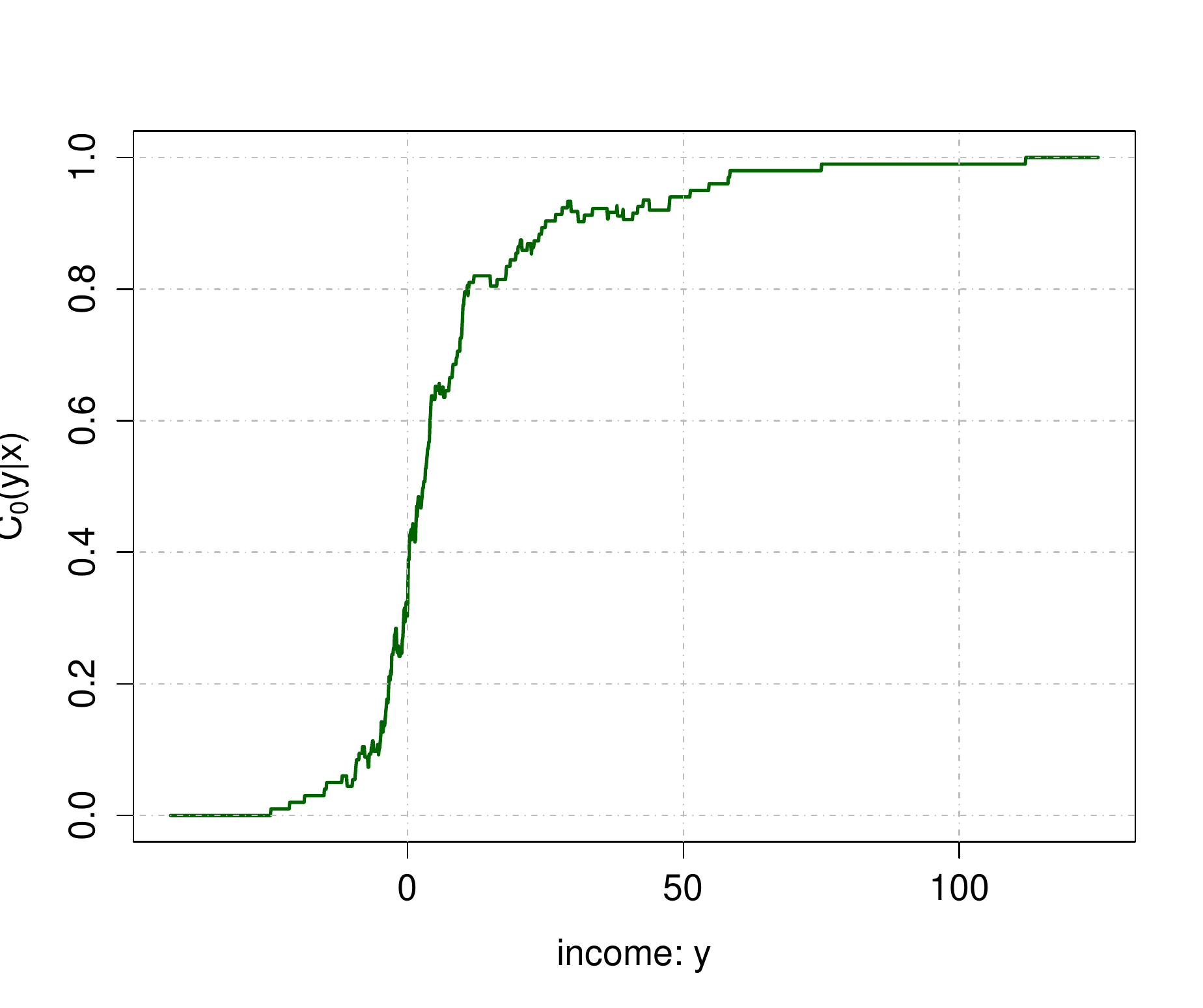} 
      \includegraphics[height=2.5in,width=3in]{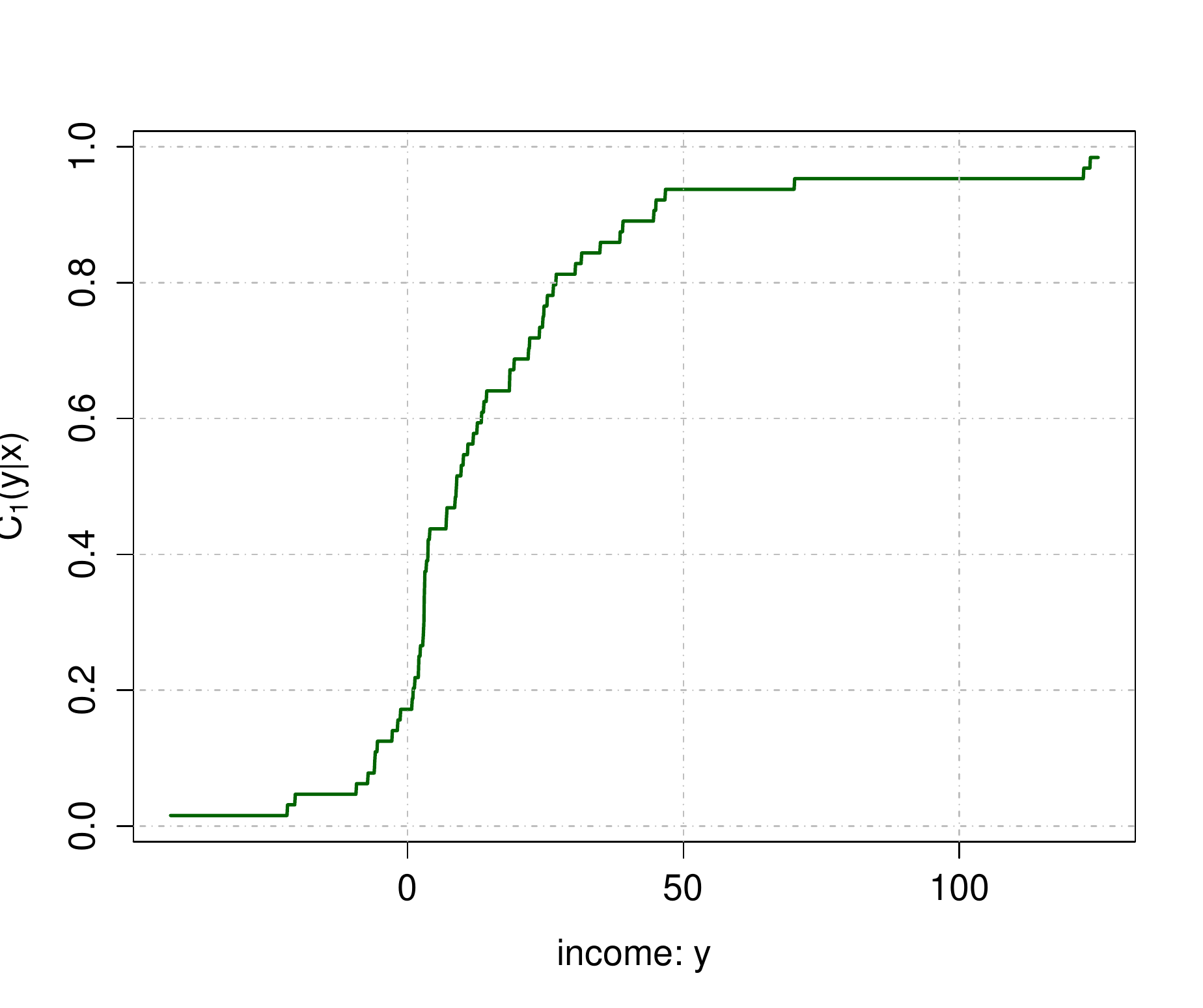} 
   \label{F0}
\end{figure}

Moreover, as shown in  \cite{vuong2014counterfactual}, the main restrictions imposed by our model require that  $C_{dx}(\cdot )$ defined by \eqref{eq4} should be monotone increasing for $d=0,1$ and all $x\in\mathscr S_X$. We plot estimates of $C_{0x}(\cdot )$ and $C_{1x}(\cdot )$ in \Cref{F0} for the subgroup of \Cref{supp}. Both of them are increasing functions globally.

\Cref{5num} reports  summary statistics of the ITE estimates in our sample. From \Cref{5num}, the ITE has a mean of \$$22.45$k and median $Q_2$ of \$$8.83$k, indicating a long right tail of the ITE distribution.  The mean of ITE  is larger than the average treatment effects (ATE) of OLS and 2SLS, which are \$13.53k and \$9.42k, respectively, while  the median of ITE turns out to be smaller than these two ATEs. The differences reflect the distortion due to the linear specification used in OLS and 2SLS, as well as the selection bias.

\begin{table}[h]
   \centering
   \caption{Summary of ITE estimates (in thousand dollars)} 
   \begin{tabular}{cccccccc} \hline\hline
& Min & Max &Mean& Std. &$Q_1$ & $Q_2$  & $Q_3$.  \\ \hline
&-918&1,533 &22.45 &102.77 &3.10&8.83&20.90 \\ \hline
        \end{tabular}
   \label{5num}
\end{table}

\begin{figure}[h] 
   \centering
    \caption{Estimated densities of ITE for full sample}
   \includegraphics[height=3.2in]{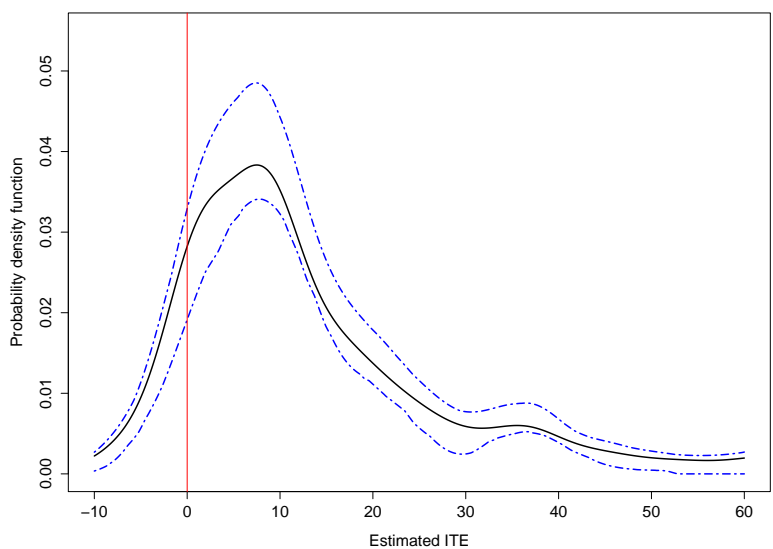} 
   \label{full_sample}
\end{figure}

\begin{figure}[h] 
   \centering
   \includegraphics[height=2.25in]{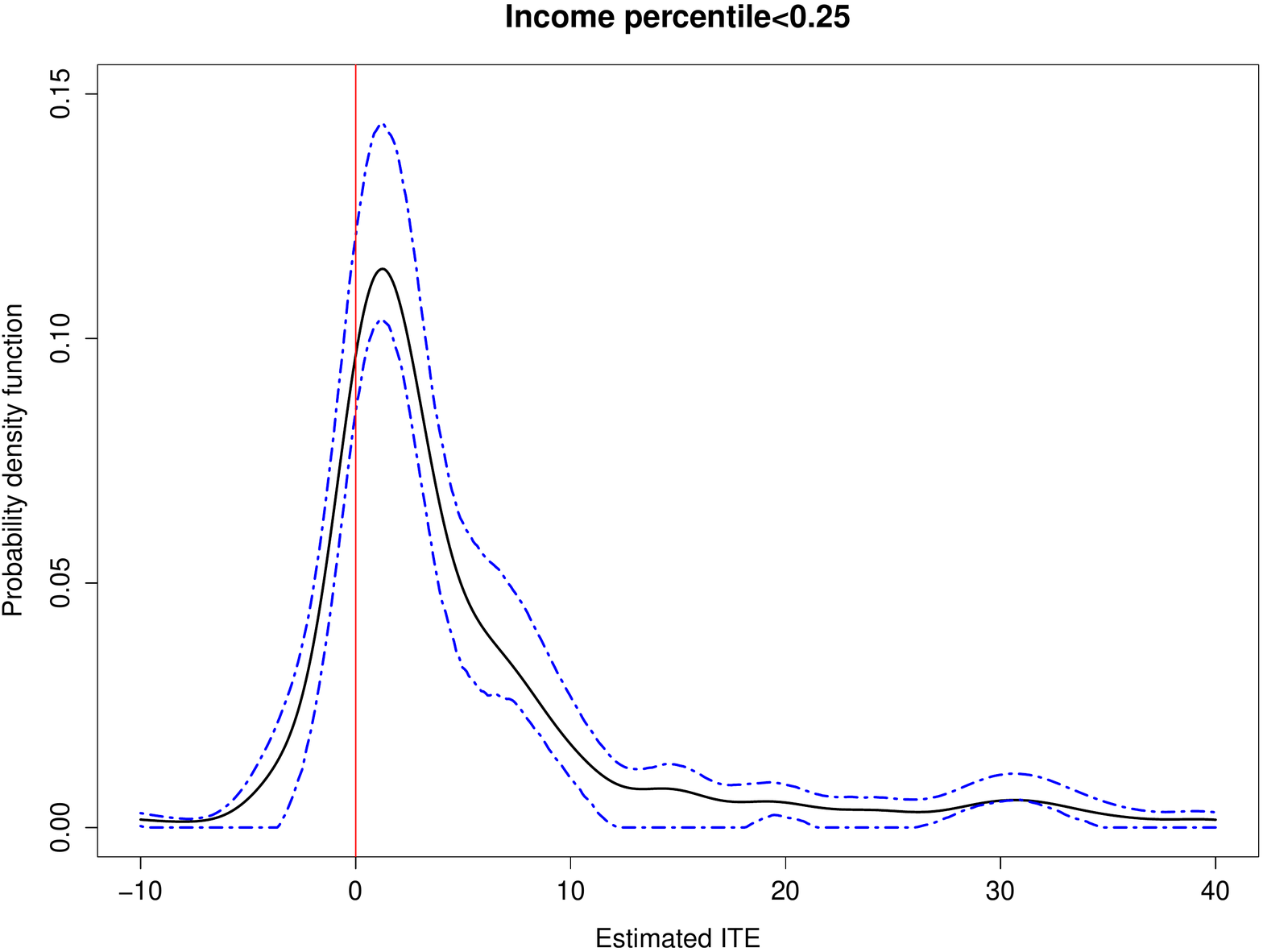}
   \includegraphics[height=2.25in]{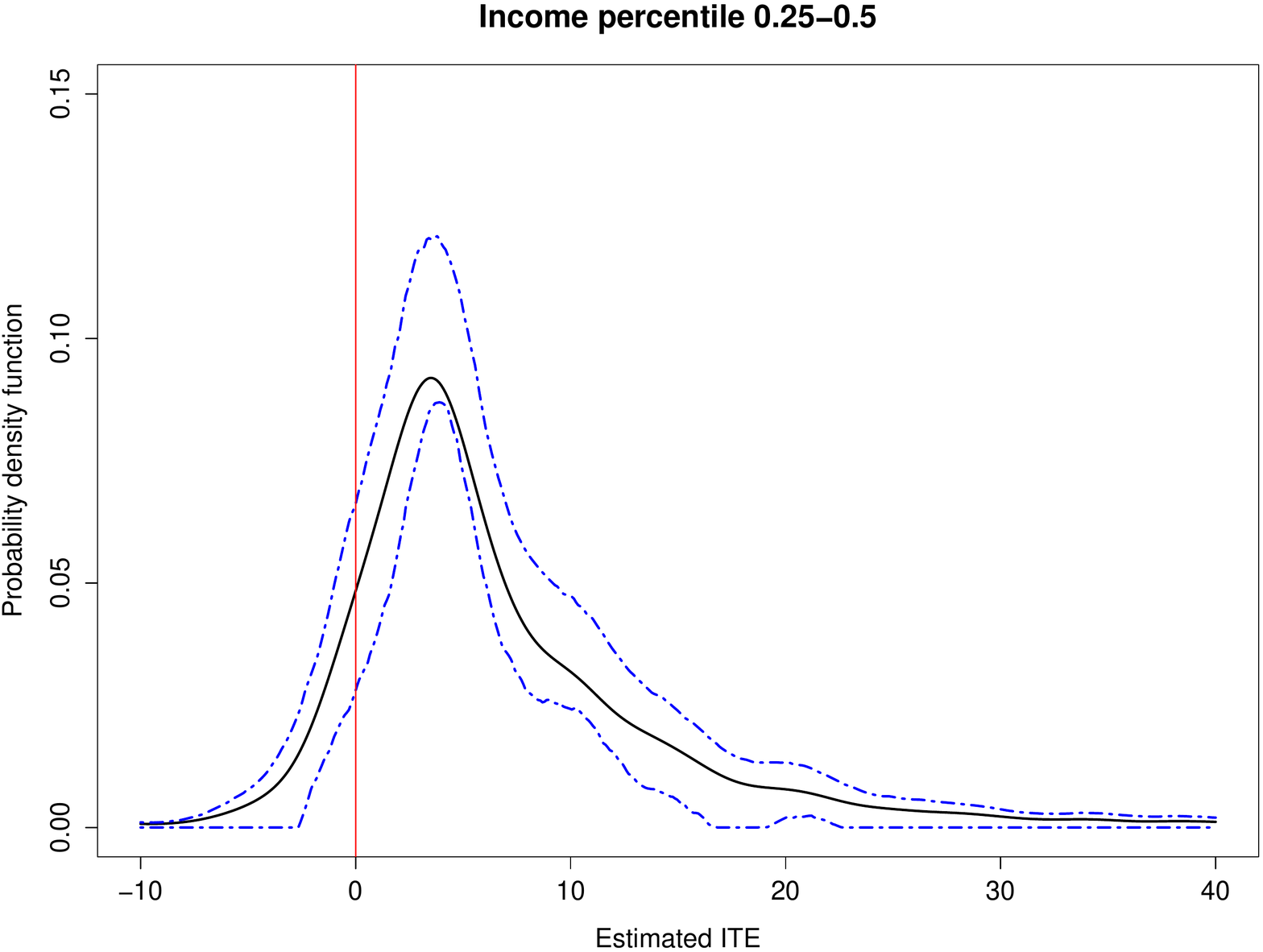}\\
    \includegraphics[height=2.25in]{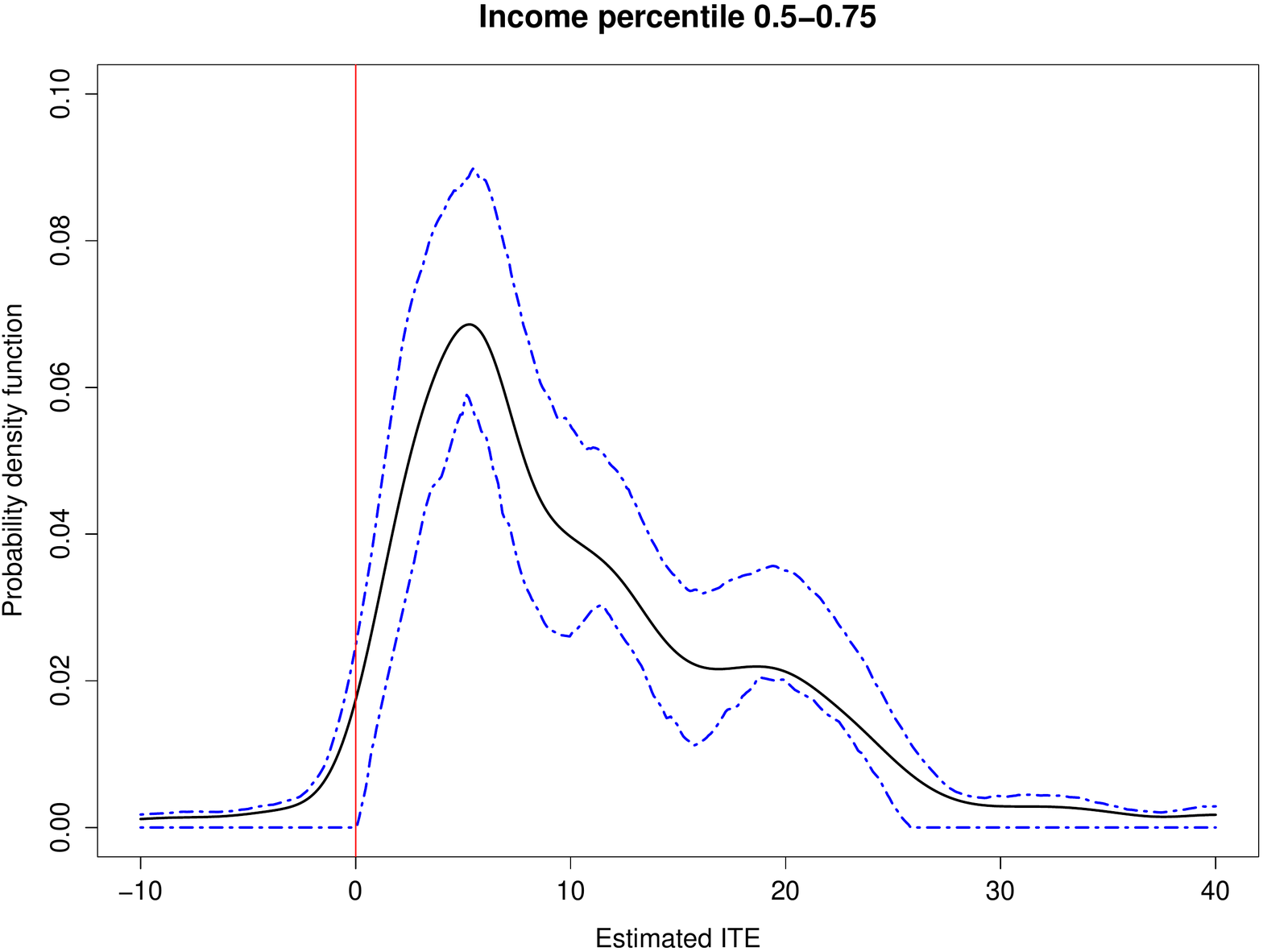}
    \includegraphics[height=2.25in]{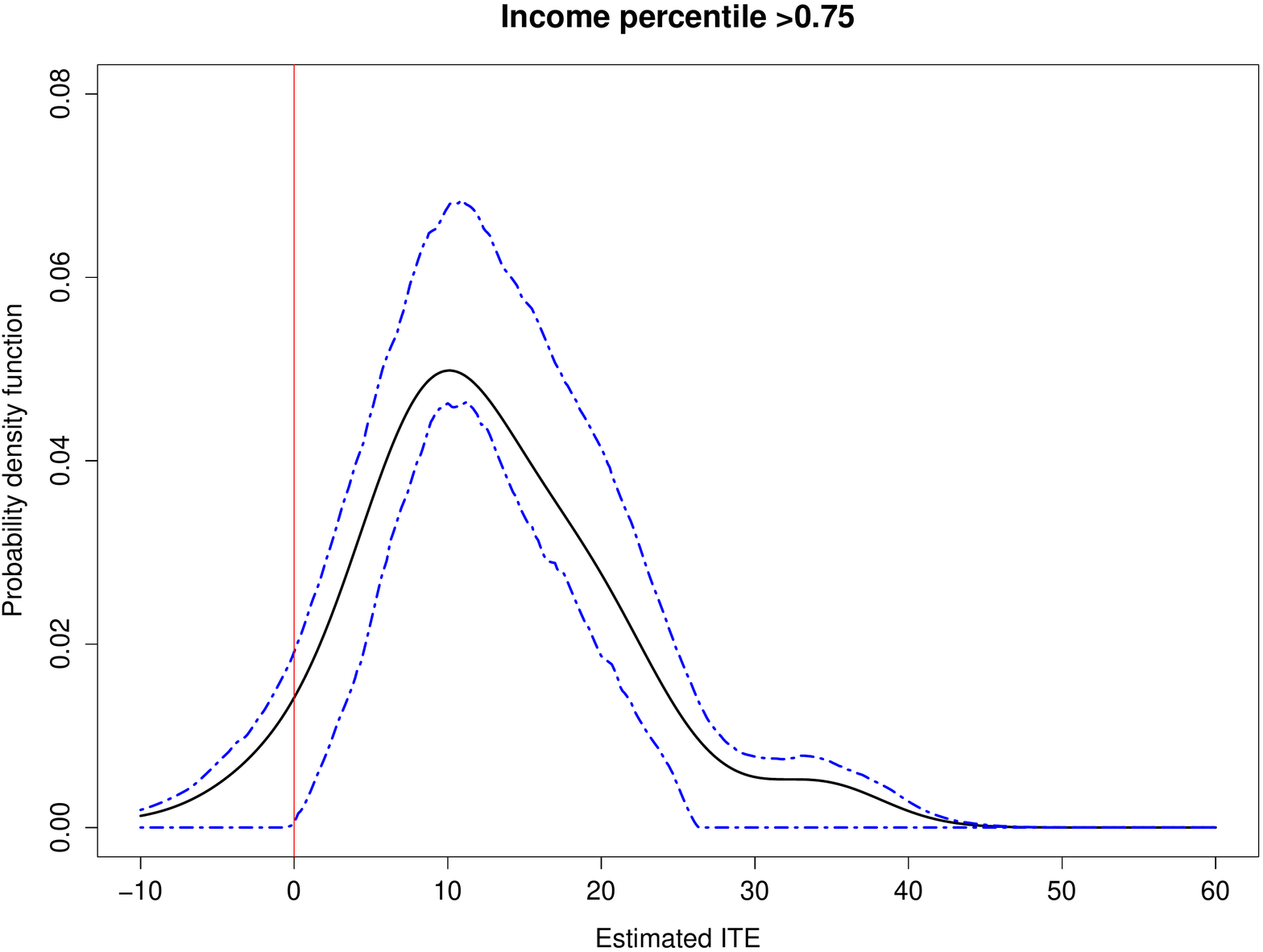}
   \caption{Estimated densities of ITE by income category}
   \label{ite_income}
\end{figure}

\begin{figure}[h] 
   \centering
   \includegraphics[height=2.25in]{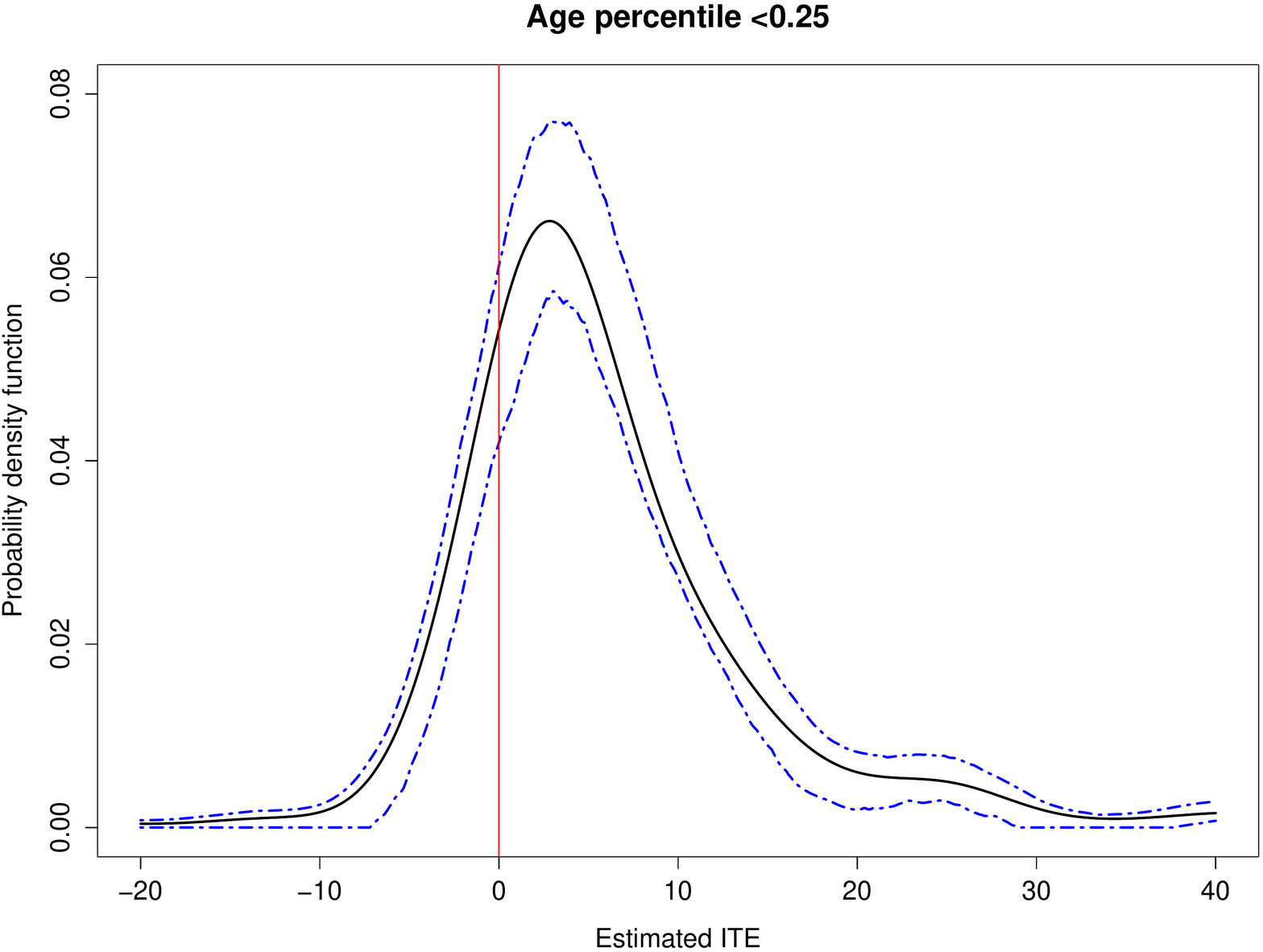}
   \includegraphics[height=2.25in]{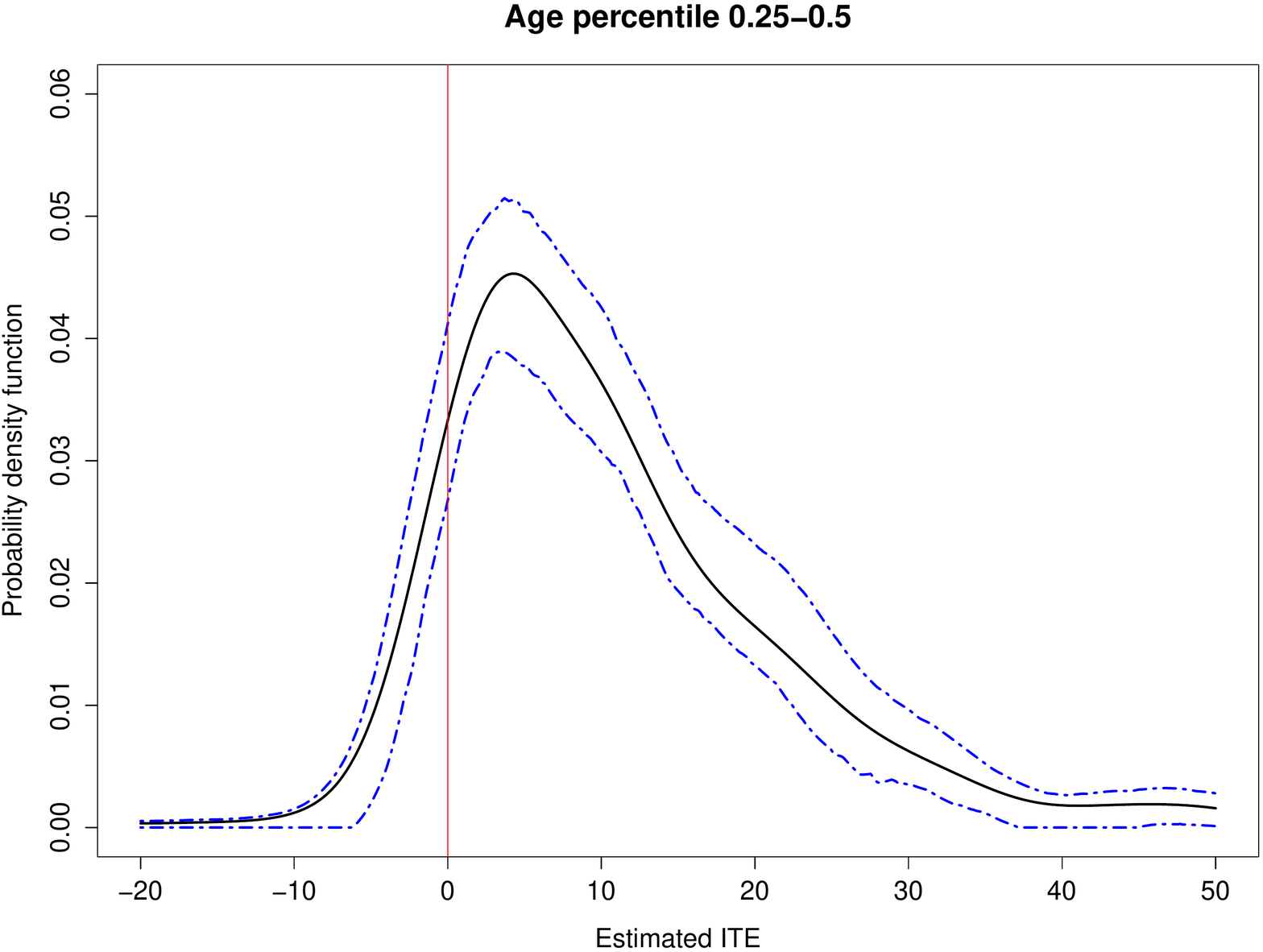}\\
    \includegraphics[height=2.25in]{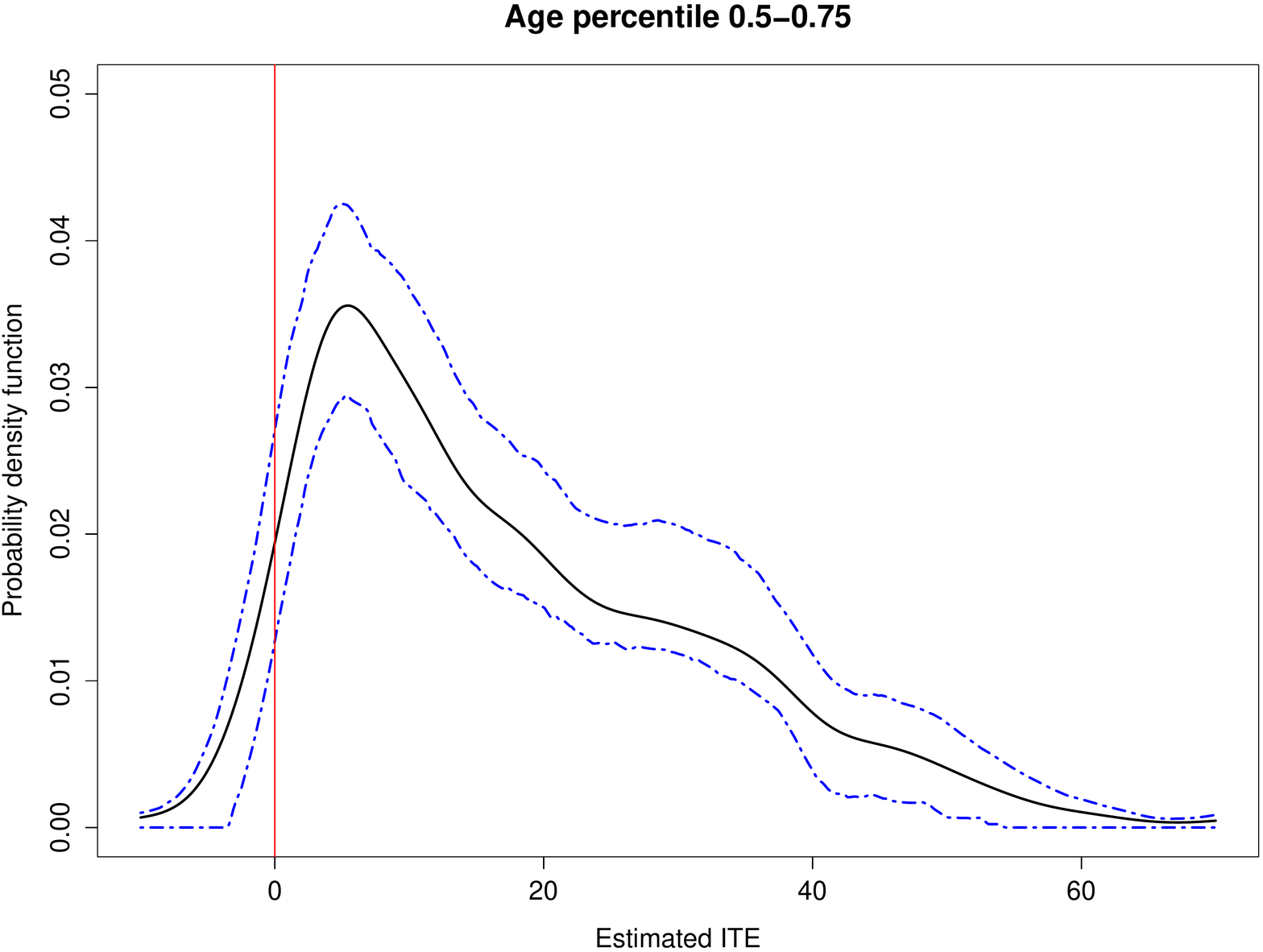}
    \includegraphics[height=2.25in]{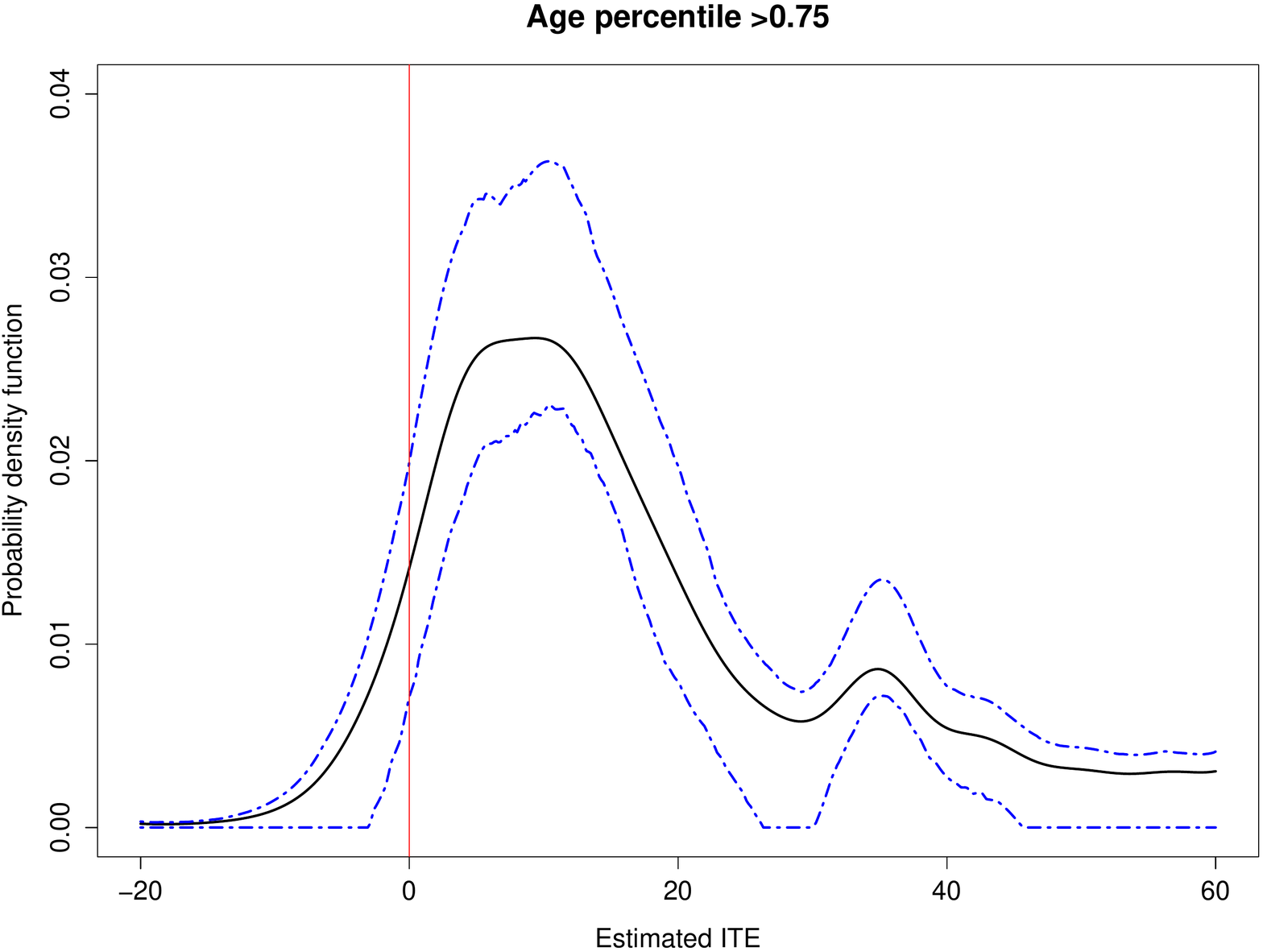}
   \caption{Estimated densities of ITE by age category}
   \label{ite_age}
\end{figure}

\begin{figure}[h] 
   \centering
   \includegraphics[height=2.25in]{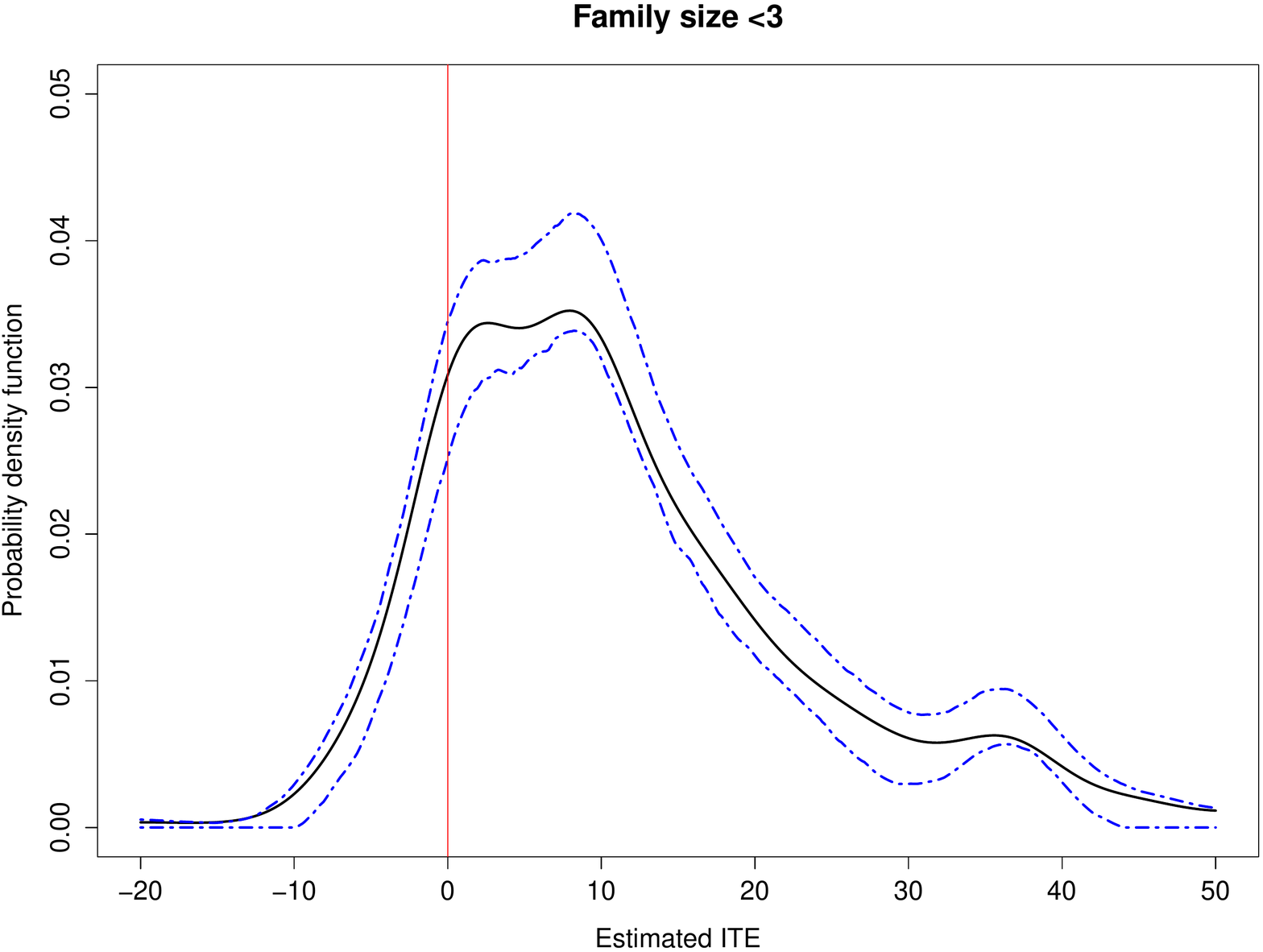}
   \includegraphics[height=2.25in]{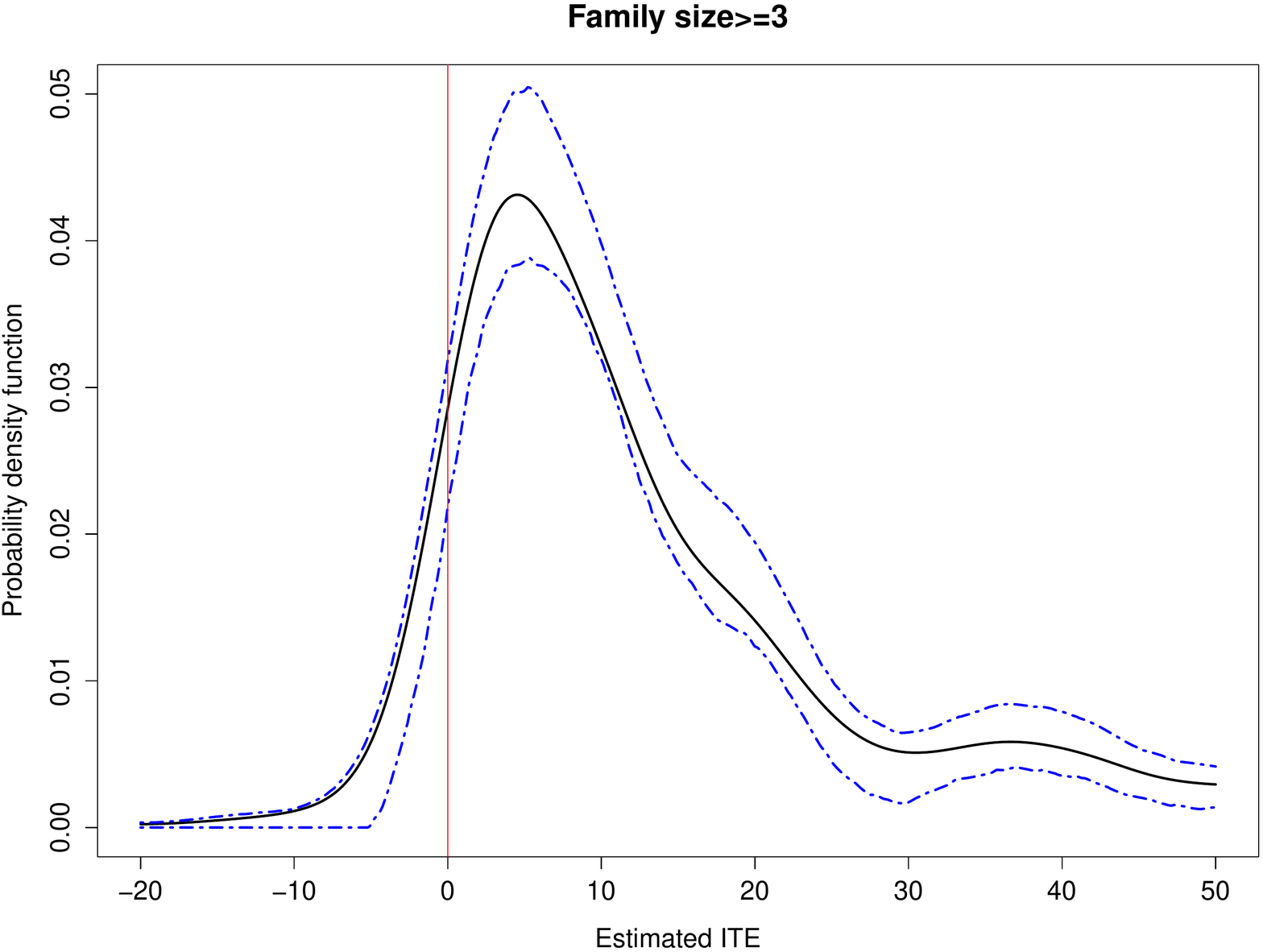}\\
   \caption{Estimated densities of ITE by family size category}
   \label{ite_fsize}
\end{figure}

\begin{figure}[h] 
   \centering
   \includegraphics[height=2.25in]{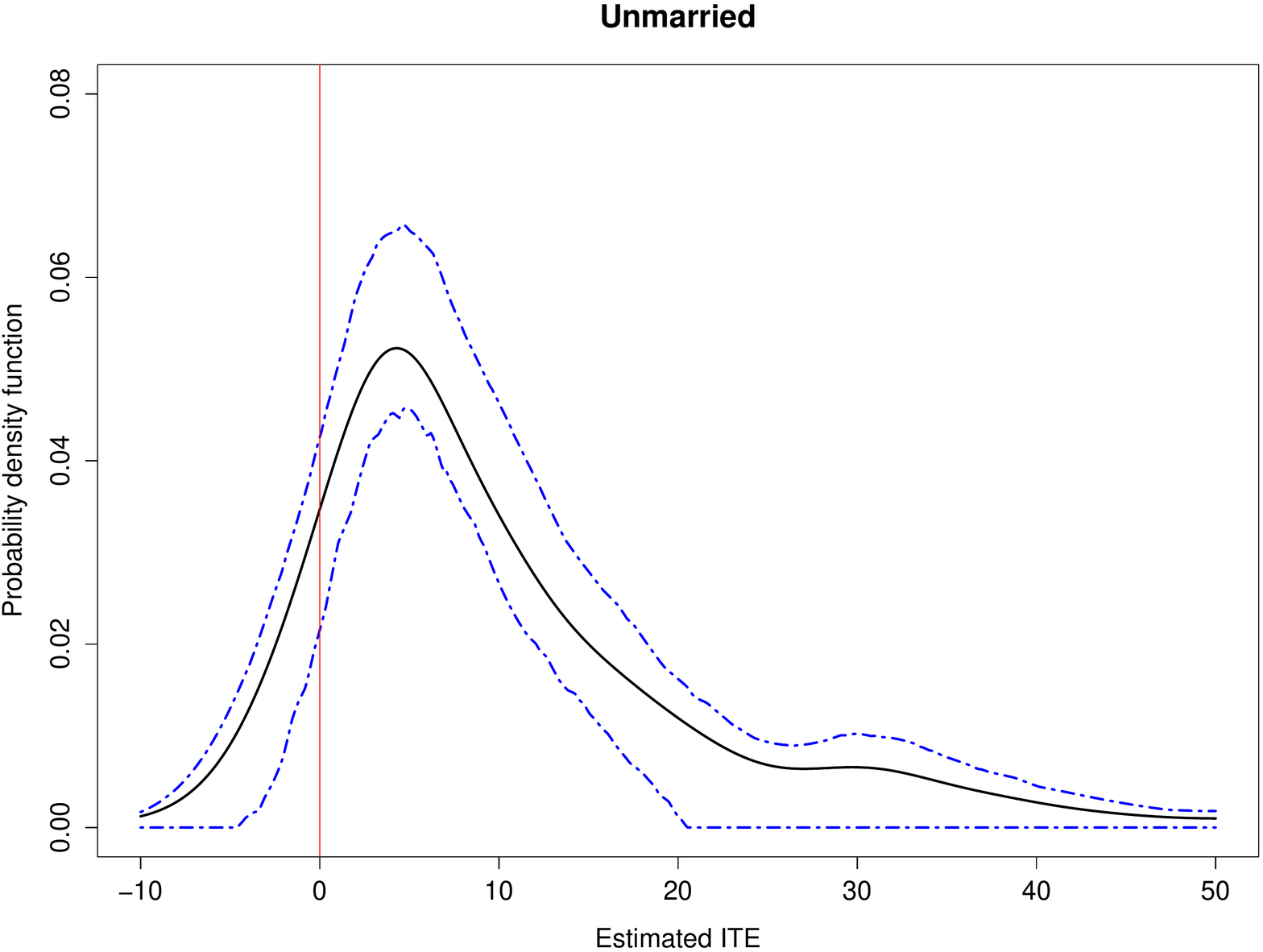}
   \includegraphics[height=2.25in]{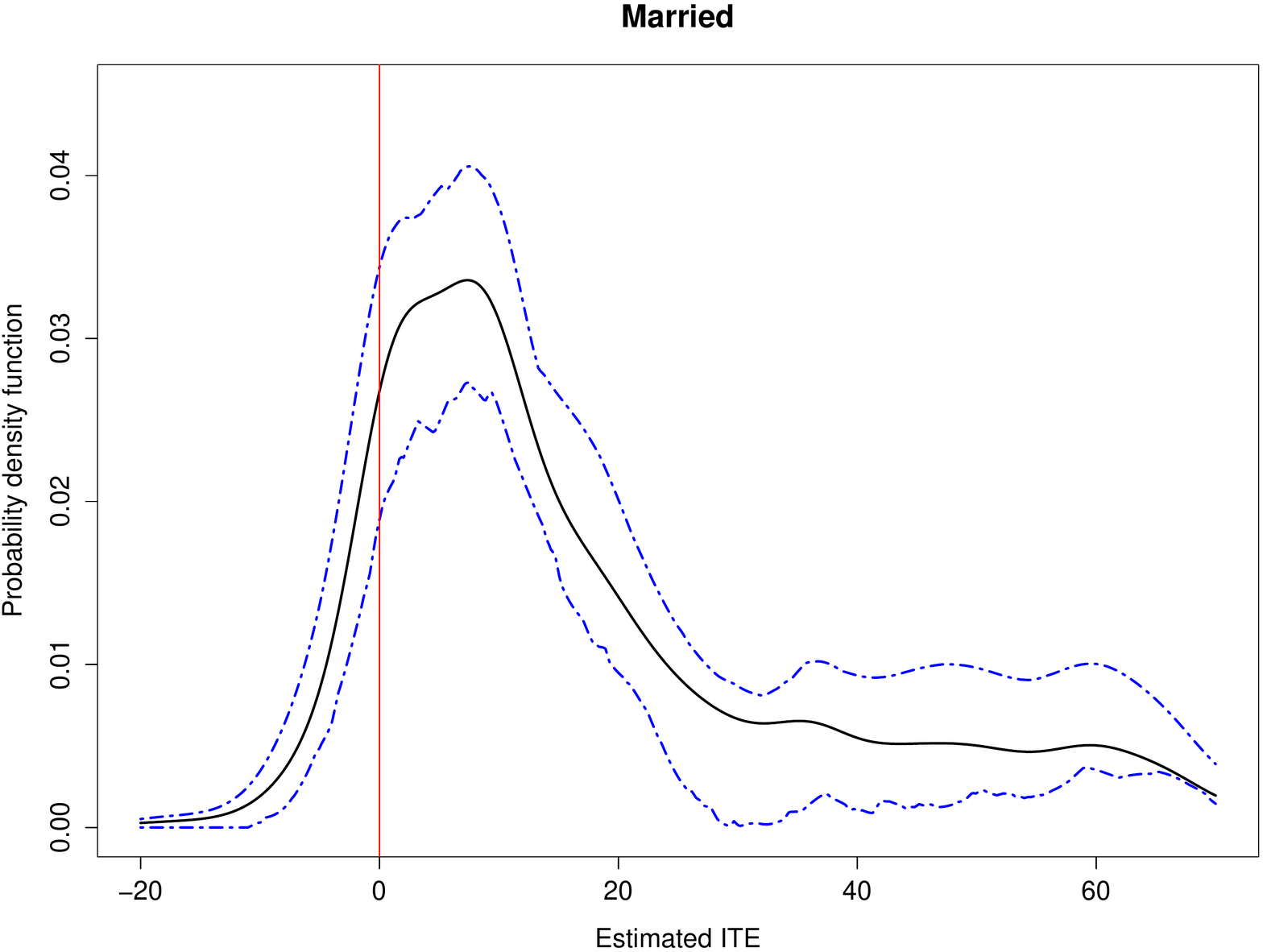}\\
   \caption{Estimated densities of ITE by marital status category}
   \label{ite_mstatus}
\end{figure}

\Cref{full_sample} provides the ITE density estimates for the full sample along with $95\%$ pointwise bootstrap confidence intervals. The participation effects of 401(k) on net financial assets are distributed on the interval [-\$10k, \$60k], with a mode around  \$4k.   As the bootstrap confidence intervals indicate, the ITE density is quite well-estimated.
\Cref{ite_income,ite_age,ite_fsize,ite_mstatus} plot the ITE density estimates conditional on income, age, family size and family status, separately. In particular, the ITE density given income shifts to the right with a slight increase in variance as income increases, revealing that ITEs for individuals with high income is larger though more heterogeneous than for those whose income are low.  Thus, the benefits from participating to  401(k) retirement programs on personal savings increase as Family Income increases. Though not as pronounced, the same trend is found when conditioning on age, family size and family status.

A striking feature of \Cref{full_sample,ite_income,ite_age,ite_fsize,ite_mstatus} is that there exists a small but statistically significant proportion (about 8.77\% in the full sample) of individuals who experience negative effects, although the majority of ITEs is positive.\footnote{For such an empirical evidence, one could investigate it alternatively by using the (conditional) quantile treatment effects for the complier group \citep[see e.g.][]{abadie2002instrumental,froelich2013unconditional} at low quantiles. We thank Isaiah Andrews for this point.} This is especially the case for young individuals (age percentile below 0.25) where such a proportion is 15.93\%.
Such a finding is new. In particular, \Cref{negITE} provides the summary statistics of the subgroup with negative ITEs, compared with the subgroup with positive ITE and the entire sample. Individuals with negative ITEs are more likely to be younger, single, and from smaller families with lower family income. 
A puzzling feature is that the subgroup with negative ITEs has a larger FNFA than the rest of the sample, though the large standard error (113.92) indicates a large heterogeneity among this group. Our conjecture is that the majority of this group use their savings to invest aggressively in their own businesses or in financial markets.

\begin{table}[h]\small
\vspace{6pt}
   \centering
      \caption{Summary statistics assorted according to ITE}
   \begin{tabular}{lccc}
\hline\hline
&Negative ITE&Positive ITE& Entire sample\\\hline
Participation in 401(k) &0.2534&0.2784&0.2762\\
&(0.4352)&(0.4482)&(0.4472)\\
FNFA (in thousand \$) &21.9558&18.7946&19.0717\\
&(113.9247)&(56.9039)&(63.9638)\\
Family income &30.5890&40.0872&39.2546\\
(in thousand \$)&(16.8846)&(24.5117)&(24.0900)\\
Age &34.8327&41.6805&41.0802\\
&(9.2949)&(10.1917)&(10.2995)\\
Married &0.5572&0.6354&0.6286\\
&(0.4970)&(0.4813)&(0.4832)\\
Family size &2.6421&2.9084&2.8851\\
&(1.4826)&(1.5280)&(1.5258)\\
Number &813&8,462&9,275\\ \hline
   \end{tabular}
   \label{negITE}
\end{table}


\Cref{tree} uses a classification tree to summarize the benefits and losses of participation decisions  for all individuals in the sample:  Among those who are eligible, 5.67\% of them participate in 401(k) but have negative ITEs, while 27.52\%  do not participate but would benefit from the 401(k) program. There are also 90.55\% of non-eligible individuals who would benefit from the program if they participate.  In monetary terms, the 401(k) program provides an average increase of \$29.62k in FNFA to the 2,356 participants  with positive ITEs and an average decrease of \$19.42k in FNFA to the 206 participants with negative ITEs.  That is a net increase of \$65.7939 million in total in FNFA for the 401(k) program  based on our sample of 9,275 households.  

From \Cref{tree}, about 93.12\% of those who are eligible but do not participate in 401(k) programs have positive ITEs. How should one interpret this empirical evidence?  Do these eligible nonparticipants have low preference for savings, or low ability for managing their financial assets? 
Our ITE estimates show that the average ITE for the group of eligible nonparticipating households is \$40.36k, which is significantly larger than \$25.68k, the average ITE of the participating group. This evidence suggests an adverse selection issue: Households who benefit more are less likely to participate. To shed some light on this second puzzling finding, \Cref{potential_outcome_0} provides density estimates of the potential outcome $\hat{\phi}_{0X}(Y)$ for not participating to the 401(k)  program for the participating group as well as the group of eligible nonparticipants. An interesting feature is that the distribution of participants' counterfactual FNFA (i.e., their savings without participating to 401(k) programs)  are bimodal: Without participating to 401(k) programs, those participants would either do quite well or extremely poorly on their savings. In contrast, for the group of eligible but not participating households, the FNFA conforms to a unimodal distribution.  

Finally,  we can consider the following counterfactuals: Given that we recover the ITE for each individual,  we can entertain a situation in which  each eligible individual chooses his/her best option regarding participation.  The 401(k) program would lead to a total increase of \$116.4681 million in FNFA coming from the  2,356 eligible households  with positive ITEs and the 1,001 eligible households with positive ITEs who did not participate.  In addition, if the 401(k) program was available to all households, under the same scenario where each household is perfectly informed and make the correct decision, the 401(k) program will gain an additional  \$120.8375 million in FNFA due to those 5,105 non-eligible households with positive ITEs.  This would lead to the maximum gain of \$237.3056 million in FNFA for the 401(k) program from the 9,275 households in our sample.

 \begin{figure}
  \caption{ Classification Tree for 401(k) Participation Decisions}
 {\small
\begin{center}
\begin{tikzpicture} [scale=1.8,>=stealth]

\draw [fill=white,blue] (1,1) circle [radius=0.03];  
\draw node [] at (1,1.2){Whole sample: $n=9,275$};
\draw[thick] (1,1) to (0,0); 
\draw[thick] (1,1) to (2,0); 
\draw [fill=white,blue] (0,0) circle [radius=0.03];  
\draw node [] at (-.5,0.1){Eligible};
\draw node [] at (-.5,-0.15){$3,637$};
\draw [fill=white,blue] (2,0) circle [radius=0.03];  
\draw node [] at (2.65,0.1){Not eligible};
\draw node [] at (2.65,-0.15){$5,638$};
\draw[thick] (0,0) to (-1,-1); 
\draw[thick] (0,0) to (1,-1); 
\draw[thick] (2,0) to (3, -1); 
\draw [fill=white,blue] (-1,-1) circle [radius=0.03];  
\draw node [] at (-1.6,-1){Participate};
\draw node [] at (-1.6,-1.25){$2,562$};
\draw [fill=white,blue] (1,-1) circle [radius=0.03];  
\draw node [] at (1.4,-0.6){Not Participate};
\draw node [] at (1.4,-.85){$1,075$};
\draw [fill=white,blue] (3,-1) circle [radius=0.03];  
\draw node [] at (3.8,-1){Not Participate};
\draw node [] at (3.8,-1.25){$5,638$};

\draw[thick] (-1,-1) to (-2,-2); 


\draw [fill=white,blue] (-2,-2) circle [radius=0.03];  
\draw node [] at (-2,-2.22){ Positive};
\draw node [] at (-2,-2.48){$2,356$};

\draw[thick] (-1,-1) to (-0.5,-2); 

\draw [fill=white,red] (-0.5,-2) circle [radius=0.03];  
\draw node [red] at (-0.7,-2.25){Negative};
\draw node [red] at (-0.7,-2.5){$206$};

\draw[thick] (3,-1) to (2.5,-2); 
\draw[thick] (3,-1) to (4,-2); 

\draw [fill=white,red] (2.5,-2) circle [radius=0.03];  
\draw node [red] at (2.7,-2.22){Positive};
\draw node [red] at (2.7,-2.48){$5,105$};

\draw [fill=white,blue] (4,-2) circle [radius=0.03];  
\draw node [] at (4,-2.22){Negative};
\draw node [] at (4,-2.48){$533$};

\draw[thick] (1,-1) to (0.4,-2); 
\draw[thick] (1,-1) to (1.6,-2); 

\draw [fill=white,red] (0.4,-2) circle [radius=0.03];  
\draw node [red] at (0.4,-2.22){Positive};
\draw node [red] at (0.4,-2.5){$1,001$};

\draw [fill=white,blue] (1.6,-2) circle [radius=0.03];  
\draw node [] at (1.6,-2.22){Negative};
\draw node [] at (1.6,-2.5){$74$};

\end{tikzpicture} 
\end{center}}

\label{tree}

\end{figure}

\begin{figure}[h] 
   \centering
   \includegraphics[height=3.6in]{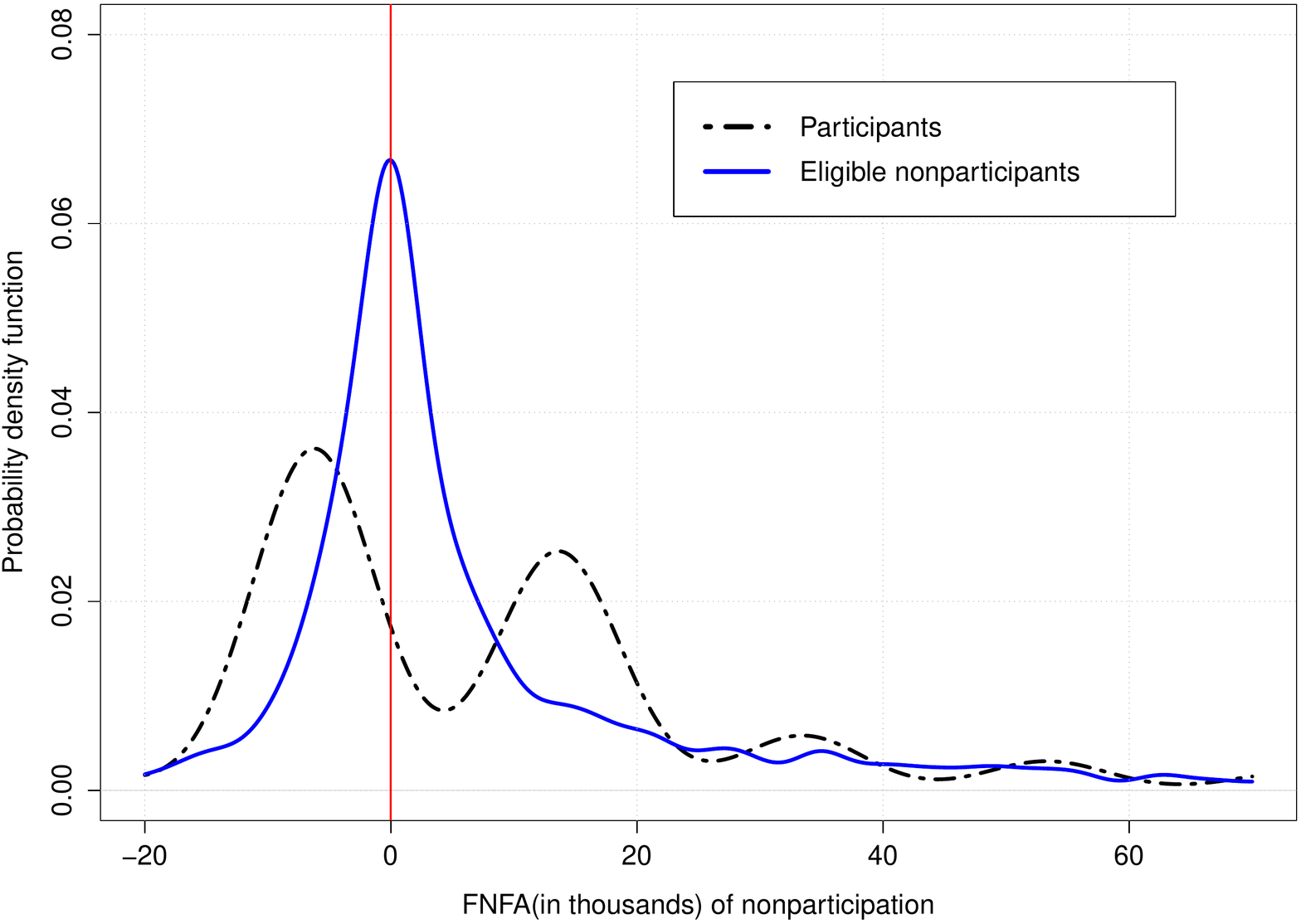}\\
   \caption{Densities of potential outcome of nonparticipation}
   \label{potential_outcome_0}
\end{figure}

\clearpage
\bibliographystyle{econometrica}
\bibliography{bibles_ASF}

\clearpage
\appendix
\small

\section{Proofs} 

\subsection{Proof of \Cref{lemma1}}  \label{prooflemma1}
\proof
First, we differentiate $Q_0(y_0,y_1)$ with respect to $y_0$. Noting that $\partial \mathbb E|W-w|/\partial w = 2F_W(w)-1$ for a continuous distribution $F_W(\cdot)$, we obtain
\begin{eqnarray*}
&&\frac{\partial }{\partial y_0}\mathbb E \big[|Y-y_0| (1-D)|X=x, Z=z\big]\\
&=& \frac{\partial }{\partial y_0}\mathbb E \big(|Y-y_0|\big|D=0,X=x, Z=z\big)
\times\Pr (D=0|X=x,Z=z) \\
&=&2\Pr(Y\leq y_0; D=0|X=x, Z=z)-\Pr(D=0|X=x,Z=z).
\end{eqnarray*} 
Moreover, we have
\begin{multline*}
\mathbb E[ \text{sign}(Y-y_1)\cdot D|X=x,Z=z]\\
= -2 \Pr(Y\leq y_1;D_1=1|X=x, Z=z)+\Pr (D=1|X=x, Z=z).
\end{multline*}
It follows that
 \begin{eqnarray*}
\frac{\partial }{\partial y_0}Q_0(y_0,y_1)&=&2[\Pr(Y\leq y_0; D=0|X=x,Z=0) -\Pr ( Y\leq y_0; D=0|X=x,Z=1) ]\\
&+&2[\Pr(Y\leq y_{1};D=1|X=x,Z=0)-\Pr(Y\leq y_{1};D=1|X=x, Z=1)]\\
&=&2[p(x,1)-p(x,0)]\times [C_{0x}(y_0)-C_{1x}(y_1)],
\end{eqnarray*}
where the last step comes from the definition of $C_{dx}$ in   \eqref{eq4}. Fix $y_1\in\mathbb R$. Note that $C_{0x}(\cdot)$ is weakly increasing on $\mathbb R$ and strictly increasing on $\mathscr C_{dx}^o=\mathscr S^\circ_{Y|D=0,X=x}$ by Theorem 1.  Moreover, because $p(x,0)<p(x,1)$,\footnote{When such a rank of $p(x,z)$ is unknown, we can modify the objective function by $\tilde Q_0(y_0,y_1)=[p(x,1)-p(x,0)]\times Q_0(y_0,y_1)$. The additional term $p(x,1)-p(x,0)$ changes the sign of $\tilde Q_0(\cdot,y_1)$ based on the relative rank of $p(x,z)$ while its scale does not matter for the optimization of $\tilde Q_0(\cdot,y_1)$. } then $Q_0(\cdot,y_1)$ has a weakly and strictly increasing derivative on $\mathbb R$ and $\mathscr C_{dx}^o$, respectively. Therefore, $Q_0(\cdot,y_1)$ is weakly and strictly convex on $\mathbb R$ and $\mathscr C_{dx}^o$, respectively, for arbitrary $y_1\in\mathbb R$. 
Furthermore, if $y_1\in\mathscr S^\circ_{Y|D=1,X=x}$, we have $C_{0x}(y_0)=C_{1x}(y_1)$ if and only if $y_0=\phi_{0x}(y_1)$  by \Cref{th1}. Thus, $y_0=\phi_{0x}(y_1)$ uniquely solves the first--order condition $\frac{\partial}{\partial y_0}Q_0(y_0,y_1)=0$ whenever $y_1\in\mathscr S^\circ_{Y|D=1,X=x}$. A similar argument also applies to the population objective function $Q_1(y_0,\cdot)$.  
\qed

\subsection{Proof of \Cref{theorem2}}  \label{prooftheorem_2}
\proof
Fix $X=x$. All the following argument is conditional on $X=x$. For simplicity, we suppress the dependence on $x$, e.g., we use $\phi_{d}$ for $\phi_{dx}$, omit the term $\mathbb 1(X_i=x)$ in the estimation,  and $X=x$ in the conditional probability $\Pr(Y\leq y; D=d|X=x;Z=z)$.  Moreover, we only show the results for $d=0$. The proof for  the case $d=1$ can be derived similarly. 

First, we  show uniform consistency. By \cite{angrist2006quantile},  it suffices to show that $\sup_{(y_0,y_1)\in\mathcal B}\|\hat Q_0(y_0,y_1)-Q_0(y_0,y_1)\|=o_p(1)$ for any compact set $\mathcal B\subset \mathbb R^2$. By the law of large number, we have  pointwise convergence, i.e., $\|\hat Q_0(y_0,y_1)-Q_0(y_0,y_1)\|=o_p(1)$. Then, it suffices to show the stochastic equicontinuity of the empirical process $\hat \rho_0(\cdot,\cdot;z)-\rho_0(\cdot,\cdot;z)$, which directly follows the general argument in \cite{koenker2002inference}.
Next, we establish the limiting distribution of the process. 

Taking the directional derivative, we have 
\begin{multline*}
\frac{d}{dt} \hat Q_0(y_0+t,y_1)\Big|_{t\downarrow 0}
=\frac{2\sum_{i=1}^n  \mathbb 1 (Y_i\leq y_0;D_i=0;Z_i=0)}{\sum_{i=1}^n \mathbb 1 (Z_i=0)}
-\frac{2\sum_{i=1}^n   \mathbb 1 (Y_i\leq y_0;D_i=0;Z_i=1)}{\sum_{i=1}^n \mathbb 1 (Z_i=1)}\\
+\frac{2\sum_{i=1}^n \mathbb 1 (Y_i\leq y_1; D_i=1;Z_i=0)}{\sum_{i=1}^n \mathbb 1 (Z_i=0)}-\frac{2 \sum_{i=1}^n\mathbb 1 ( Y_i\leq y_1; D_i=1;Z_i=1)}{\sum_{i=1}^n \mathbb 1 (Z_i=1)}+\xi_n(y_0).
\end{multline*}
where the remainder term $\xi_n(y_0)$ is bounded by 
\[
\
\frac{n\cdot \sum_{i=1}^n\mathbb 1 (Y_i=y_0)}{\sum_{i=1}^n \mathbb 1 (Z_i=0)\times \sum_{i=1}^n \mathbb 1 (Z_i=1)}.
\] 
By the computational properties of linear programming in \citet[Theorem 3.3]{koenker1978regression}, we have  $\xi_n(y_0)=O_p(n^{-1})$ uniformly in $y_0\in\mathbb R$. We can derive a similar expression for $\frac{d}{dt} \hat Q_0(y_0-t,y_1)\Big|_{t\downarrow 0}$. Note that $\frac{d}{dt} \hat Q_0(\hat \phi_0(y_1)+t,y_1)\Big|_{t\downarrow 0}\geq 0$ and $\frac{d}{dt} \hat Q_0(\hat \phi_0(y_1)-t,y_1)\Big|_{t\downarrow 0}\geq 0$ as $\hat \phi_0(y_1)$ minimizes $\hat Q_0(\cdot,y_1)$. Hence, we have
\begin{multline*}
\frac{\sum_{i=1}^n  \mathbb 1 (Y_i\leq \hat\phi_0(y_1);D_i=0;Z_i=0)}{\sum_{i=1}^n \mathbb 1 (Z_i=0)}
-\frac{\sum_{i=1}^n   \mathbb 1 (Y_i\leq \hat\phi_0(y_1);D_i=0;Z_i=1)}{\sum_{i=1}^n \mathbb 1 (Z_i=1)}\\
+\frac{\sum_{j=1}^n \mathbb 1 (Y_i\leq y_1; D_i=1;Z_i=0)}{\sum_{i=1}^n \mathbb 1 (Z_i=0)}-\frac{ \sum_{j=1}^n\mathbb 1 ( Y_i\leq y_1; D_i=1;Z_i=1)}{\sum_{i=1}^n \mathbb 1 (Z_i=1)}=O_p(n^{-1})
\end{multline*}
uniformly in $y_1$. 

Following the convention, we introduce some notation from the empirical process literature: For $W=(Y,D,Z)'$ and a generic function $g$, let $\mathbb E_n [g(W)]=n^{-1}\sum_{i=1}^n g(W_i)$ and $\mathbb G_n[g(W)]=n^{-1/2}\sum_{i=1}^n \big\{g(W_i)-\mathbb E[g(W_i)]\big\}$. 
Hence, the above condition can be rewritten as
\begin{multline*}
\sqrt n\left\{ \frac{\mathbb E_n\mathbb 1 (Y\leq \hat\phi_0(y_1); D=0; Z=0)}{\mathbb E_n \mathbb 1 (Z=0)}+ \frac{ \mathbb E_n \mathbb 1 (Y\leq y_1; D=1;Z=0)}{\mathbb E_n \mathbb 1 (Z=0)}\right\}\\
-\sqrt n\left\{ \frac{\mathbb E_n\mathbb 1 (Y\leq \hat\phi_0(y_1); D=0; Z=1)}{\mathbb E_n \mathbb 1 (Z=1)}+ \frac{\mathbb E_n \mathbb 1 (Y\leq y_1; D=1;Z=1)}{\mathbb E_n \mathbb 1 (Z=1)}\right\}=o_p(1)
\end{multline*}
uniformly in $y_1\in\mathbb R$. It follows that
\begin{multline}
\label{proof_eq_1}
\frac{\sqrt n \ \mathbb E\Big\{ \mathbb 1 (Y\leq \hat\phi_0(y_1); D=0;Z=0)+\mathbb 1 (Y\leq y_1; D=1;Z=0)\Big\}}{\mathbb E_n\mathbb 1 (Z=0)}\\
-\frac{\sqrt n\ \mathbb E\Big\{ \mathbb 1 (Y\leq \hat\phi_0(y_1); D=0;Z=1)+\mathbb 1 (Y\leq y_1; D=1;Z=1)\Big\}}{\mathbb E_n\mathbb 1 (Z=1)}\\
+ \frac{\mathbb G_n\left[\mathbb 1 (Y\leq \hat\phi_0(y_1); D=0; Z=0)+ \mathbb 1 (Y\leq y_1; D=1;Z=0)\right]}{\mathbb E_n \mathbb 1 (Z=0)}\\
- \frac{\mathbb G_n\left[\mathbb 1 (Y\leq \hat\phi_0(y_1); D=0; Z=1)+ \mathbb 1 (Y\leq y_1; D=1;Z=1)\right]}{\mathbb E_n \mathbb 1 (Z=1)}=o_p(1).
\end{multline}
Because $\mathbb E_n \mathbb 1 (Z=z)=\Pr(Z=z)+O_p(n^{-1/2})$, then by Taylor expansion, 
\[
\frac{1}{\mathbb E_n \mathbb 1 (Z=z)}= \frac{1}{\Pr(Z=z)}-\frac{1}{\Pr^2(Z=z)}\times [\mathbb E_n \mathbb 1 (Z=z)-\Pr(Z=z)] + O_p(n^{-1}).
\]
Thus, 
\begin{eqnarray*}
&&\frac{\sqrt n\ \mathbb E\Big\{ \mathbb 1 (Y\leq \hat\phi_0(y_1); D=0;Z=z)+\mathbb 1 (Y\leq y_1; D=1;Z=z)\Big\}}{\mathbb E_n\mathbb 1 (Z=z)}\\
&=&\sqrt n\ \mathbb E\big\{ \mathbb 1 (Y\leq \hat\phi_0(y_1); D=0)+\mathbb 1 (Y\leq y_1; D=1)\big|Z=z\big\}\\
&-& \mathbb E\Big\{ \mathbb 1 (Y\leq \hat \phi_0(y_1); D=0)+\mathbb 1 (Y\leq y_1; D=1)\big|Z=z\Big\}\times \frac{\mathbb G_n\mathbb 1 (Z=z)}{\Pr(Z=z)}+o_p(1)\\
&=&\sqrt n \ \mathbb E\big\{ \mathbb 1 (Y\leq \hat\phi_0(y_1); D=0)+\mathbb 1 (Y\leq y_1; D=1)\big|Z=z\big\}\\
&-& \mathbb E\Big\{ \mathbb 1 (Y\leq  \phi_0(y_1); D=0)+\mathbb 1 (Y\leq y_1; D=1)\big|Z=z\Big\}\times\frac{\mathbb G_n\mathbb 1 (Z=z)}{\Pr(Z=z)}+o_p(1)
\end{eqnarray*}
where the last $o_p(1)$ term is uniform in $y_1$ due to  the uniform convergence of $\hat\phi_0$ to $\phi_0$.

Let $\varphi(\cdot,y_1)=\mathbb 1 (Y\leq \cdot; D=0)+ \mathbb 1 (Y\leq y_1; D=1)$. 
Therefore, \eqref{proof_eq_1} implies 
\begin{eqnarray*}
&&\sqrt n\ \mathbb E\big[\varphi(\hat\phi_0(y_1),y_1)|Z=0\big]-\sqrt n\ \mathbb E\big[\varphi(\hat\phi_0(y_1),y_1)|Z=1\big]\\
&=&- \frac{\mathbb G_n\left[\varphi(\hat \phi_0(y_1),y_1)\times \mathbb 1(Z=0)\right]}{\mathbb E_n \mathbb 1 (Z=0)}+ \frac{\mathbb G_n\left[\varphi(\hat\phi_0(y_1),y_1)\times \mathbb 1 (Z=1)\right]}{\mathbb E_n \mathbb 1 (Z=1)}\\
&+& \frac{\mathbb E \left[\varphi(\phi_0(y_1),y_1)|Z=0\right]}{\Pr (Z=0)} \times \mathbb G_n\mathbb 1 (Z=0)- \frac{\mathbb E \left[\varphi(\phi_0(y_1),y_1)|Z=1\right]}{\Pr (Z=1)} \times \mathbb G_n\mathbb 1 (Z=1)+o_p(1).
\end{eqnarray*}
Note that $\mathbb E  \left[\varphi(\phi_0(y_1),y_1)|Z=z\right]=R_1(y_1)$ which does not depend on $z$. Hence, 
\begin{multline*}
\sqrt n\ \mathbb E\big[\varphi(\hat\phi_0(y_1),y_1)|Z=0\big]-\sqrt n\ \mathbb E\big[\varphi(\hat\phi_0(y_1),y_1)|Z=1\big]\\
=- \frac{\mathbb G_n\left[\varphi(\hat \phi_0(y_1),y_1)\times \mathbb 1(Z=0)\right]}{\mathbb E_n \mathbb 1 (Z=0)}+ \frac{\mathbb G_n\left[\varphi(\hat\phi_0(y_1),y_1)\times \mathbb 1(Z=1)\right]}{\mathbb E_n \mathbb 1 (Z=1)}\\
+ \frac{R_1(y_1)}{\Pr (Z=0)}\times \mathbb G_n \mathbb 1 (Z=0) - \frac{R_1(y_1)}{\Pr (Z=1)}\times \mathbb G_n \mathbb 1 (Z=1)+o_p(1).
\end{multline*}
Moreover, the derivative of $\mathbb E  \left[\varphi(\cdot,y_1)|Z=z\right]$ is  the derivative of $\Pr(Y\leq \cdot; D=0|Z=z)$.  Thus, using \eqref{eq4} and the definition of $c^*_{dx}(\cdot)$, a Taylor expansion gives 
\[
\sqrt n\ \mathbb E\big[\varphi(\hat\phi_0(y_1),y_1)|Z=0\big]-\sqrt n \ \mathbb E\big[\varphi(\hat\phi_0(y_1),y_1)|Z=1\big]
=c^*_0(\tilde \phi_0(y_1))\times \sqrt n\ [\hat \phi_0(y_1)-\phi_0(y_1)]
\]
where $\tilde \phi_0(y_1)$ is between $\phi_0(y_1)$ and $\hat \phi_0(y_1)$. Note that $c^*_0(\tilde \phi_0(y_1))=c^*_0(\phi_0(y_1))+o_p(1)$ uniformly in $y_1$.  It follows that
\begin{multline*}
[c^*_0(\phi_0(y_1))+o_p(1)]\times \sqrt n \ [\hat \phi_0(y_1)-\phi_0(y_1)]\\
=- \frac{\mathbb G_n\left[\varphi(\hat \phi_0(y_1),y_1)\times \mathbb 1 (Z=0)\right]}{\mathbb E_n \mathbb 1 (Z=0)}+ \frac{\mathbb G_n\left[\varphi(\hat\phi_0(y_1),y_1)\times \mathbb 1 (Z=1)\right]}{\mathbb E_n \mathbb 1 (Z=1)}\\
+ \frac{R_1(y_1)}{\Pr (Z=0)}\times \mathbb G_n\left[\mathbb 1 (Z=0)\right]- \frac{R_1(y_1)}{\Pr (Z=1)}\times \mathbb G_n\left[\mathbb 1 (Z=1)\right]+o_p(1).
\end{multline*} 
Because $\varphi$ is Donsker, by the empirical process theorem \citep[see e.g.][]{van1996weak}, we have the equicontinuity of the function class  $\varphi(\cdot,\cdot)$. Hence, uniformly in $y_1$,
\[
\mathbb G _n\left[\varphi(\hat \phi_0(y_1),y_1)\times \mathbb 1(Z= z)\right]= \mathbb G _n\left[\varphi( \phi_0(y_1),y_1)\times \mathbb 1(Z= z)\right]+o_p(1),
\]
which converges to a zero-mean Gaussian process. 
Thus, we obtain
\begin{multline}
\label{eq9}
[c^*_0(\phi_0(y_1))+o_p(1)]\times \sqrt n \ [\hat \phi_0(y_1)-\phi_0(y_1)]\\
=- \mathbb G_n\left\{ \left[\varphi(\phi_0(y_1),y_1)-R_1(y_1)\right]\times \left[\frac{\mathbb 1 (Z=0)}{\Pr (Z=0)}-\frac{\mathbb 1 (Z=1)}{\Pr (Z=1)}\right]\right\}
+o_p(1)
\end{multline}
where the right--hand side  converges  to a zero-mean Gaussian process. Therefore,  
$c^*_0(\phi_0(\cdot))\times \sqrt n [\hat \phi_0(\cdot)-\phi_0(\cdot)]$ converges in distribution to a zero-mean Gaussian process.  

Its covariance kernel $\Sigma_0(y,y')$ for $y\leq y'$ is obtained as
\begin{eqnarray*}
\Sigma_0(y,y') 
&=& \mathbb E\left\{ \left[\varphi(\phi_0(y),y)-R_1(y)\right] 
\times \left[\varphi(\phi_0(y'),y')-R_1(y')\right] \times \left[\frac{\mathbb 1 (Z=0)}{\Pr (Z=0)}-\frac{\mathbb 1 (Z=1)}{\Pr (Z=1)}\right]^2 \right\} \\
&=& \mathbb E\left\{ \left[\varphi(\phi_0(y),y)-R_1(y)\right] 
\times \varphi(\phi_0(y'),y') \times \left[\frac{\mathbb 1 (Z=0)}{\Pr (Z=0)}-\frac{\mathbb 1 (Z=1)}{\Pr (Z=1)}\right]^2 \right\} \\
&=& \mathbb E\left\{ \left[\varphi(\phi_0(y),y) - R_1(y)\varphi(\phi_0(y'),y')\right] \times \left[\frac{\mathbb 1 (Z=0)}{\Pr (Z=0)}-\frac{\mathbb 1 (Z=1)}{\Pr (Z=1)}\right]^2 \right\} \\
&=&  \left[ R_1(y) - R(y)R_1(y') \right] \times\mathbb E \left[\frac{\mathbb 1 (Z=0)}{\Pr (Z=0)}-\frac{\mathbb 1 (Z=1)}{\Pr (Z=1)}\right]^2 
\end{eqnarray*}
where the second and third equalities use the definition of $\varphi(\phi_0(y_1),y_1)$, and the fourth equality uses $\mathbb E  \left[\varphi(\phi_0(y_1),y_1)|Z=z\right]=R_1(y_1)$.  The expression for $\Sigma_0(y,y')$ given in the theorem follows upon noting that 
\[
\mathbb E\left[ \left(\frac{\mathbb 1 (Z=0)}{\Pr (Z=0)}-\frac{\mathbb 1 (Z=1)}{\Pr (Z=1)}\right)^2 \right]=\frac{1}{\Pr (Z=0)\Pr (Z=1)}.\qed
\]

\vspace{6pt}
\subsection{Proof of \Cref{theorem3}}  \label{prooftheorem_3}
\proof
We have  $\hat f_{\Delta}(\delta)- f_{\Delta}(\delta) = [\hat f_{\Delta}(\delta)- \tilde f_{\Delta}(\delta)] + \tilde f_{\Delta}(\delta)- f_{\Delta}(\delta)$, where
\[
\tilde f_{\Delta}(\delta)= \frac{1}{nh}\sum_{i=1}^n K\big(\frac{ \Delta_i-\delta}{h}\big), \ \ \forall \delta\in [\underline{\delta}+h,\overline{\delta}-h],
\]
is the infeasible kernel estimator of $f_\Delta(\delta)$. From standard kernel estimation, we have 
\[
\sup_{\delta \in [\underline \delta+h, \overline \delta-h]}|\tilde f_{\Delta}(\delta)-f_{\Delta}(\delta)|=O_p\big(h^P\big)
\]
since $h=(\ln n/n)^{\frac{1}{2P+2}}$ leads to oversmoothing.  Thus, it suffices to show that the same uniform convergence rate holds for $|\hat f_{\Delta}(\delta)- \tilde f_{\Delta}(\delta)|$.  We actually show that
\[
\sup_{\delta \in [\underline \delta+h, \overline \delta-h]}|\hat f_{\Delta}(\delta)-\tilde f_{\Delta}(\delta)|=o_p\big(h^P\big)
\]
so that the first step estimation error is negligible given our choice of bandwidth.

From a second-order Taylor expansion we have
\[
\hat f_\Delta(\delta) - \tilde f_\Delta(\delta)=\frac{1}{nh^2}\sum_{i=1}^n K'\big(\frac{ \Delta_i-\delta}{h}\big) (\hat \Delta_i-\Delta_i) + \frac{1}{2nh^3}\sum_{i=1}^n K''\big(\frac{ \Delta^\dag_i-\delta}{h}\big)(\hat \Delta_i-\Delta_i)^2
\]
where $\Delta^\dag_i$ is between $\hat \Delta_i$ and $\Delta_i$. Since $\sup_i |\hat \Delta_i-\Delta_i|= O_p(n^{-1/2})$ from \Cref{theorem3}, we have
\[
\Big|\frac{1}{nh^2}\sum_{i=1}^n K'\big(\frac{ \Delta_i-\delta}{h}\big) (\hat \Delta_i-\Delta_i)\Big|\leq O_p(n^{-\frac{1}{2}} h^{-1} )\times \frac{1}{nh}\sum_{i=1}^n \left|K'\big(\frac{ \Delta_i-\delta}{h}\big)\right|
\]
where the summation is a nonparametric estimator of $f_\Delta(\delta)\times \int |K'(u)| du$. Therefore, 
\[
\frac{1}{nh^2}\sum_{i=1}^n K'\big(\frac{ \Delta_i-\delta}{h}\big) (\hat \Delta_i-\Delta_i)= O_p(n^{-\frac{1}{2}} h^{-1}) = O_p(h^P/(\ln n)^{1/2})
\]
which is an $o_p\big(h^P\big)$.  Furthermore, because $K''$ is bounded, we have
\[
\Big|\frac{1}{nh^3}\sum_{i=1}^n K''\big(\frac{ \Delta^\dag_i-\delta}{h}\big) (\hat \Delta_i-\Delta_i)^2\Big|=O_p(n^{-1}h^{-3})
\]
which is also an $o_p\big(h^P\big)$ provided $P\geq 1$. Therefore, the first-step estimation error is negligible. 
\qed

\end{document}